\documentclass{aa}
\usepackage{mathrsfs}
\usepackage{amsmath,amsfonts,amssymb}
\usepackage{graphicx} 
\usepackage{xspace}
\usepackage{booktabs}
\usepackage{txfonts}
\usepackage{multirow}
\usepackage{array}
\usepackage[nolist,nohyperlinks]{acronym}
\usepackage{orcidlink}

\newcommand{\afterglowpy}{\textsc{afterglowpy}\xspace}
\newcommand{\pbag}{\textsc{pyblastafterglow}\xspace}
\newcommand{\fiesta}{\textsc{fiesta}\xspace}
\newcommand{\NMMA}{\textsc{nmma}\xspace}
\newcommand{\flowMC}{\textsc{flowMC}\xspace}
\newcommand{\jax}{\textsc{jax}\xspace}
\newcommand{\jim}{\textsc{jim}\xspace}
\newcommand{\jester}{\textsc{jester}\xspace}
\newcommand{\possis}{\textsc{possis}\xspace}

\begin{document}
\title{Efficient Bayesian analysis of kilonovae and gamma ray burst afterglows with \fiesta}

\author{H. Koehn\inst{1}\orcidlink{0009-0001-5350-7468}
        \and
        T. Wouters\inst{2,3}\orcidlink{0009-0006-2797-3808}
        \and
        P. T. H. Pang\inst{3,2}\orcidlink{0000-0001-7041-3239}
        \and
        M. Bulla\inst{4,5,6}\orcidlink{0000-0002-8255-5127}
        \and
        H. Rose\inst{1}\orcidlink{0009-0009-2025-8256}
        \and
        H. Wichern\inst{7}\orcidlink{0009-0004-1442-619X}
        \and
        T. Dietrich\inst{1,8}\orcidlink{0000-0003-2374-307X}
          }

   \institute{Institut f\"ur Physik und Astronomie, Universit\"at Potsdam, Haus 28, Karl-Liebknecht-Str. 24/25, 14476, Potsdam, Germany,  
   \email{\href{mailto:hauke.koehn@uni-potsdam.de}{hauke.koehn@uni-potsdam.de}}
         \and
    Institute for Gravitational and Subatomic Physics (GRASP), Utrecht University, Princetonplein 1, 3584 CC Utrecht, The Netherlands,  
    \email{\href{mailto:t.r.i.wouters@uu.nl}{t.r.i.wouters@uu.nl}}
        \and
    Nikhef, Science Park 105, 1098 XG Amsterdam, The Netherlands
        \and
    Department of Physics and Earth Science, University of Ferrara, via Saragat 1, I-44122 Ferrara, Italy
        \and
        INFN, Sezione di Ferrara, via Saragat 1, I-44122 Ferrara, Italy
        \and
    INAF, Osservatorio Astronomico d’Abruzzo, via Mentore Maggini snc, 64100 Teramo, Italy 
        \and
    DTU Space, National Space Institute, Technical University of Denmark, Elektrovej 327/328, DK-2800 Kongens Lyngby, Denmark
        \and
    Max Planck Institute for Gravitational Physics (Albert Einstein Institute), Am M{\"u}hlenberg 1, Potsdam 14476, Germany
             }

   \date{Received 28 July 2025 / Accepted 20 October 2025}
 
\abstract{Gamma-ray burst (GRB) afterglows and kilonovae (KNe) are electromagnetic transients that can accompany binary neutron star (BNS) mergers. 
Therefore, studying their emission processes is of general interest for constraining cosmological parameters or the behavior of ultra-dense matter.
One common method to analyze electromagnetic data from BNS mergers is to sample a Bayesian posterior over the parameters of a physical model for the transient.
However, sampling the posterior is computationally costly and because of the many likelihood evaluations required in this process, detailed models are too expensive to be used directly in Bayesian inference.
In this paper, we address the problem by introducing \fiesta, a \textsc{python} package to train machine learning (ML) surrogates for GRB afterglow and kilonova models that have the capacity to accelerate likelihood evaluations.
Specifically, we introduce extensive ML surrogates for the state-of-the-art GRB afterglow models \textsc{afterglowpy} and \textsc{pyblastafterglow}, along with a new surrogate for KN emission based on the \possis code.
Our surrogates enable evaluation of the light-curve posterior within minutes.
We also provide built-in posterior sampling capabilities in \fiesta that rely on the \flowMC package, which  efficiently scale to higher dimensions when adding up to tens of nuisance sampling parameters.
Because of its use of the \jax framework, \fiesta also allows for GPU acceleration during both surrogate training and posterior sampling. 
We applied our framework to reanalyze AT2017gfo/GRB170817A and GRB211211A with our surrogates, thus employing the new \pbag model for the first time in Bayesian inference.}


\maketitle

\section{Introduction} \label{sec:intro}

Gamma-ray bursts (GRBs) and kilonovae (KNe) are types of electromagnetic transients that can originate from a binary neutron star (BNS) merger, as confirmed most prominently by the observation of AT2017gfo~\citep{J-GEM:2017tyx, Coulter:2017wya, Andreoni:2017ppd, Shappee:2017zly, DES:2017kbs, Lipunov:2017dwd, Valenti:2017ngx, TOROS:2017pqe, Tanvir:2017pws} and GRB170817A~\citep{Goldstein:2017mmi, LIGOScientific:2017zic, Savchenko:2017ffs} in the wake of the gravitational wave (GW) event GW170817~\citep{LIGOScientific:2017vwq, LIGOScientific:2017ync}. 
Additionally, several GRBs have been associated with potential subsequent KNe~\citep{Tanvir:2013pia, Ascenzi:2018mbh, Troja:2019ccb, Rastinejad:2022zbg, Troja:2018ybt, JWST:2023jqa, Yang:2023mqt, Levan:2023qdh, Stratta:2024kbs}.
To analyze these electromagnetic transients with Bayesian inference, a physical model that links the source and observational parameters of the system (e.g., ejecta masses, isotropic energy equivalent of the jet, observation angle) to the observed data is needed. 
This model can then be employed in a sampling procedure to obtain a posterior distribution.

The KN emission arises from quasi-thermal radiation produced by the BNS ejecta heated from the radioactive decay of nuclei synthesized through the r-process~\citep{Thielemann:2011} and is observed on the timescale of days after the merger.
This process has been investigated through various methods and approaches in the literature~\citep[e.g.,][]{Kasen:2017sxr, Villar:2017wcc, Kawaguchi:2018ptg, Metzger:2019zeh, Breschi:2021tbm, Wollaeger:2021qgf, Curtis:2021guz, Nicholl:2021rcr, Bulla:2022mwo}.
The GRB prompt emission, on the other hand, takes place on scales of seconds to minutes and is observed as bursts of highly energetic gamma and X-ray radiation. 
Its emission mechanism is largely uncertain and, thus, the associated data are typically not taken into account for Bayesian inference of BNS mergers. 
However, the prompt emission is followed by an afterglow of broadband radiation spanning from radio to $\gamma$-ray frequencies, observable on timescales ranging from minutes to years~\citep{Miceli:2022efx}.
Unlike the GRB prompt emission, the afterglow physics are comparatively well understood and can thus be used to infer properties of the jet and the progenitor.
Specifically, the afterglow emission arises from the interaction of the GRB jet with the surrounding cold interstellar medium and various afterglow models are available in the literature~\citep[e.g.,][]{vanEerten:2011yn, Ryan:2014nea, Lamb:2018ohw, Ryan:2019fhz, Zhang:2020tem, Pellouin:2024gqj, Wang:2024wbt, Nedora:2024vrv, Wang:2025ccz}.

Such models have been successfully applied in joint Bayesian analysis of GW170817, its KN, and the GRB afterglow to place constraints on the equation of state (EOS) for neutron star (NS) matter~\citep{Radice:2017lry, Dietrich:2020efo, Raaijmakers:2021uju, Nicholl:2021rcr, Pang:2022rzc, Guven:2020dok, Annala:2021gom, Breschi:2024qlc, Koehn:2024set} or to determine the Hubble constant~\citep{LIGOScientific:2017adf, Hotokezaka:2018dfi, Dietrich:2020efo, Mukherjee:2019qmm, Wang:2020vgr, Gianfagna:2023cgk}.
However, these joint multi-messenger inferences pose certain computational challenges, because exploring their high-dimensional parameter space requires many likelihood evaluations.
Therefore, a single likelihood evaluation needs to be as cheap as possible in order to keep the total sampling time manageable. 
However, to evaluate the likelihood function at a given parameter point, we have to determine the expected emission for these given parameters from the physical model.
Due to the cost limit on the likelihood function, any direct, physical calculation of the emission on-the-fly is only viable with computationally cheap, semi-analytical models. 
Yet, even relatively efficient models can become prohibitive, when considering multi-messenger inferences of BNS mergers signals, where GW and electromagnetic signals are analyzed jointly in a large parameter space. 
Accordingly, applying more expensive and involved models in multi-messenger inference necessitates the development of accurate surrogate models, often through machine learning (ML) techniques to enable likelihood evaluation at sufficient speeds~\citep{Pang:2022rzc, Almualla:2021znj, Kedia:2022onl, Ristic:2021ksz, Boersma:2022qtg}.
Furthermore, joint multi-messenger BNS inferences also provide motivation for the GPU compatible light-curve surrogates, since the analysis of BNS GW data can be significantly accelerated on GPU hardware~\citep{Wysocki:2019grj, Wouters:2024oxj, Hu:2024lrj, Dax:2024mcn}.

In the present article, we introduce \fiesta, a \jax-based \textsc{python} package for training ML surrogates of KN and GRB afterglow models and for the Bayesian analysis of photometric transient light curves.
With \fiesta, we provide extensive ML surrogates that effectively replace the costly evaluation of the physical base model, enabling rapid prediction of the expected light curve given the model's parameters.
Specifically, we present surrogates that are built upon GRB afterglow models from the popular \afterglowpy model~\citep{Ryan:2019fhz} and the recently developed \pbag~\citep{Nedora:2024vrv}.
Additionally, we introduce a new KN surrogate based on the 3D Monte Carlo radiation transport code \possis~\citep{Bulla:2019muo, Bulla:2022mwo}.
In contrast to many previous works, our surrogates have not been trained on the magnitudes in a specific photometric passband; instead, they predict the entire spectral flux density, providing maximal flexibility.
Given photometric transient data, \fiesta's surrogates can be used in stochastic samplers to achieve fast evaluation of the likelihood, which opens the door for swift transient analysis.
Such analyses can be conducted using the established inference framework \NMMA~\citep{Pang:2022rzc}, which our surrogates are compatible with. 
In addition, \fiesta also contains its own sampling implementation that relies on the \flowMC package~\citep{Wong:2022xvh} to generate the posterior Markov chain Monte Carlo (MCMC) chain with normalizing flows and the Metropolis-adjusted Langevin algorithm (MALA).
These advanced sampling techniques reduce the number of likelihood evaluations needed and thus sample the posterior more efficiently, thereby improving the scaling of \fiesta as we consider additional nuisance parameters to account for systematic uncertainties.
Because \fiesta uses the \jax~framework, surrogate training as well as posterior sampling can be GPU-accelerated.

The approach implemented in \fiesta enables sampling the full light-curve posterior within minutes, which, depending on the base model, would previously have either taken several hours to days or proven prohibitively expensive from the start.
Therefore, the \fiesta surrogates make previously intractable models available for Bayesian inference, which allows us to present the first Bayesian analyses of GRB afterglows with \pbag.
The \fiesta code together with the surrogates is publicly available\footnote {\url{https://github.com/nuclear-multimessenger-astronomy/fiestaEM}} and all the data used in the present article can be accessed as well\footnote{\url{https://github.com/nuclear-multimessenger-astronomy/paper_fiesta/}}. 

The present article is organized as follows: 
In Sect.~\ref{sec:sampling_methods}, we briefly review Bayesian inference of photometric transient data. 
We then continue in Sect.~\ref{sec:surrogates} discussing machine learning approaches to create surrogate models for the KN and GRB afterglow emission and present our flagship surrogates for \afterglowpy, \pbag, and \possis.
From there, we verify that our surrogates are able to accurately recover the posterior and discuss their performance in Sect.~\ref{sec:inference}.
In Sect.~\ref{sec:applications}, we apply \fiesta to analyze the data from AT2017gfo/GRB170817A and GRB211211A before concluding in Sect.~\ref{sec:conclusion}.
Throughout, we use the AB magnitude system to convert interchangeably between flux densities, $F_\nu$, and magnitudes,
\begin{align}
    m=-2.5\ \log_{10}\left(\frac{\int F_\nu \frac{e(\nu)}{h\nu} d\nu}{\int 3631~\text{Jy}\ \frac{e(\nu)}{h\nu} d\nu}\right),
\end{align}
where $e(\nu)$ is the detector response function. 
In this way, even X-ray or radio flux density measurements can be expressed as magnitudes within \fiesta.

\section{Bayesian inference of transients} \label{sec:sampling_methods}
Given some photometric light curve data $d$, the goal is to find the posterior $P(\vec{\theta}|d)$, where $\vec{\theta}$ denotes the parameters of the model that describes the physical emission process.
The posterior can be obtained by using Bayes' theorem,
\begin{equation}\label{eq: Bayes theorem}
    P(\vec{\theta}|d) = \frac{\mathcal{L}(\vec{\theta}|d) \pi(\vec{\theta})}{Z} \, ,
\end{equation}
where $\mathcal{L}(\vec{\theta}|d)$ is called the likelihood function, $\pi(\vec{\theta})$ the prior distribution, and $Z$ is the Bayesian evidence. 
The latter is an important quantity for model selection (e.g., jet geometries in the context of GRBs). 
Since the posterior is often analytically intractable, stochastic sampling methods are used.
One technique is nested sampling~\citep{Skilling:2004pqw, Skilling:2006gxv}, which computes the Bayesian evidence (i.e., the normalization constant of the posterior distribution, see Eq.~\eqref{eq: Bayes theorem}), from which posterior samples can be obtained as a byproduct. 
Alternatively, we can look to MCMC methods~\citep{Neal:2011mrf}, which directly generate samples from the posterior.
Direct MCMC methods can only provide an estimate of the evidence if they are supplemented with additional techniques such as parallel tempering~\citep{Marinari:1992qd} or the learned harmonic mean estimator~\citep{Newton:1994, mcewen2023machinelearningassistedbayesian, Polanska:2024arc}.
However, we do not consider these approaches in the present article.

In the context of photometric light curve analyses, we denote the observed data, $d$, as a time series of magnitudes $\{m(t_j)| j = 1,2,3 \dots\}$, and the corresponding predictions of the model as $m^{\star}(t_j, \vec{\!\theta}\,)$.
The data are taken with some measurement uncertainty, $\sigma(t_j)$. 
In \fiesta, we assume that this uncertainty is Gaussian and hence the corresponding likelihood function can be written as
\begin{align}
\begin{split}
    \ln\mathcal{L}&(\vec{\theta}|d) = \\
     &- \sum_{t_j} \biggl( \frac{1}{2} \frac{(m(t_j) - m^{\star}(t_j, \vec{\!\theta}\,))^2}{\sigma(t_j)^2 + \sigma_{\text{sys}}(t_j)^2} 
    + \ln\bigl(2\pi (\sigma(t_j)^2 + \sigma_{\text{sys}}(t_j)^2)\bigr) \biggr)\ .
    \label{eq:likelihood}
\end{split}
\end{align}
Here, $\sigma_{\text{sys}}(t_j)$ is the model systematic uncertainty that accounts for both the surrogate error and the systematic offset caused by simplifying assumptions in the physical base model.
Moreover, detection limits can also be incorporated into \fiesta's likelihood. 
For every detection limit $m^{\blacktriangledown}(t_j)$, the likelihood in Eq.~\eqref{eq:likelihood} is simply multiplied by
\begin{align}
    \mathcal{L}&(\vec{\theta}|d) = \mathcal{L}(\vec{\theta}|d) \times \int^{\infty}_{m^{\blacktriangledown}(t_j)} \frac{\exp\left(-\frac{1}{2}\bigl(\frac{x-m^{\star}(t_j, \vec{\theta})}{\sigma_{\rm sys}(t_j)}\bigr)^2\right)}{\sqrt{2\pi \sigma_{\rm sys}(t_j)^2}}\ dx\ .
\end{align}
We note that depending on the detector and measurement, other likelihood statistics might be more appropriate. 
For instance, for low-flux X-ray data, the Poisson distribution is generally better suited to describe the measurement uncertainty~\citep{Ryan:2023pzk, Humphrey:2008xz}.
In \fiesta, we stick to a Gaussian likelihood and assume equal upper and lower error bars.

Existing inference frameworks such as \NMMA~\citep{Pang:2022rzc}, \textsc{bajes}~\citep{Breschi:2024qlc}, \textsc{redback}~\citep{Sarin:2023khf}, or \textsc{MOSFiT}~\citep{Guillochon:2017bmg} evaluate a model or a surrogate to determine the value of Eq.~\eqref{eq:likelihood}. 
They either use nested sampling or MCMC methods to sample the posterior.
\fiesta provides surrogates that can be used with \NMMA and potentially other inference frameworks,
but also contains its own sampling implementation based on the \flowMC sampler~\citep{Wong:2022xvh}.
While \NMMA relies on nested sampling through the \textsc{PyMultinest}~\citep{Feroz:2008xx} and \textsc{dynesty}~\citep{Speagle:2019ivv} packages, 
\fiesta has sampling functionalities that utilize \flowMC. 
The latter is an MCMC sampler enhanced by gradient-based sampling (in particular, the Metropolis-adjusted Langevin algorithm~\citep{grenander1994representations}) and normalizing flows, which are a class of generative ML methods that act as neural density estimators~\citep{Rezende:2015ocs, Papamakarios:2019fms, kobyzev2020normalizing}.
The flows are trained on the fly from the MCMC chains and subsequently used as proposal distributions in an adaptive MCMC algorithm~\citep{Gabrie:2021tlu}. 

In both \fiesta and \NMMA, the systematic error $\sigma_{\text{sys}}(t)$ in Eq.~\eqref{eq:likelihood} can either be fixed to a constant value or be sampled freely from a prior.
Moreover, \fiesta and \NMMA support time- and filter-dependent systematic uncertainties by sampling the nuisance parameters $\sigma_{\text{sys}}^{(k)}$ at specific time nodes $t_k$.
The systematic error for a filter, $f$, at a specific data point, $t_j$, is then determined via a linear interpolation,
\begin{align}
    \sigma_{\text{sys}}(t_j, f) = \sigma_{\text{sys}}^{(k)}(f)+ \frac{t_j-t_k}{t_{k+1}-t_k} (\sigma_{\text{sys}}^{(k+1)}(f) - \sigma_{\text{sys}}^{(k)}(f)),
    \label{eq:systematic_uncertainty}
\end{align}
which is then placed into Eq.~\eqref{eq:likelihood}.
The nuisance parameters, $\sigma_{\text{sys}}^{(k)}(f)$, can be sampled separately for different filters.
This implementation for a data-driven inference of the systematic uncertainty essentially follows~\cite{Jhawar:2024ezm}.

\section{Surrogate training} \label{sec:surrogates}
\begin{table*}[t]
    \centering
    \tabcolsep=0.3cm
    \caption{Overview of \fiesta surrogates for different models.}
    \begin{tabular}{>{\centering\arraybackslash}p{2.5 cm} >{\centering\arraybackslash}p{6 cm}
    >{\centering\arraybackslash}p{1.5 cm} >{\centering\arraybackslash}p{2 cm} >{\centering\arraybackslash}p{3 cm}}
    \toprule
    \toprule
     Physical base model & Parameters & Symbol & Range & Surrogate architecture \\
     \midrule
     & inclination [rad] & $\iota$ & $[0, \pi/2]$ &\\
     & log isotropic kinetic energy~[erg] & $\log_{10}(E_0)$ & $[47, 57]$ &\\
     & jet core angle~[rad] & $\theta_{\text{c}}$ & $[0.01, \pi/5]$ &\\
     \textsc{afterglowpy} (Gaussian jet) & wing factor & $\alpha_{\text{w}}$ & $[0.2, 3.5]$  & \textsc{FluxModel} (MLP or \textbf{cVAE}) \\
     & log interstellar medium density~[cm$^{-3}$] & $\log_{10}(n_{\text{ism}})$ & $[-6, 2]$ &\\
     & electron power index & $p$ & $[2, 3]$ &\\
     & log electron energy fraction & $\log_{10}(\epsilon_e)$ & $[-4, 0]$ &\\
     & log magnetic energy fraction & $\log_{10}(\epsilon_B)$ & $[-8, 0]$ &\\
     \midrule
     \textsc{pyblastafterglow} (Gaussian jet) & initial Lorentz factor & $\Gamma_0$ & $[100, 1000]$ & \textsc{FluxModel} (MLP or \textbf{cVAE}) \\
      & same as above for \textsc{afterglowpy} & " & " & \\
     \midrule
     & inclination [rad] & $\iota$ & $[0, \pi/2]$ &\\     
     & log dynamical ejecta mass~[M$_\odot$] & $\log_{10}(m_{\text{ej, dyn}})$ & $[-3, -1.3]$ & \\
     & average dynamical ejecta velocity~[$c$] & $\bar{v}_{\text{ej, dyn}}$ & $[0.12, 0.28]$ & \\
     \possis & average dynamical ejecta electron fraction & $\bar{Y}_{\text{e, dyn}}$ & $[0.15, 0.35]$ & \textbf{\textsc{FluxModel}} (\textbf{MLP} or cVAE) or\\
     & log wind ejecta mass [M$_\odot$] & $\log_{10}(m_{\text{ej, wind}})$ & $[-2, -0.9]$ & \textsc{LightcurveModel} (MLP) \\
     & average wind ejecta velocity [$c$] & $\bar{v}_{\text{ej, wind}}$ & $[0.05, 0.15]$ & \\
     & wind ejecta electron fraction & $Y_{\text{e, wind}}$ & $[0.2, 0.4]$ & \\
     \bottomrule
     \end{tabular}
     \tablefoot{We list the parameters, their ranges in the training data set, as well as the architectures employed for the ML surrogates.
    The models marked in bold are the best-performing surrogate architectures and are used in Secs.~\ref{sec:inference} and~\ref{sec:applications}.
    The bar above the kilonova parameters indicates mass-averaged quantities.}
    \label{tab:surrogate_models}
\end{table*}
To evaluate the likelihood function in Eq.~\eqref{eq:likelihood} efficiently, \fiesta provides surrogates that determine the expected magnitudes, $m^{\star}(t_j, \vec{\!\theta}\,)$, for a given parameter point $\vec{\theta}$ through ML surrogates.
 For KN models, previous works have demonstrated the capability of ML techniques to replace the expensive light curve generation of radiative transfer models, which enabled their use in stochastic samplers~\citep{Almualla:2021znj, Ristic:2021ksz, Kedia:2022onl, Lukosiute:2022tmd, Ristic:2023ywr, Ford:2023bdi, Saha:2023zse, King:2025tqo}. 
For GRB afterglows, many fast semi-analytic models are available, so the use of ML surrogates has not been as essential; however, an increasing number of recent works have introduced ML techniques to this area. 
In~\citet{Lin:2021auf}, the authors linearly interpolated a fixed table of GRB afterglow light curves from the prescription in~\citet{Lamb:2018ohw} to accelerate the likelihood evaluation, although the interstellar medium density and other microphysical parameters were kept fixed. 
A surrogate model for the X-ray emission in \afterglowpy was trained in~\citet{Sarin:2021yya} to analyze the Chandra transient CDF-S XT1.
In~\citet{Boersma:2022qtg}, \textsc{DeepGlow} was introduced, a \textsc{python} package that emulates \textsc{boxfit}~\citep{vanEerten:2011yn} light curves through a neural network (NN).
In~\citet{Wallace:2024pmx}, the authors trained an NN for the afterglow model of~\citet{Lamb:2018ohw}. 
\citet{Rinaldi:2024xjz} suggested developing an ML surrogate based on the afterglow model developed in~\cite{Warren:2021whb}, but postponed the implementation to future work.
In~\citet{Aksulu:2020hnl, Aksulu:2021crt}, the authors used Gaussian processes to model the likelihood function in Bayesian analysis, although the expected light curves were computed directly with \textsc{scalefit}~\citep{Ryan:2014nea}.

In the present work, we introduce the first large-scale surrogates for the state-of-the-art afterglow models \afterglowpy and \pbag, covering the radio to hard X-ray emission over a timespan of $10^{-4}-2\times10^3$~days.
Additionally, we extend the work of \citet{Almualla:2021znj, Anand:2023jbz} and train a new KN surrogate with an updated version of the \possis code.
Our surrogates provide a large speed-up, since the \possis Monte Carlo radiative transfer code~\citep{Bulla:2019muo, Bulla:2022mwo} takes on the order of $\sim 1000$ CPU-hours to predict a light curve.
Likewise, depending on the settings, a GRB afterglow simulation in \afterglowpy takes on the order of 0.1 - 10 seconds; whereas for \pbag, the computation time may exceed several minutes.
In the following subsections, we provide details about the implemented ML approaches in \fiesta and present the surrogates we obtained for the GRB afterglow and KN models.

\subsection{Surrogate types and architectures}
To create a surrogate for \fiesta, a NN learns the relationship between the input $\vec{\theta}$, i.e., the parameters of the model, and the output $\vec{y}$ (i.e., the flux).
The surrogate models thus interpolate a precomputed training data set that consists of many evaluations of the physical base model (e.g., \possis or \pbag) on various combinations of input parameters.

Specifically, the output $\vec{y}$ could either represent the magnitudes in predefined frequency filters, or yield the spectral flux density $F_\nu$ across a continuum of frequencies.
In previous works, surrogate models typically predicted magnitudes for a set of predefined filters with fixed wavelength~\citep{Almualla:2021znj, Pang:2022rzc, Peng:2024jqe}. 
Training on the spectral flux densities provides more flexibility, as the surrogate does not need to be retrained if a new filter becomes available, and the surrogate's output can be further processed to account for arbitrary redshifts or extinction effects.
Surrogates trained on $F_\nu$ will thus return a 2D-array containing the flux density across time along one dimension, and across frequency along the other dimension.
\fiesta implements both types of surrogates, which we refer to as \textsc{FluxModel} class for the latter kind of surrogate, whereas the surrogates that are trained in the traditional approach on passband magnitudes are referred to as the \textsc{LightcurveModel} class.

Regardless of the particular type of surrogate model, \fiesta employs two kinds of NN architectures, namely the simple feed-forward multilayer perceptron (MLP) and the conditional variational autoencoder (cVAE).
The feed-forward NN, in this work containing three hidden layers, is used to train the relationship between the input parameters $\vec{\theta}$ and the coefficients $\vec{c}$ of the principal component analysis (PCA) of the training data.
The training simply minimizes the mean squared error on the PCA coefficients as loss function,
\begin{align}
    L(\vec{\phi}) =  \frac{1}{n_{\text{train}}}\sum_{j=1}^{n_{\text{train}}}\,  \lVert \vec{c}^{\text{ (train)}}_j - \vec{c}^{\text{ (predict)}}_j \rVert^2\,,
\end{align}
where $\vec{\phi}$ are the NN weights, $\vec{c}^{\text{ (train)}}_j$ are the PCA coefficients of the training data $\vec{y}$, and $\vec{c}^{\text{ (predict)}}_j$ are the coefficients the NN would predict.
The passband magnitude or flux density $\vec{y}$ is then determined by applying the inverse PCA decomposition to $\vec{c}$. 
The feed-forward architecture is thus used both for \textsc{LightcurveModel} and \textsc{FluxModel} surrogates.

The other implemented architecture is the cVAE~\citep{Kingma:2013hel, Rezende:2014azh}.
This approach is inspired by \cite{Lukosiute:2022tmd}, where cVAEs were used to predict KN spectra. 
In contrast to their work, our setup predicts the flux density across a fixed grid of times and frequencies, and thus time is not a training parameter. 
Moreover, we extended this approach to GRB afterglows.

In the cVAE architecture, an encoder and decoder are trained simultaneously on the spectral flux densities directly. 
The encoder takes the parameters $\vec{\theta}$ and the flux $\vec{y}$ as inputs and maps them to the latent parameters $\vec{\mu}$ and $\vec{\sigma}$.
These parameters represent the variational distribution of the latent space from which the decoder reconstructs $\vec{y}$.
Specifically, the latent vector is drawn according to $\vec{z} \sim \mathcal{N}(\vec{\mu},\,\text{diag}(\vec{\sigma}))$ and serves as input to the decoder, together with $\vec{\theta}$. 
The encoder and decoder are trained on minimizing the joint loss function, 
\begin{align}
\begin{split}
    & L(\vec{\phi}_\text{encoder}, \vec{\phi}_\text{decoder}) = \\
    & \frac{1}{n_{\text{train}}}\sum_{j=1}^{n_{\text{train}}} \bigg( 
    \frac{1}{2} \left(1 + \ln(\vec{\sigma}_j)^2 - \vec{\sigma}_j^2 - \vec{\mu}_j^2 \right) +  \lVert \vec{y}^{\text{ (train)}}_j - \vec{y}^{\text{ (predict)}}_j \rVert^2 \bigg)\,,
\label{eq:loss_cVAE}
\end{split}
\end{align}
where $\vec{\phi}$ represents the NN weights, $\vec{\sigma}_j$, $\vec{\mu}_j$, and $\vec{y}^{\text{ (predict)}}_j$ are the predicted parameters for the variational distribution and the flux, and $\vec{y}^{\text{ (train)}}_j$ is the training data.
The first term in Eq.~\eqref{eq:loss_cVAE} represents the Kullback-Leibler divergence between the variational distribution and the standard normal distribution, while the second term is the reconstruction loss.
Since the decoder is conditioned with the parameters $\vec{\theta}$, its output after training will not depend much on the latent vector, and for the actual flux prediction we set $\vec{z} = (0,\dots,0)$. 
Since the cVAE is computationally more expensive to train, it was only implemented for \textsc{FluxModel} surrogates, where just a single surrogate is trained, in contrast to the \textsc{LightcurveModel} where each photometric filter constitutes its own surrogate.
All NNs were implemented through the \textsc{Flax} API~\citep{flax2020github}.

\subsection{GRB afterglow surrogates}
\label{subsec:GRB_afterglow_surrogates}

\begin{figure}
    \centering
    \includegraphics[width=1\linewidth]{"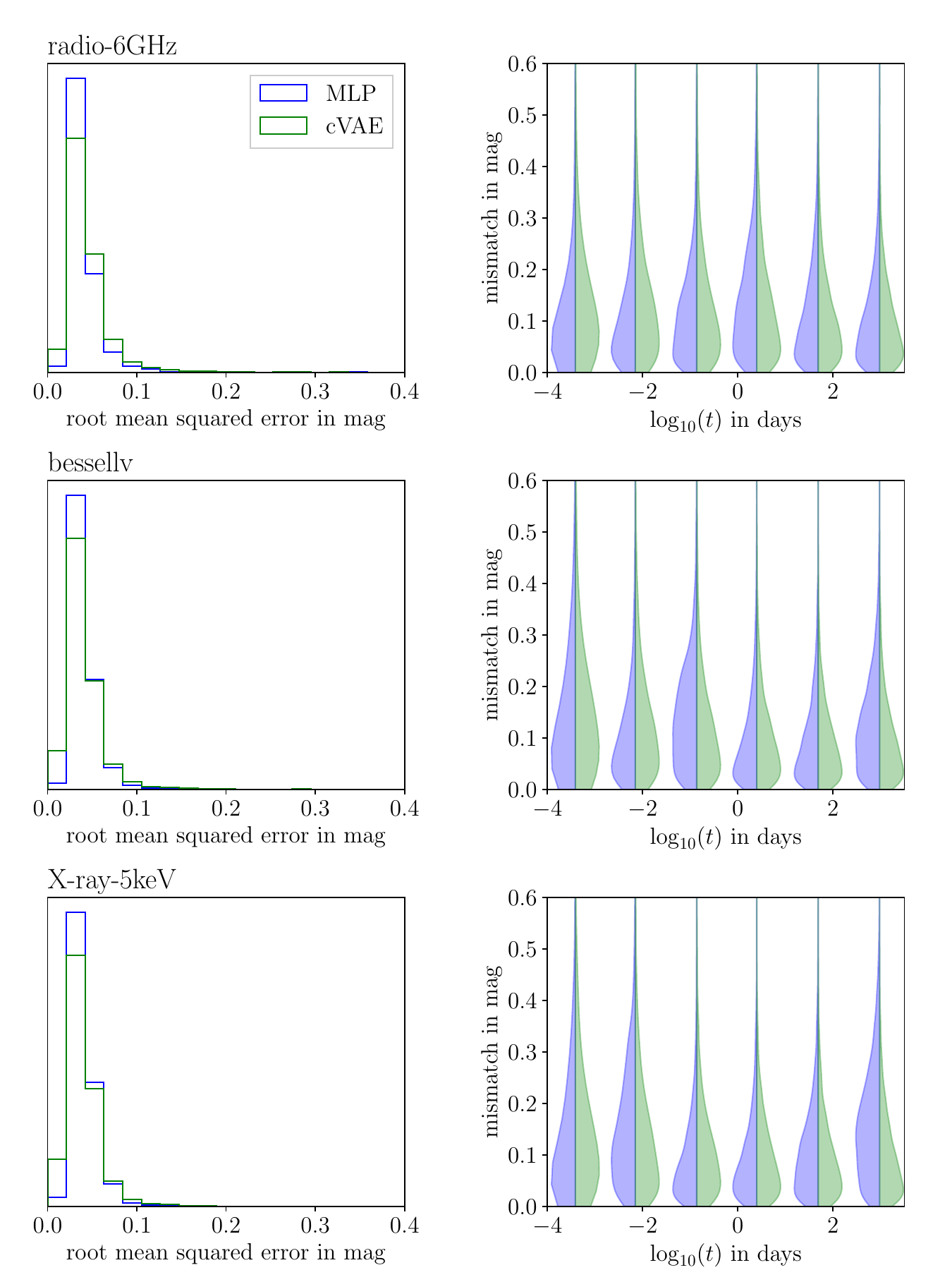"}
    \caption{Benchmarks of the two surrogates for the \afterglowpy Gaussian jet model. 
    We show the error distributions of the surrogate predictions against a test data set of size $n_{\text{test}}=7500$.
    The different rows show the error across different passbands.
    The left panels show the distribution of the mean squared error as defined in Eq.~\eqref{eq:mse_error}.
    The right panels show the mismatch distribution across the test data set as defined in Eq.~\eqref{eq:mis_error}.
    The figure compares two different surrogates: one using the MLP architecture (blue) and the other a cVAE (green).}
    \label{fig:benchmark_afgpy_gaussian}
\end{figure}
\begin{figure}
    \centering
    \includegraphics[width=1\linewidth]{"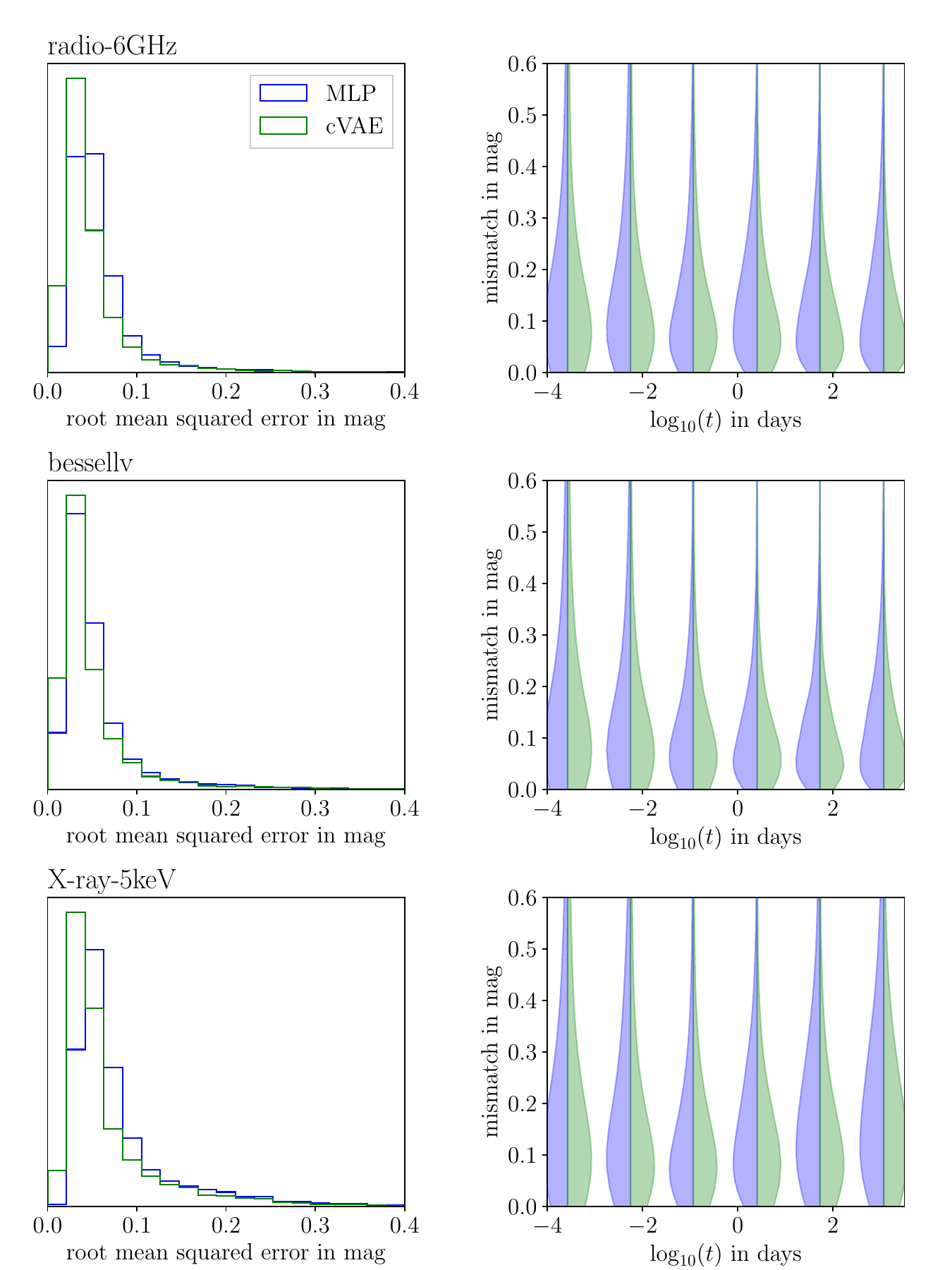"}
    \caption{Benchmarks of the two surrogates for the \pbag Gaussian jet model. 
    We show the deviations of surrogate predictions against a test data set of size $n_{\text{test}}=7232$.
    Figure layout is the same as in Fig.~\ref{fig:benchmark_afgpy_gaussian}.}
    \label{fig:benchmark_pbag_gaussian}
\end{figure}

In \fiesta, we include surrogates for the GRB afterglow models \afterglowpy~\citep{Ryan:2019fhz} and \pbag~\citep{Nedora:2024vrv}.
Table~\ref{tab:surrogate_models} shows the parameter ranges and trained architectures for these models.
We note that these surrogates are specific to a structured Gaussian jet.
While we also created surrogates for the tophat jet models, their relevance in real-life applications is limited~\citep{Ryan:2019fhz, Salafia:2022dkz} and we discuss their performance in Appendix \ref{app:tophat_surrogates}.
We simply note here that the tophat jet surrogates generally outperform the surrogates for the Gaussian jet in terms of accuracy due to their simpler physical behavior.

Both \afterglowpy and \pbag assume that the jet of the GRB can be modeled as a relativistic fluid shell that propagates trough a cold, ambient interstellar medium with constant density $n_{\rm ism}$ as a shockwave.
The shock jumping conditions can then be used to determine the shell dynamics analytically given the initial kinetic energy $E_0$ and bulk Lorentz factor $\Gamma_0$~\citep{Nava:2012hq, Ryan:2019fhz}.
All \afterglowpy and the \pbag runs in the present work ignore reverse shock contributions and only include the forward shock.
If the jet has a Gaussian structure, then the jet is evolved as a collection of individual, independent, annular shells each of them assigned an energy according to 
\begin{align}
    E(\theta) = \begin{cases}
        E_0 \exp\left(-\frac{\theta^2}{2\theta_c^2}\right)\qquad \text{if $\theta\leq\theta_{\rm w}$} \\
        0 \qquad \qquad \qquad \qquad \text{otherwise}
    \end{cases}\ ,
\end{align}
where $\theta$ is the polar angle from the jet's center axis and $\theta_{\rm c}$ the core angle parameter of the jet. 
The wing angle $\theta_{\rm w}$ is parameterized in Table~\ref{tab:surrogate_models} through the factor $\alpha_{\text{w}} = \frac{\theta_{\rm w}}{\theta_{\rm c}}$. 
Once the dynamics have been determined, the emission is modeled as synchrotron radiation from the electrons accelerated at the shock front, which receive a fraction $\epsilon_e$ of the shock energy.
The magnetic field in the downstream shock is given through a fraction $\epsilon_B$ of the shock energy.
While employing a similar semi-analytical framework, \afterglowpy and \pbag have notable differences in their specific implementation. 
Most notably, \afterglowpy assumes that the electron distribution follows a broken power law in which newly shocked electrons are injected with $N_{e, \text{inj}}(\gamma_e) \propto \gamma_e^{-p}$, where $p$ is the electron power index. 
\pbag assumes the same power law for the newly shocked electrons, but instead numerically evolves the existing electron distribution.
Further differences regard the implementation of the jet spreading and initial coasting phase and the methods used to compute the synchrotron radiation spectra.

For both GRB afterglow models, we trained two surrogates of the \textsc{FluxModel} type: one with an MLP architecture and the other with the cVAE architecture.
Each surrogate was trained with input parameters as listed in Table \ref{tab:surrogate_models}.
The training data was randomly drawn from the ranges specified there.
The training data set for the \afterglowpy Gaussian jet surrogate encompasses $n_{\text{train}}=80000$ flux density calculations, the set for the \pbag surrogate $n_{\text{train}}=91670$.
The surrogates were trained on standardized $\ln(F_\nu)$, and standardized parameter samples $\vec{\theta}$.
After different attempts with similar results, the number of PCA coefficients for the MLP training was set to 50 and the cVAE trained on a down-sampled flux density array of size $42\times57$.
On a H100 GPU, training took about 2.2~h for the cVAE and 0.3~h for the MLP architecture.

We used two different metrics to compare the light curves $m^{\rm pred}(t)$  predicted by the surrogate against a set of test light curves $m^{\rm test}(t)$. 
These test light curves are from the physical base model and were not part of the training set. 
The one metric is the mean squared error MSE,
\begin{align}
    \text{MSE}^2 = \int_{\log(t_{\rm min})}^{\log(t_{\rm max})} \frac{(m^{\rm test}(t) - m^{\rm pred}(t))^2}{\log(t_{\rm max}) - \log(t_{\rm min})} d\log(t)\ ,
    \label{eq:mse_error}
\end{align}
and the other is the mismatch MIS,
\begin{align}
    \text{MIS}(t) = |m^{\rm test}(t) - m^{\rm pred}(t)|\ .
    \label{eq:mis_error}
\end{align}
In Fig.~\ref{fig:benchmark_afgpy_gaussian}, we show the performance of the surrogates for the \afterglowpy Gaussian jet model when trying to predict the magnitudes in different passbands.
Overall, the squared error is confined to $\lesssim 0.1$ mag across different photometric filters and does not vary notably between the MLP or cVAE architectures.
The panels on the right hand-side show the distribution of the absolute mismatch over time.
While this mismatch is mostly within 0.3~mag, some outlier predictions show stronger deviations from the test data.
Specifically, for the cVAE architecture, $\sim$6\% of all test data samples exceed a mismatch of 1 mag at some point along the test light curve. 

In Fig.~\ref{fig:benchmark_pbag_gaussian}, we show the performance of the corresponding \pbag surrogates for the Gaussian jet.
The deviation from the test data set is typically larger than for the \afterglowpy surrogate, which we attribute to higher variability in the training data arising from the additional features of \pbag.
This is also represented in the mismatch, which typically falls in the range of $\lesssim 0.4$~mag, though for $\sim$10\% of the test data samples the mismatch exceeds 1~mag at least once along the light curve.
We also note that the X-ray filter in the bottom right panel shows somewhat higher mismatches from the predictions.
This is due to the fact that we cropped the training data below $2\cdot10^{-22}$~mJys at 10 pc (corresponding roughly to the 70th absolute magnitude) due to numerical noise in the \pbag flux computation at very low brightness.
This cutoff mostly affects the high frequencies, and since the surrogates struggle to reproduce this hard cut, the typical mismatch in the X-ray filter is higher than for the lower frequency filters.
As this concerns only light curves far below the detection limit, this is no issue in real-life applications.

The MLP and cVAE architectures perform very similarly for the GRB afterglow surrogates, though the cVAEs tend to have a slightly smaller absolute mismatch in the light curve at later times. 
This is especially true for the \pbag model.
Overall, the cVAEs also tend to have the smaller average square error as shown by the distributions in the left panels of Figs.~\ref{fig:benchmark_afgpy_gaussian} and \ref{fig:benchmark_pbag_gaussian}.
For these reasons, we set the cVAE as the default architecture to be used for the analyses later in Sects.~\ref{sec:inference} and \ref{sec:applications}.

\subsection{KN surrogates}
\label{subsec:KN_surrogates}
\begin{figure}
    \centering
    \includegraphics[width=1\linewidth]{"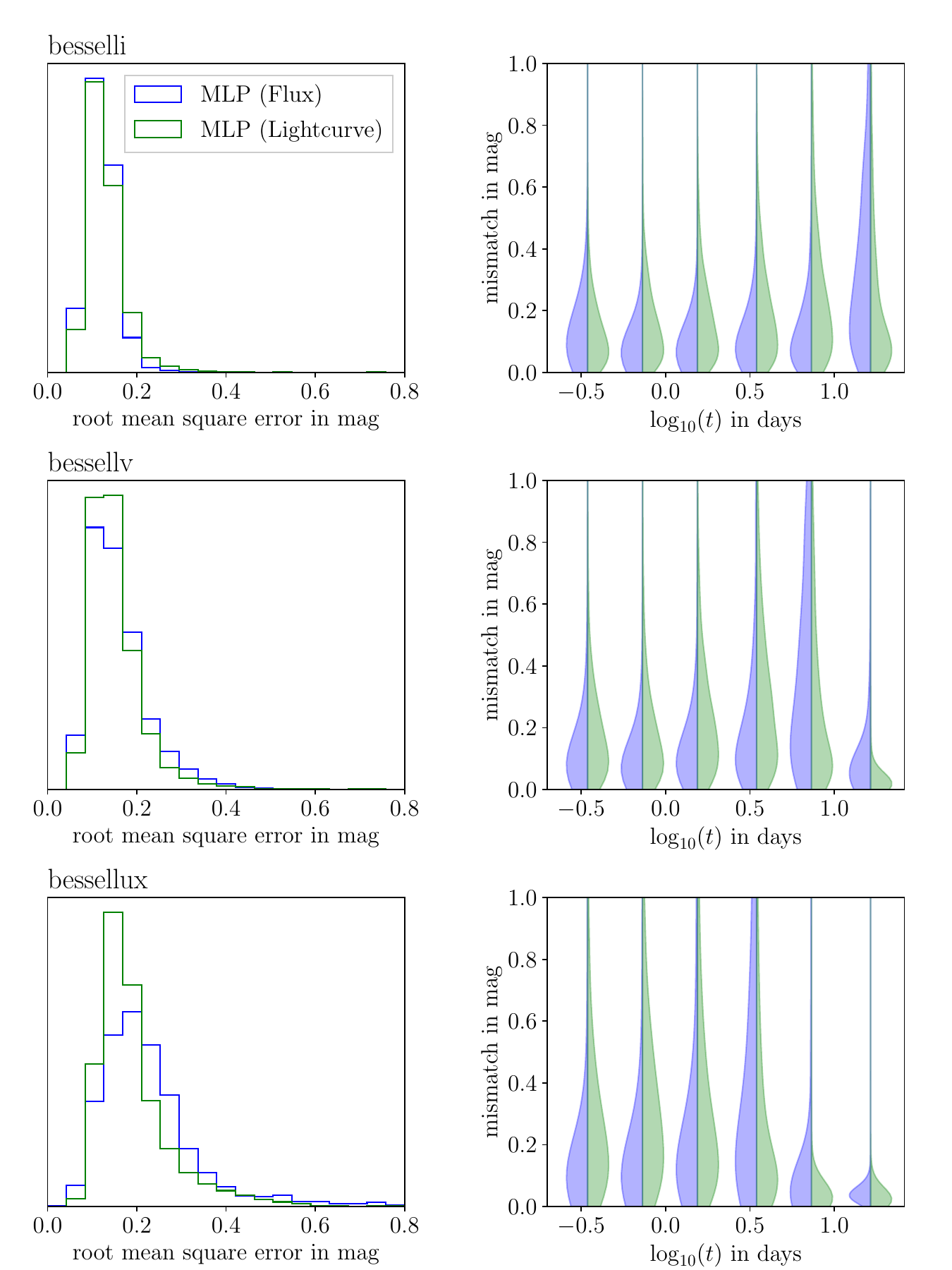"}
    \caption{Benchmarks of two surrogates for the KN \possis model. 
    We show the deviations of surrogate predictions against a test data set of size $n_{\text{test}}=2238$.
    Figure layout as in Fig.~\ref{fig:benchmark_afgpy_gaussian}.
    The figure compares two different surrogates: one using the MLP architecture (blue) and the other a \textsc{LightcurveModel}, where an MLP is trained for each passband separately (green).}
    \label{fig:benchmark_Bu2025}
\end{figure}
\fiesta includes surrogates for the KN model from the \possis code~\citep{Bulla:2019muo, Bulla:2022mwo}.
\possis is a 3D Monte Carlo radiation transport code that assumes homologously expanding ejecta and determines the emitted flux and polarization at each timestep from the Monte Carlo propagation of photon packets.
As the photon packets move through the matter cells in the ejecta profile, they can interact with the matter through bound-bound and electron-scattering transitions, where the prescriptions take into account temperature-, density-, and electron-fraction dependent opacities~\citep{Tanaka:2019iqp}.
The BNS ejecta are assumed to follow a certain geometry inspired by numerical relativity simulations~\citep{Kiuchi:2017pte, Radice:2018pdn, Kawaguchi:2019nju, Hotokezaka:2018dfi}.
In particular, the ejecta have two components, the dynamical ejecta with mass $m_{\rm ej, dyn}$ and the wind ejecta with mass $m_{\rm ej, wind}$.
The velocity in the dynamical component ranges from 0.1$c$ to a cutoff value that is determined such that the mass-averaged velocity is $\bar{v}_{\rm ej, dyn}$.
Likewise, the wind component has the mass-averaged velocity $\bar{v}_{\rm ej, wind}$ with a minimum velocity of at least 0.02$c$.
The dynamical electron fraction varies with the polar angle $\theta$ according to 
\begin{align}
    Y_{e, \text{dyn}} = a \cos(\theta)^2 +b\ ,
\end{align}
where $a=0.71 b$ \citep{Setzer2023} and b is scaled to achieve the desired mass-averaged electron fraction $\bar{Y}_{e, \text{dyn}}$.
This setup was also used in \citet{Anand:2023jbz, Ahumada:2025ubp}.
The electron fraction for the wind ejecta, however, is assumed to be constant and thus we do not mark its symbol $Y_{e,\text{wind}}$ with a bar.

We trained three \possis surrogates, two as a \textsc{FluxModel} with an MLP and cVAE architecture, respectively, and the other as a \textsc{LightcurveModel}.
These surrogates constitute an update for the \textsc{Bu2019} surrogate implemented in \NMMA that was based on an older \possis version.
Our training data consists of $n_{\text{train}}=17899$ flux densities calculated from \possis.
For each of these computations we set the number of Monte Carlo photon packets to $10^7$.
Parameters for the training data set are randomly drawn within the ranges from Table~\ref{tab:surrogate_models}, though we note that the ejecta masses were uniformly in linear space, not in log-space.
Furthermore, the inclination, $\iota$, is confined to a fixed grid from the \possis output that is spaced uniformly in $\cos(\iota)$ between 0 (edge-on) and 1 (face-on).
We also mention that the \textsc{LightcurveModel} additionally receives the redshift $z$ sampled randomly between 0 and 0.5 as a training parameter (the luminosity distance is fixed at 10~pc), so that the surrogate automatically incorporates the K-correction to the passband magnitudes.
We find no significant performance difference for various hyperparameter settings, so the final number of PCA coefficients for the MLP architecture in the \textsc{FluxModel} and \textsc{LightcurveModel} was set to 100, the cVAE was trained on a down-sampled flux array of dimensions $64\times40$.
Training took about 4 minutes for the MLP architectures and 27 minutes for the cVAE on an H100 GPU.

In Fig.~\ref{fig:benchmark_Bu2025}, we benchmark the surrogate models for the \possis KN model. 
In particular, we compare the MLP \textsc{FluxModel} against the MLP \textsc{LightcurveModel}, since the \textsc{FluxModel} with the cVAE architecture underperforms both.
This is because, in contrast to the GRB afterglow training data, the training data from \possis contains inherent Monte Carlo noise, which is more difficult for the cVAE to ``average out,'' as it is trained directly on the flux output.
The MLP architecture uses the coefficients of the principal components as training input and, thus, it is  less sensitive to small noise fluctuations.
Still, even for those architectures, we generally find higher deviations in the predictions from the test data set than for the afterglow surrogates.
In particular, the mismatch rises drastically above 0.5~mag for many test data samples when the flux brightness suddenly drops.
This can be seen in Fig.~\ref{fig:benchmark_Bu2025}, where the mismatch in the right panels is typically confined to $\lesssim 0.5$~mag, but then spikes around the time the KN starts to fade (i.e., after around 1~day) in the UV band and after around $\gtrsim$ 10~days in the $i$-band.
In general, the dimmer the light curve, the higher the Monte Carlo noise contribution and the larger the mismatch between the surrogate and the test data becomes.
However, inspection of random test samples reveals that the surrogate prediction matches quite well at early times, when the absolute magnitude is still brighter than $-10$~mag.
At very late times, we truncated the training data below $10^{6.5}$~mJys (corresponding roughly to the 0th absolute magnitude), which causes the mismatch to decrease again when the surrogates pick up on this trend.
The \textsc{LightcurveModel} performs slightly better at these later times, however, the \textsc{FluxModel} performs better at earlier times when the emission is still bright.
We thus set the MLP \textsc{FluxModel} as the default KN surrogate for the subsequent analyses in Sects.~\ref{sec:inference} and \ref{sec:applications}, as the early and bright parts of the light curve are most relevant for real-life applications.

While we generally find that the surrogates presented here are well trained, the typical prediction error still exceeds the observation accuracy $\sigma$ of most GRB afterglow or KN observations. 
However, the similar performance across the different architectures, as well as consistent performance across training runs with different hyperparameters indicates that improving the surrogates further might be challenging. 
Thus, when fitting to light curve data, we need to offset this surrogate error through the systematic uncertainty in the likelihood in Eq.~\eqref{eq:likelihood}.

We also note that the surrogates from Table~\ref{tab:surrogate_models} are rather extensive in their scope in which they can predict the spectral flux density. Specifically, the GRB afterglow surrogates have been trained across $10^9$--$5\cdot10^{19}$~Hz, i.e., from the radio to hard X-ray, and over a time interval of $10^{-4}$--$2000$~days. 
The KN surrogates can predict the expected emission between $0.2$--$26$~days in the frequency range of $10^{14}$--$2\cdot10^{15}$~Hz, i.e., from the far infrared to the far ultraviolet (UV). 
In certain use cases, the surrogates could easily be retrained on a smaller frequency or time interval, to potentially deliver even better performance.

\section{Bayesian inference with \fiesta}
\label{sec:inference}
Using the surrogate for the determination of the expected light curve $m^{\star}(t, \vec{\theta})$ in Eq.~\eqref{eq:likelihood} is an approximation to the real likelihood function.
In this section, we demonstrate that this approximation is still capable of recovering the correct posterior when accounting for the surrogate uncertainty.
We do so by using the best-performing surrogates presented in Sect.~\ref{sec:surrogates}, namely, the \textsc{FluxModel} instances with cVAE architecture for the afterglow models, and the \textsc{FluxModel} with the MLP architecture for the KN model.
We also discuss the performance of the \flowMC implementation in \fiesta and how it scales when the dimension of the parameter space is increased to include more systematic nuisance parameters in the sampling.

\subsection{Injection recoveries}
\begin{figure*}
    \centering
    \includegraphics[width=1\linewidth]{"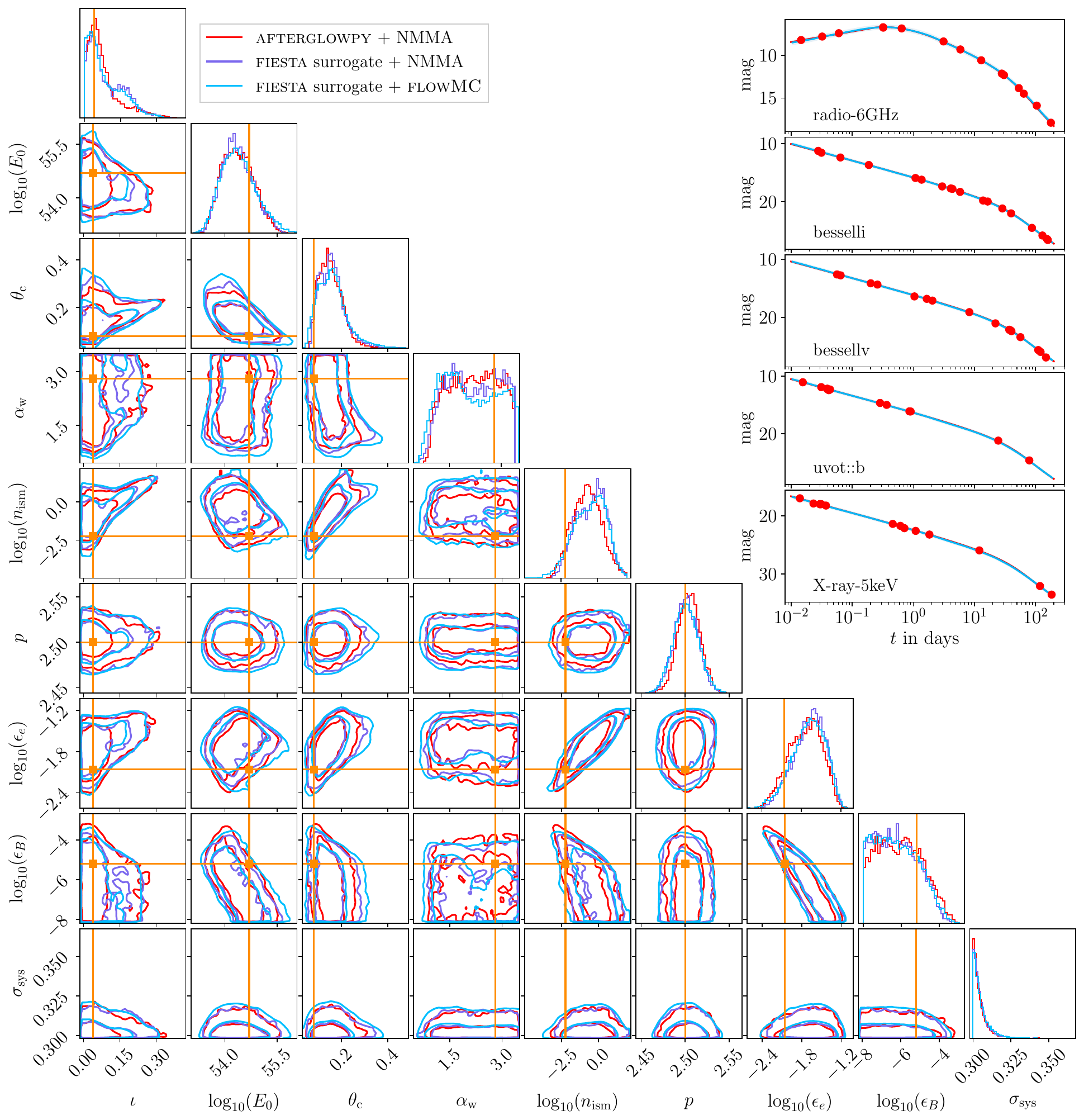"}
    \caption{Parameter recovery for an injected mock light curve from the gaussian \afterglowpy jet model.
    The corner plot shows the posterior contours at 68\% and 95\% credibility. 
    Parameters correspond to the symbols in Table~\ref{tab:surrogate_models}, $\sigma_{\text{sys}}$ is the freely sampled systematic uncertainty. 
    Different colors compare posteriors obtained with different sampling methods.
    The posterior in red is based on likelihood evaluations from the proper \afterglowpy model with the \NMMA sampler. 
    The purple posterior relies on the \fiesta surrogate for the likelihood evaluation but uses the \NMMA sampler.
    The light blue posterior uses the \fiesta surrogate as well but is sampled within \fiesta's own inference framework that relies on \flowMC.
    The injection parameters used to generate the mock light curve data are indicated by the orange lines.
    The insets on the upper right side show the injection data across the photometric filters and the best-fit light curve (i.e., highest likelihood) of the \fiesta posterior (lightblue) and the actual \afterglowpy light curve used to generate the mock data (red). The latter lies almost completely underneath the former.}
    \label{fig:injection_afgpy_gaussian}
\end{figure*}

\begin{figure}
    \centering
    \includegraphics[width=1\linewidth]{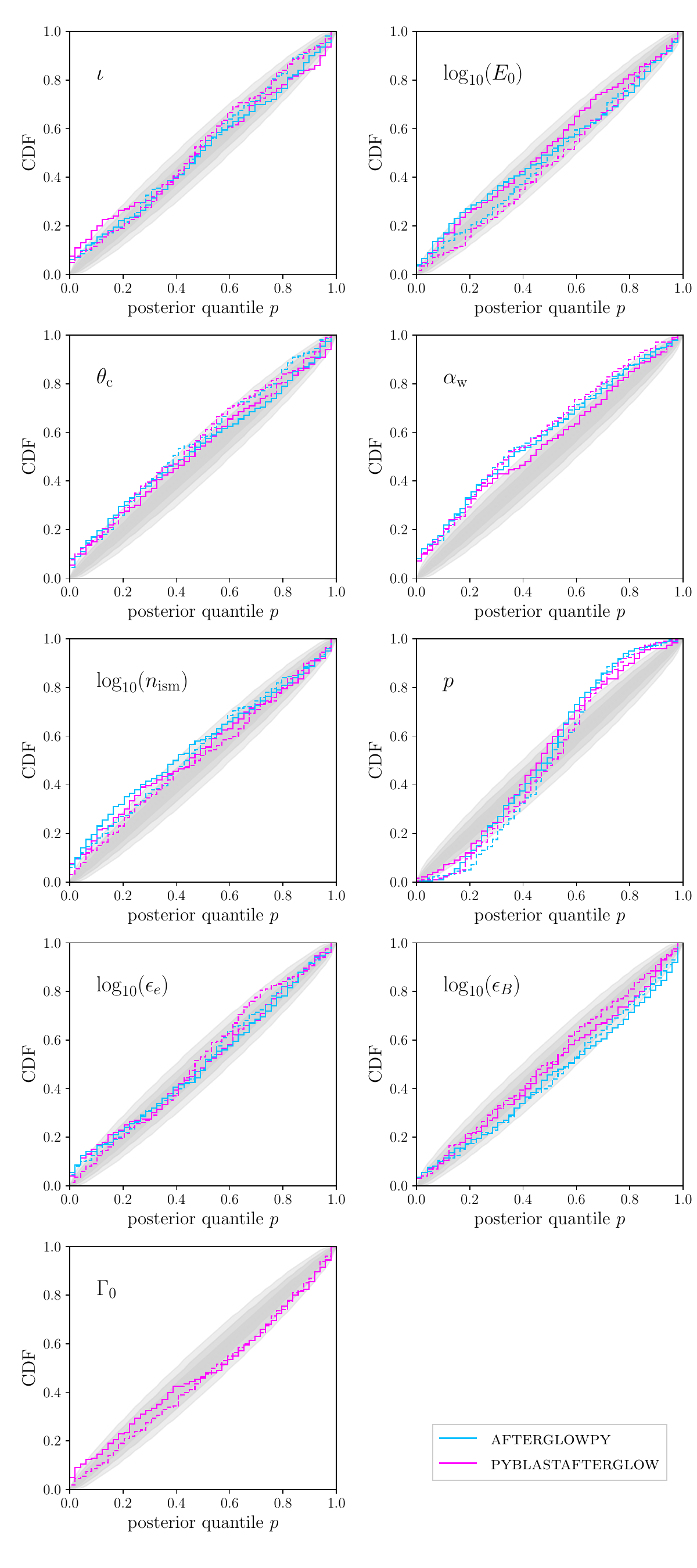}
    \caption{P-P plots for GRB afterglow injections. 
    Each panel shows a P-P plot for the recovery of the parameter displayed in its top left corner. 
    The P-P plots show the cumulative distribution of the injected values' posterior quantiles for 200 injections. 
    The lightblue curves indicate injection recoveries with \afterglowpy, the magenta ones for \pbag.
    The solid lines signify that the injections stem from physical base model, the dashed lines indicate an injection with the surrogate itself. 
    The gray areas mark the 68\%-95\%-99.7\% confidence range in which the quantile distribution should fall if it was uniformly distributed.}
    \label{fig:pp_plot_GRB}
\end{figure}

\begin{figure}
    \centering
    \includegraphics[width=1\linewidth]{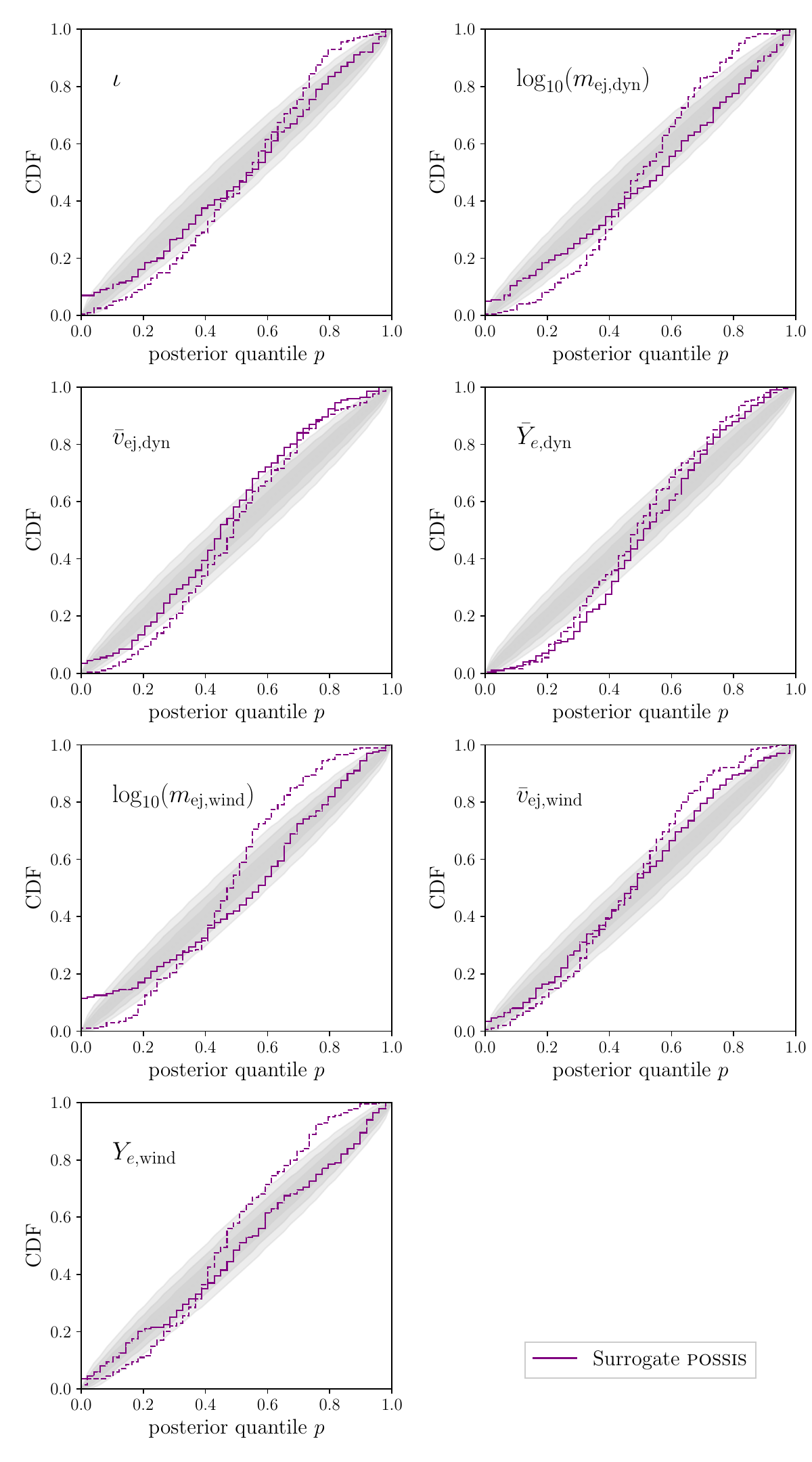}
    \caption{P-P plots for the \possis surrogate model. 
    Figure layout as in Fig.~\ref{fig:pp_plot_GRB}.}
    \label{fig:pp_plot_KN}
\end{figure}
To evaluate whether \fiesta's surrogates are capable of recovering the correct posterior, we created mock light curve data with the physical base model using randomly drawn model parameters.
We were then able to obtain the posterior using the surrogate.
The injection data always contain 75 mock magnitude measurements across multiple bands, encompassing the frequency and time range of the model and representing a well-sampled light curve.
Specifically, for GRB afterglows, the injection data span from 0.01 to 200~days and contains mock observations from the radio to the X-ray bands. 
The KN injections reach from the infrared to UV and contain data points between 0.5 to 20~days.
For the KN injections, we also apply a detection limit of 24 apparent mag at 40~Mpc, which prevents the surrogates of being used in regions with high prediction error due to high Monte Carlo noise in the \possis training data.
We add Gaussian noise to these mock measurements, where the measurement errors $\sigma(t_j)$ are drawn from a $\chi^2$-distribution with one degree of freedom and scaled to lie around $0.1$~mag.
To recover the posterior with the surrogate, we use uniform priors across the parameter ranges specified in Table~\ref{tab:surrogate_models}. 
The luminosity distance is fixed to $40$~Mpc for each injection.

In Fig.~\ref{fig:injection_afgpy_gaussian}, we show how the parameters of one particular injection from the \afterglowpy Gaussian jet model are recovered with \fiesta's surrogates, using either \textsc{PyMultinest} in \NMMA or \textsc{flowMC} as sampler.
Since the injection also incorporates a random mock measurement error, the posterior is not always centered around the true injected parameters indicated by the orange lines, but the marginalized posteriors contain the injected values in the 95\% credible interval.

Of all the models presented in Sects.~\ref{subsec:GRB_afterglow_surrogates} and \ref{subsec:KN_surrogates}, \afterglowpy is the only one where the execution time of a single likelihood call is sufficiently fast to be used directly in Bayesian inference. 
Thus, we can compare the approximate posterior obtained with the \fiesta surrogate to the posterior obtained using the actual physical base model for the likelihood evaluation.
The latter is shown in red in Fig.~\ref{fig:injection_afgpy_gaussian}.
We find good agreement with the posteriors obtained with the \fiesta surrogate, though some small deviations in the posterior tails for the inclination and jet opening angle exist.

We also ran four additional \afterglowpy injection recoveries (similar to those in Fig.~\ref{fig:injection_afgpy_gaussian}) and compared them to the surrogate posteriors, finding good agreement.
For certain degenerate parameters, we observed that even the \flowMC sampling algorithm with the surrogate recovers the true injected values better; namely, the true value lies more at the center of the posterior than with \NMMA's \textsc{PyMultinest} sampler and when using the actual \afterglowpy model for the likelihood.
This can be attributed to broader exploration in the parameter space for degenerate parameters such as $\epsilon_B$ and $\log_{10}(n_{\rm ism})$, indicating an advantage in \flowMC's dependence on gradient-based samplers and global proposals.

To systematically evaluate the disagreement between the posteriors caused by the surrogates, we resorted to the Kullback-Leibler (KL) divergence $D_{KL}$.
\cite{Bevins:2025srk} assessed a theoretical link between the root mean square error (RMSE) of an emulator and the impact on the posterior in terms of $D_{KL}$ if it was obtained with the base model or with the surrogate.
For a linear model, they derive the upper-bound on $D_{KL}$,
\begin{align}
    D_{KL} \leq \frac{N_d}{2} \frac{\rm{RMSE}^2}{\sigma^2}\ ,
    \label{eq:DKL_upper_limit}
\end{align}
where $N_d$ is the number of data points and $\sigma$ their typical error scale.
While this upper bound assumes a linear relation $d(\vec{\theta})$, we can still examine this relationship using our \afterglowpy injections and the respective surrogate and base model posteriors.
We then determine $D_{KL}$ between the posteriors through a kernel density estimate.
Using our 5 injection-recoveries with $N_d=75$, we find that $D_{KL}$ ranges between 2 -- 8~nats and the upper limit is indeed obeyed when setting $\rm{RMSE} = 0.1$~mag and taking the largest error from the injected mock data as a conservative value for $\sigma$. 
In fact, Eq.~\eqref{eq:DKL_upper_limit} is a factor of 1.4 -- 2.8 above our estimated value for $D_{KL}$.
It should be noted however, that this only serves as a limited sanity check, since the bound in Eq.~\eqref{eq:DKL_upper_limit} is derived under simplifying assumptions, and determining $D_{KL}$ through kernel density estimates might not be numerically accurate.

We also point out that in order to achieve this agreement between the surrogate posterior and the posterior based on \afterglowpy directly, we set a minimum threshold on the systematic uncertainty $\sigma_{\text{sys}}$. 
Specifically, $\sigma_{\text{sys}}$ was sampled as a free parameter with a uniform prior $\sigma_{\text{sys}} \sim \mathcal{U}(0.3, 1)$. 
Lowering the limit on $\sigma_{\text{sys}}$ can lead to biases in the posteriors recovered with the surrogate, since sampling the systematic uncertainty mainly accounts for potential tension between model and data, but inherent surrogate errors need to be incorporated a priori.
When we set $\sigma_{\text{sys}} =0$, i.e., turn off the systematic uncertainty entirely, we observe that the posteriors using the surrogate diverge from the posteriors using the direct \afterglowpy evaluations.
Although the surrogate posteriors still find values that are close the injection parameters, the latter are not contained in the 95\% credible limits.
Likewise, the credibility contours between the surrogate and direct \afterglowpy posterior do not overlap anymore.
Thus, adjusting for the surrogate uncertainty remains crucial.

In Appendix \ref{app:pbag_possis_injections}, we show similar injection recoveries for the \pbag and \possis surrogates.
For these models, we could not compare the \fiesta posterior to a posterior that evaluates the likelihood with the physical base model, yet we still found a good recovery of the injected parameters.

However, a systematic assessment of whether injections are generally well recovered requires going beyond individual examples.
For this reason, we ran 200 injection recoveries each for \afterglowpy, \pbag, and \possis.
This way, we obtained the distribution of the injection values' posterior quantiles across multiple inferences.
The cumulative distribution of these quantiles can be visualized in a P-P plot.
Figure~\ref{fig:pp_plot_GRB} shows the resulting P-P plot for the GRB afterglow inferences, in Fig.~\ref{fig:pp_plot_KN} we provide the P-P plot for the \possis injection recoveries.
Overall, we find that the injection values are recovered well, with the injected values lying within the posterior samples in 98.7\% of the \afterglowpy and \pbag injections and in 94.8\% the \possis injections.
However, if the posteriors were unbiased, then, according to the probability integral transform, the quantiles would adhere to a uniform distribution.
In Figs.~\ref{fig:pp_plot_GRB} and \ref{fig:pp_plot_KN}, it is apparent that this is not always the case and the cumulative distributions sometimes fall outside the grey 68\%, 95\%, or 99.7\% confidence ranges in which they would lie if they were uniform.

There are several reasons for this behavior.
In certain cases, the surrogate error might introduce biases in the recovery.
However, in Figs.~\ref{fig:pp_plot_GRB} and \ref{fig:pp_plot_KN} we also show P-P plots for cases where the injection data is generated with the surrogate instead of the base model.
These are shown as dashed lines in Figs.~\ref{fig:pp_plot_GRB} and \ref{fig:pp_plot_KN} and display the same trends as the P-P plots with the injections from the base model.
Hence, we conclude that the suboptimal recovery of certain parameters is primarily due to our inclusion of the systematic uncertainty and the way we generate the mock data.

For the GRB afterglow injection recoveries, it is the wing angle $\alpha_w$, the interstellar density $\log_{10}(n_{\rm ism})$, and the electron power index $p$ that show the most notable deviation from uniformity.
In the case of $\alpha_w$, for instance, we note a high degeneracy with the output data.
The light curve does not change noticeably when $\alpha_w$ goes from 2.5 to 3.5, unless perhaps the alignment with the observer changes, i.e., $\theta_w$ suddenly becomes larger than $\iota$.
This leads to broad posterior support for $\alpha_w$, therefore causing the more extreme quantiles for $\alpha_w$ to be overrepresented in the P-P plot.
This also applies to $\log_{10}(n_{\rm ism})$, which mostly affects the early part of the light curve prior to the jet-break.
When $\log_{10}(n_{\rm ism})$ is low, this late part of the light curve will also match a jet with somewhat higher interstellar density and the posterior will have significant support above the injected value, overrepresenting low quantiles.
Our mock injection data are spaced log-uniformly from 0.01 days to 200 days and thus in most cases will contain a large data segment from the post-jet break which may cause a degeneracy in $\log_{10}(n_{\rm ism})$.
On the other hand, for the electron power index $p$, the injection value's posterior quantile is too often too close to 0.5, i.e., $p$ is overdetermined.
This can be attributed to the fact that the value for $p$ strongly influences the post-jet break slope from the data.
The large segment of post-jet break data enables the sampler to infer $p$ with good accuracy, but the addition of the systematic uncertainty artificially broadens the posterior, which causes the injected value to lie too often at the center of the posterior.

A similar effect can be seen for some of the KN parameters in Fig.~\ref{fig:pp_plot_KN}. 
The parameters $\bar{v}_{\rm ej, dyn}$, $\bar{Y}_{\rm e, dyn}$, and $\bar{v}_{\rm ej, wind}$ show the same over-determination, i.e., their quantiles lie too often close to $0.5$. 
However, the corresponding P-P plots where injection data is constructed with the surrogate show similar or even stronger over-determination across all parameters. 
When the injection is directly with \possis, the parameters $\iota$, $\log_{10}(m_{\rm ej, dyn})$, $\log_{10}(m_{\rm ej, wind})$ also have an overrepresentation of lower quantiles.
However, this can be attributed to the fact that when we inject \possis data, we do so from randomly selected test data light curves in our set of fixed \possis simulations.
These were run on a discrete set of parameters and therefore, as seen for instance in the case of the inclination, the injected value will be exactly $0$~rad instead of some small value if the injection value was drawn from a continuous distribution.
Since $0$~rad is the prior bound in the inference, the $0$th quantile is overrepresented in the distribution of posterior quantiles.
This also offers a partial explanation for the 5.1\% of injections mentioned above, where the posterior samples lie exclusively above or below the injection value.

\subsection{Performance}
\begin{figure}
    \centering
    \includegraphics[width=1\linewidth]{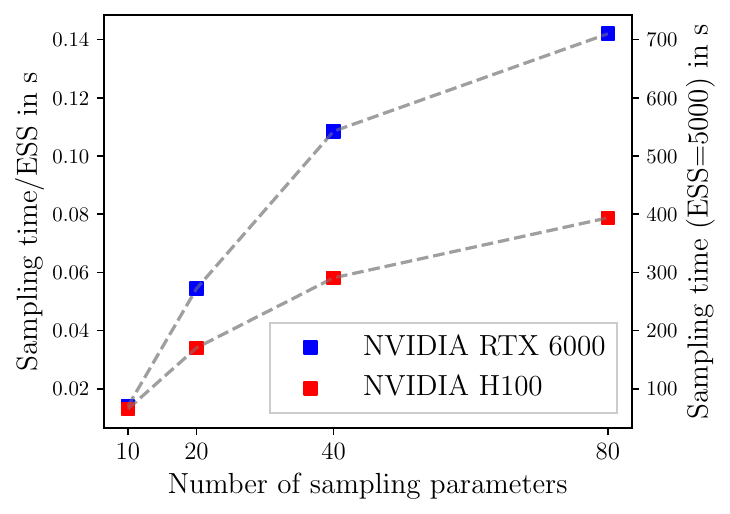}
    \caption{Sampling run time of a \fiesta inference as a function of the parameter space dimension. 
    The plot shows the runtime of a \pbag injection recovery per effective sample size (ESS) when different numbers of nuisance parameters for the time-dependent systematic uncertainty are added. The performance test was conducted on two different GPU types as indicated by the colors in the legend.
    }
    \label{fig:performance_plot.pdf}    
\end{figure}
The computational cost of sampling the posterior, $P(\vec{\theta}|d),$ is mainly determined by the cost of the likelihood function.
Using ML surrogates reduces the evaluation time of the likelihood function by several orders of magnitude.
This is best illustrated by comparing the total runtime of the inferences shown in Fig.~\ref{fig:injection_afgpy_gaussian}, where the posterior for an \afterglowpy injection was obtained with and without the surrogate.
The total sampling time with \fiesta amounts to $96$~s on an NVIDIA~H100~GPU, whereas sampling with the actual \afterglowpy model in \NMMA takes $19,700$~s ($\approx 5.5$~h) on 24~Intel$^\text{\textregistered}$~Xeon$^\text{\textregistered}$~Silver~CPUs.
Using the power consumption values for the GPUs and CPUs used~\citep{nvidia_rtx6000, nvidia_h100_2024, intel_xeon_silver_4310_2022}, this implies that inferences with \fiesta consume around $124$, respectively, $168$ less energy than the equivalent CPU-based run when using the NVIDIA H100, respectively NVIDIA RTX 6000 GPU. 
This difference is mainly due to the speed-up in the likelihood evaluation and not related to the sampling algorithm, since the run that uses the \fiesta surrogate with the \NMMA sampler takes just 203 seconds on the same 24 CPUs.
Further, it should be noted that the \flowMC sampler consists of several stages. 
First, the likelihood function and the associated neural networks are just-in-time compiled with \jax, which in our inferences takes around $60$~s. 
Then, a training loop takes place which concurrently runs MCMC sampling and training of the normalizing flow proposal, which takes another $30$~s.
The samples produced during the training loop are considered burn-in samples and are discarded.
Generating the final set of posterior samples then only takes about $5$~s.
The exact length of these different segments depends on the number of datapoints, the hyperparameters of \textsc{flowMC}, and the number of photometric filters in the data.
The more filters there are, the more often the \textsc{FluxModel} output needs to be converted to AB magnitudes, which involves the evaluation of an integral, thereby decelerating the likelihood evaluation.

Besides the cost of the likelihood function, the size of the parameter space also influences the sampling time.
The parameter space at least contains the base model parameters listed in Table~\ref{tab:surrogate_models}, but can be extended with parameters to model the systematic uncertainty. 
As mentioned above, in \fiesta, the systematic uncertainty, $\sigma_{\text{sys}}$, can either be set to a constant value or it can be constructed from a set of sampling parameters. 
By introducing these nuisance parameters $\sigma^{(k)}_{\rm sys}(f)$, the systematic uncertainty can even become time- and filter-dependent through Eq.~\eqref{eq:systematic_uncertainty}.
However, when adding parameters to the sampler, this will impact the sampling time.
In Fig.~\ref{fig:performance_plot.pdf}, we show how the sampling time (i.e. total runtime minus just-in-time compilation, which remains roughly constant over the cases considered here) per effective sample size of a \fiesta inference evolves when more systematic nuisance parameters are added.
While initially the sampling time per effective sample increases notably when going from 10 to 20 sampling dimensions, at a higher dimensionality, the increase remains limited.
This is opposite to the behavior of conventional samplers. 
For instance, in Table II of \citet{Jhawar:2024ezm} it is shown that when using the \textsc{PyMultinest} nested sampler in \NMMA, the sampling time increases from $11$~minutes when sampling $6$ parameters to $107$ minutes for a posterior of dimension $21$.
We of course note that a setting with more than $20$ nuisance parameters seems superfluous for real-life data analysis, however, Fig.~\ref{fig:performance_plot.pdf} emphasizes the capability of the \flowMC sampler to handle large parameter spaces efficiently.
This is also useful when combining different models at once, for instance, in joint analyses of GRB afterglow and KN emission.

\section{Applications}
\label{sec:applications}
In this section, we apply our newly developed inference framework to two instances of real observations, namely, AT2017gfo/GRB170817A and GRB211112A.
We use the best-performing surrogates from Sect.~\ref{sec:surrogates}, namely, the \textsc{FluxModel} cVAE's for the GRB afterglow surrogates and the \textsc{FluxModel} with MLP architecture for the KN surrogate, to jointly analyze the KN emission and GRB afterglow emission in these events.

\subsection{AT2017gfo/GRB170817A}

\begin{figure}
    \centering
    \includegraphics[width=1\linewidth]{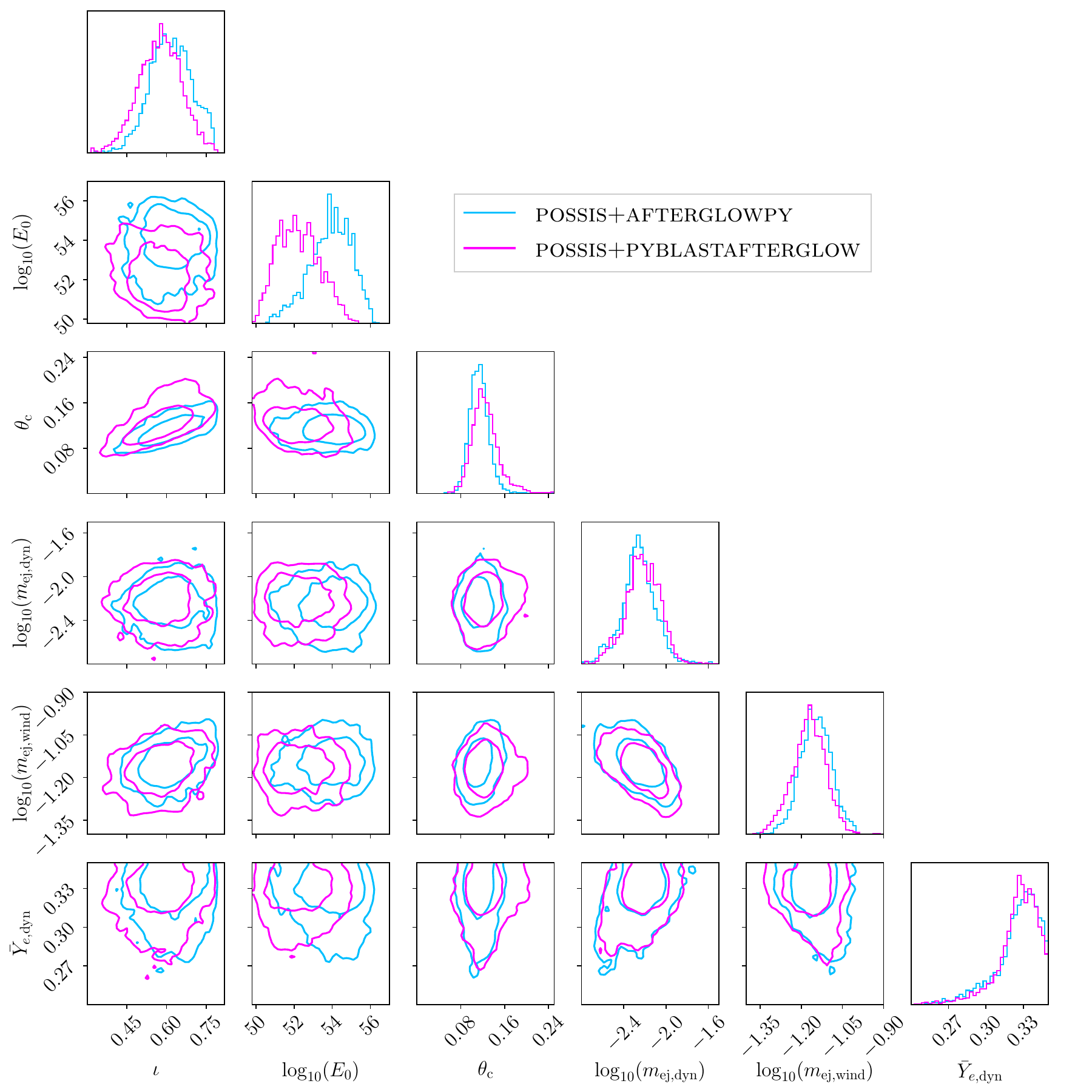}
    \caption{Posterior of the joint KN+GRB afterglow analyses of AT2017gfo/GRB170817A. Selected parameters are shown in the corner plot.
    The full corner plot can be accessed in our \href{https://github.com/nuclear-multimessenger-astronomy/paper_fiesta}{data repository}.
    The lightblue contours indicate the posterior where the GRB afterglow part is fitted with \afterglowpy, magenta is the posterior from \pbag.
    Both inferences use the KN surrogate from \possis.}

    \label{fig:posterior_GRB170817A}
\end{figure}

\begin{figure}
    \centering
    \includegraphics[width=1\linewidth]{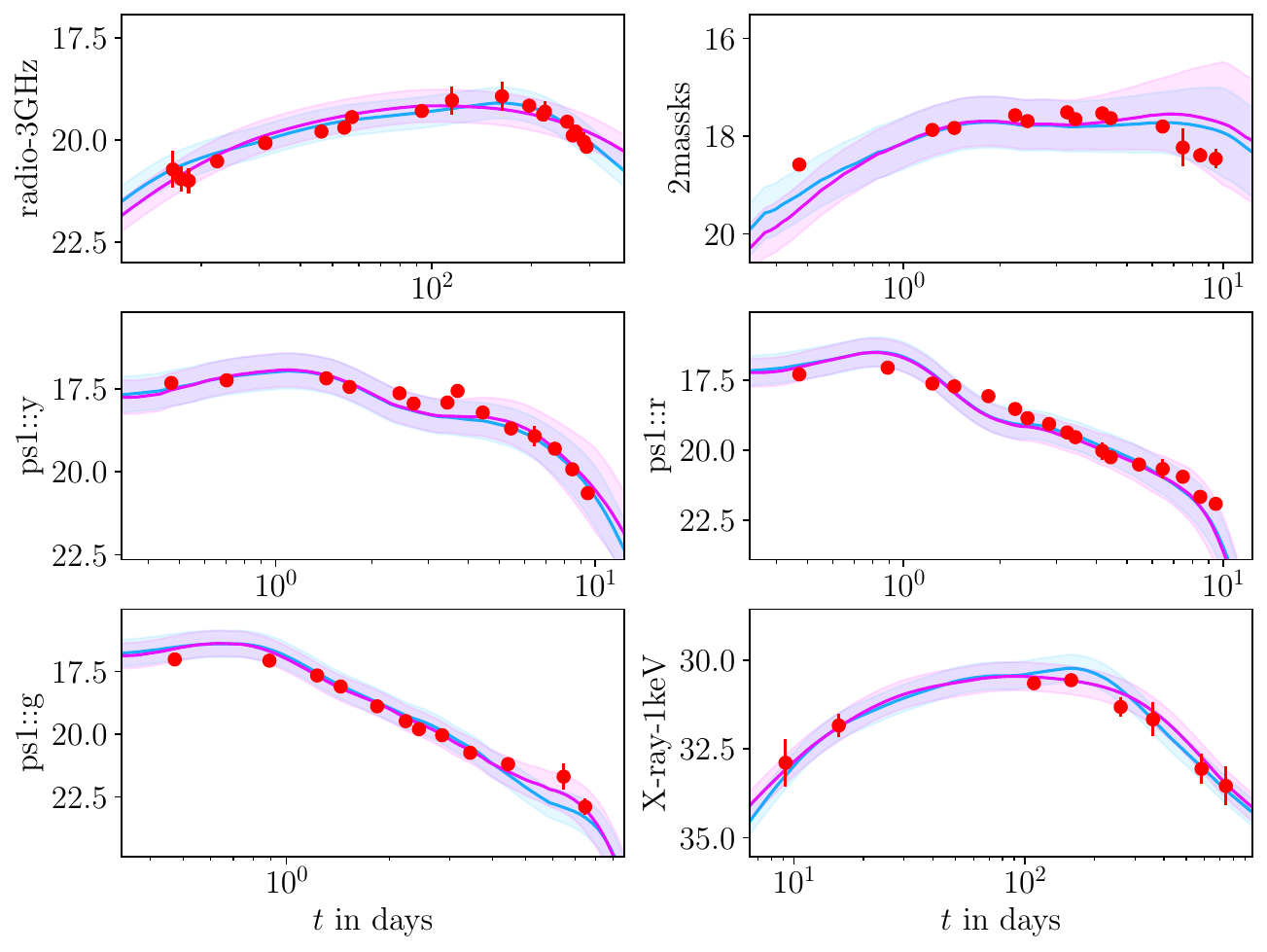}
    \caption{Best-fit light curves from the joint analyses of AT2017gfo/GRB170817A for selected photometric filters.
    The red data points show their 1$\sigma$ error bars.
    The best-fit light curves from the analysis with \afterglowpy (\pbag) are drawn as solid lines in light-blue (magenta).
    The colored bands indicate the $1\sigma$ systematic uncertainty as determined from the systematic nuisance parameters sampled for this light curve.}
    \label{fig:lightcurves_GRB170817A}
\end{figure}
As mentioned in Sect.~\ref{sec:intro}, the most prominent instance of a KN and GRB afterglow occurring together is the electromagnetic counterpart to the BNS merger associated with GW170817~\citep{LIGOScientific:2017vwq, LIGOScientific:2017ync}.
We used \fiesta to reanalyze the joint light curve of these events, performing two analyses. 
One analysis uses our KN surrogate from \possis together with the surrogate for the \afterglowpy Gaussian jet, while the other analysis uses the same KN surrogate but together with the \pbag Gaussian jet surrogate. 
Our priors are uniform within the ranges specified in Table~\ref{tab:surrogate_models}, except for the inclination, where we set $\iota\sim\mathcal{U}(0, \pi/4)$ to avoid a second mode at $\iota\approx \pi/2$.
We confirmed that this second mode is not an artifact from the surrogate, but instead a solution of \afterglowpy to the GRB170817A data, although an inclination of $\iota > \pi/4$ seems implausible given the observation of the GRB prompt emission and the radio interferometry measurement of the jet inclination angle~\citep{Mooley:2018qfh, Ghirlanda:2018uyx, Mooley:2022uqa}.
The handling of the systematic uncertainty is split between those filters containing the KN observations and those filters containing the GRB afterglow.
In particular, we find that the systematic uncertainty around the KN model needs to be represented by four parameters spaced linearly across the KN time interval, each of which is sampled from a uniform prior $\mathcal{U}(0.5, 2)$.
On the other hand, for the GRB afterglow observations only a single systematic uncertainty parameter with a uniform prior $\mathcal{U}(0.3, 2)$ is needed.
The available KN data reach from 0.3 to 24~days, the GRB afterglow data span from 9 to 742~days.
We only fit KN data points up to 10 days, since we find that later data points are not well-represented by the \possis model.
The reason for this behavior is potentially rooted in the breakdown of the local thermodynamical equilibrium in the by-then low-density environment~\citep{Waxman:2019png, Pognan:2022pix}, which causes the \possis predictions to become less applicable.
We confirmed that adding the available KN data points at $t>10$~days does not significantly alter the posterior.
We fixed the luminosity distance at $d_L=43.6$~Mpc and redshift at $z=0.009727$~\citep{Chornock:2017sdf}.

In Fig.~\ref{fig:posterior_GRB170817A}, we show the posterior of our joint KN+GRB afterglow inferences.
For the GRB afterglow parameters, we find good agreement between the \afterglowpy and \pbag models, as well as good agreement compared to previous analyses~\citep{Ryan:2019fhz, Pang:2022rzc, Ghirlanda:2018uyx, Troja:2018uns}. 
The estimated isotropic kinetic energy from \afterglowpy at $\log_{10}(E_0) = 54.0^{+1.7}_{-2.5}$ is higher than for \pbag $\log_{10}(E_0) = 52.2^{+2.3}_{-1.8}$, as is the ambient density with $\log_{10}(n_{\rm ism}) = {-0.29}^{+1.64}_{-2.86}$ from \afterglowpy and $\log_{10}(n_{\rm ism}) = {-2.25}^{+1.9}_{-2.33}$ from \pbag.
All values are quoted at the 95\% level.
We attribute this difference to the fact that generally \pbag light curves are brighter than \afterglowpy light curves due to the different microphysics and radiation scheme.
Both analyses find a jet opening angle of about $\theta_c=4-10\,^\circ$ and the inclination $\iota=23^\circ - 44^\circ$.
This value is partially consistent with analyses that include the displacement of the apparent superluminal centroid to the multiband light curve data~\citep{Mooley:2018qfh, Ghirlanda:2018uyx, Mooley:2022uqa}, where a range of $\iota \approx 14^\circ - 28^\circ$ is inferred.
Instead, more recent analyses also find $\iota = 17.2^\circ - 21.2^\circ$~\citep{Govreen-Segal:2023wvs} and $\iota = 18^\circ - 24^\circ$~\citep{Ryan:2023pzk}, which is lower than our estimate. 
Yet, our credible interval remains consistent with $\iota \approx 0^\circ -40^\circ$ inferred from the GW data alone~\citep{LIGOScientific:2017adf}, using current estimates for the Hubble constant.

The KN parameters offer a more peculiar picture than the afterglow parameters.
While the parameters for the ejecta masses broadly agree with estimates from previous works~\citep{Anand:2023jbz, Breschi:2024qlc, Pang:2022rzc, Sarin:2023khf}, we find that, in particular, the electron fractions converge to rather extreme values.
For both analyses, we find $\bar{Y}_{e, \text{dyn}} = 0.33^{+0.02}_{-0.05}$, and $Y_{e, \text{wind}} = 0.24^{+0.02}_{-0.02}$ at 95\% credibility.
These values contradict the general standard picture in which the dynamical ejecta are neutron-rich, i.e. $Y_{e, \text{dyn}}\lesssim 0.25$, and the polar wind ejecta has a higher electron fraction~\citep{Metzger:2014ila, Metzger:2019zeh, Shibata:2019wef, Kasen:2017sxr, Nedora:2020hxc}.
It should be noted, however, that our parameter $\bar{Y}_{e,\text{dyn}}$ refers to the mass-averaged electron fraction according to the distribution from Eq.~(2) in \cite{Anand:2023jbz}.
Hence, the dynamical ejecta would still contain some portion with low electron fraction. 
Nevertheless, this value seems high compared to the (uniform) electron fraction in the wind component.
The posterior distributions also indicate very low values for $v_{\text{ej, wind}}$ at the prior edge of 0.05~$c$, which is in line with previous works~\citep{Breschi:2021tbm, Breschi:2024qlc, Anand:2023jbz}.
By comparing the \possis training and test data to the AT2017gfo light curve, we confirm that the aforementioned issues are not linked to the performance of our ML surrogate, but instead point towards a systematic issue.
Specifically, it seems hard to reconcile the slow descent of the $g$ band with the steep decline of the infrared magnitudes in the 2mass filters.
This can be seen in Fig.~\ref{fig:lightcurves_GRB170817A} where we show the best-fit light curves from our analyses for selected filters.
Some tension between the bluer and redder components in AT2017gfo has been noted for instance in \cite{Breschi:2021tbm} and \cite{Hussenot-Desenonges:2025gik} as well.
As mentioned, this might be related to the breakdown of local thermodynamical equilibrium in the late-time ejecta~\citep{Waxman:2019png, Pognan:2022pix}, or other systematic reasons such as dominance of a few nucleonic decays in the heating process~\citep{Kasen:2018drm}, unexpected opacity evolution~\citep{Kasen:2018drm, Tanaka:2019iqp, Pognan:2022pix, Gillanders:2023jpd}, or ejecta geometry~\citep{Collins:2023btn, King:2025tqo}, and emphasizes the need for further research on the early and late KN emission mechanisms.

\subsection{GRB211211A}
\begin{figure}
    \centering
    \includegraphics[width=1\linewidth]{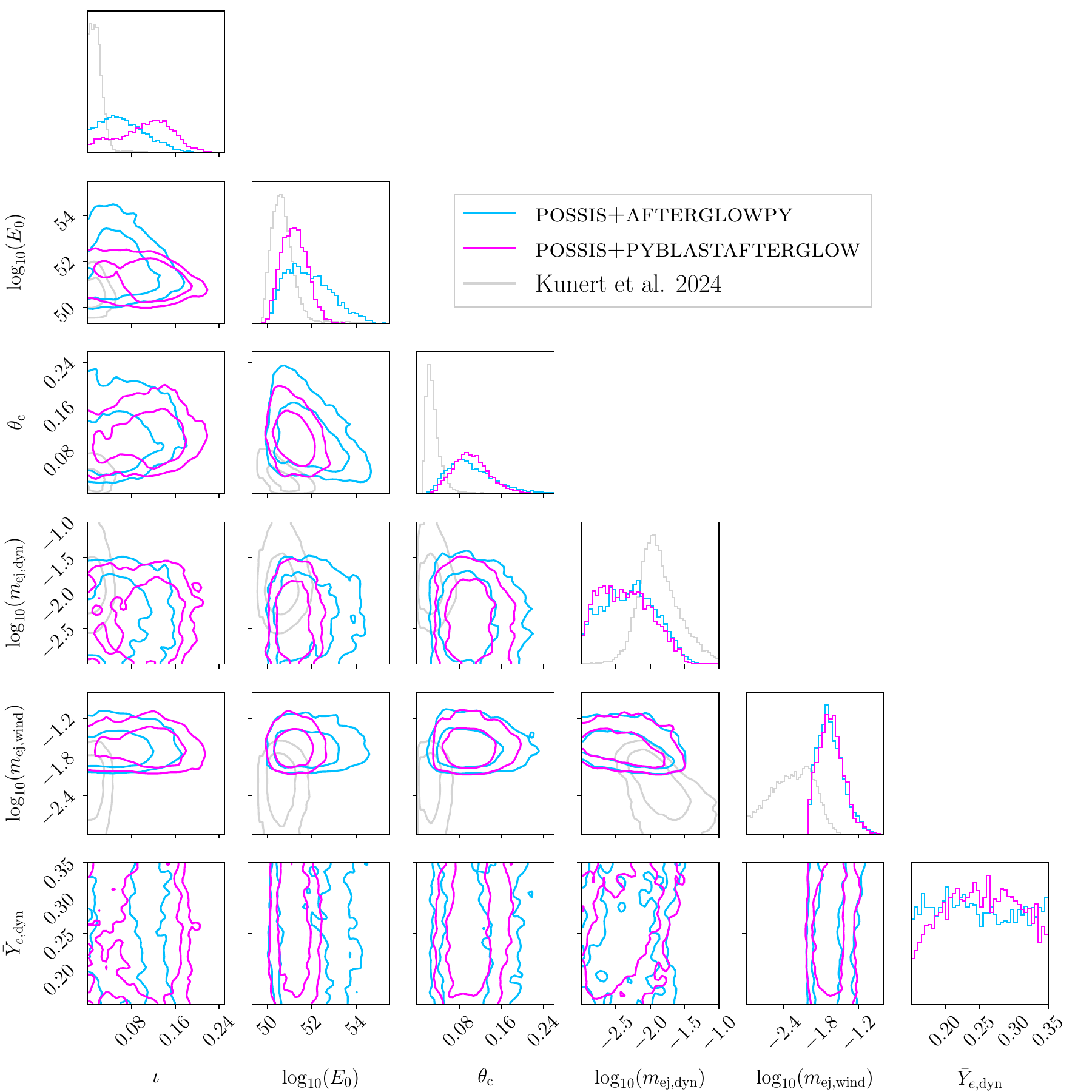}
    \caption{Posterior of the joint KN+GRB afterglow analyses of GRB211211A. Selected parameters are shown in the corner plot.
    The full corner plot can be accessed in our \href{https://github.com/nuclear-multimessenger-astronomy/paper_fiesta}{data repository}.
    The lightblue contours indicate the posterior when the GRB afterglow part is fit with \afterglowpy, magenta is the posterior from \pbag.
    Both inferences use the KN surrogate from \possis.
    The lightgrey contours show the analysis from \cite{Kunert:2023vqd} as a reference.}
    \label{fig:posterior_GRB211211A}
\end{figure}
\begin{figure}
    \centering
    \includegraphics[width=1\linewidth]{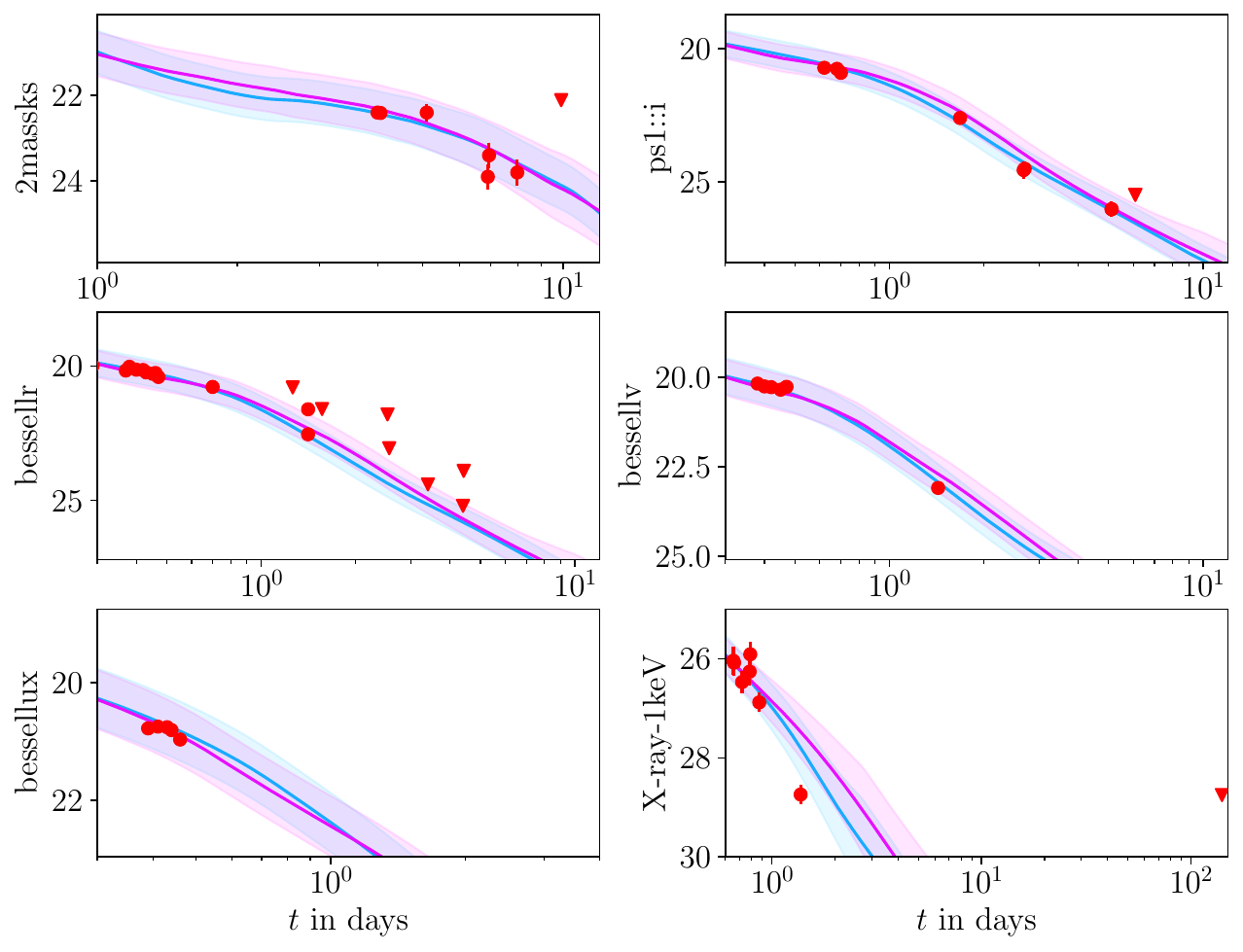}
    \caption{Best-fit light curves from the joint KN+GRB afterglow analyses of GRB211211A for selected photometric filters.
    Layout as in Fig.~\ref{fig:lightcurves_GRB170817A}.
    Detection limits are shown as triangles.}
    \label{fig:lightcurves_GRB211211A}
\end{figure}
GRB211211A was a long GRB observed with the Swift observatory and later associated with a relatively nearby host at $\sim350$~Mpc~\citep{Rastinejad:2022zbg}.
Subsequent analyses of the near-optical and X-ray emission indicated that a KN component was contributing to the emission~\citep{Rastinejad:2022zbg, Troja:2022yya, Mei:2022ncd, Yang:2022qmy, Kunert:2023vqd}.
We reanalyze the light curve data compiled in \cite{Kunert:2023vqd} with \fiesta, using the same surrogates as above for AT2017gfo/GRB170817A. 
Since the time range of the \possis surrogate is restricted to the timespan of 0.2--26 days and the validity of the \possis base model might be restricted to an even smaller time range, we exclude early data points before 0.2~days in the X-ray and uvot filters.
We set the priors to be uniform within the ranges from Table~\ref{tab:surrogate_models} and fix the luminosity distance at $d_L=358$~Mpc and the redshift at $z=0.0763$, using Planck18 cosmology~\citep{Planck:2018vyg}.
Unlike in AT2017gfo/GRB170817A, the GRB afterglow and KN emission are not featured separately in the light curve data and are superimposed in the UV, optical, and infrared due to the GRB afterglow being seen on axis. 
Therefore, we sample four systematic nuisance parameters spaced linearly between 0.2--10~days, sampled from a uniform prior $\mathcal{U}(0.5, 2)$. 
For the radio and X-ray band, we set four separate systematic uncertainty parameters that cover the time range from 0.2 to 10~days, sampled from a uniform prior $\mathcal{U}(0.3, 2)$.
This time range is sufficient, even though the X-ray data contain a late data point at 150 days.
However, this data point is a detection limit and is fulfilled by our fit regardless of the systematic uncertainty.
The posteriors are shown in Fig.~\ref{fig:posterior_GRB211211A}, the best-fit light curves in Fig.~\ref{fig:lightcurves_GRB211211A}.
As in the case for GRB170817A, the inferred GRB afterglow parameters do not differ significantly between the analyses with \afterglowpy and \pbag.
The energy from \afterglowpy is again slightly higher as the energy inferred from \pbag, $\log_{10}(E_0) = {51.7}^{+2.3}_{-1.4}$ compared to $\log_{10}(E_0) = {51.2}^{+1.2}_{-1.0}$, which is also the case for the interstellar densities from \afterglowpy $\log_{10}(n_{\rm ism})={0.5}_{-3.4}^{+1.4}$ compared to \pbag at $\log_{10}(n_{\rm ism})={1.4}_{-3.2}^{+0.6}$.
The microphysical parameters $p$, $\epsilon_e$, and $\epsilon_B$ are consistent between \afterglowpy and \pbag.

The KN parameters are relatively unconstrained and some of their marginalized posterior distributions reach down to the lower prior bound.
The KN parameters for the dynamical component, i.e., its velocity and electron fraction, essentially recover the prior.
The wind ejecta parameters are somewhat better constrained, though still relatively broad with $v_{\rm ej, wind}= {0.12}_{-0.05}^{+0.03}$ and $Y_{e, \text{wind}} = {0.27}_{-0.07}^{+0.09}$.
Despite this fact, we find that the KN component is likely needed to explain the data of GRB211211A, as a simple GRB afterglow analysis results in poor fits.
Specifically, the reduced chi-squared statistic $\chi^2/\nu$ is 0.24 (0.28) in the joint inference with \afterglowpy (\pbag) and the KN model, which increases to 1.0 (0.98) when the KN contribution is not included.
This is in line with the findings based on Bayesian model selection from~\cite{Kunert:2023vqd}.

In Fig.~\ref{fig:posterior_GRB211211A} we also show the results from the analysis of \cite{Kunert:2023vqd}. 
Though this analysis also uses the \afterglowpy Gaussian jet model, unlike our analysis it does not discard early data points before 0.2~days and relies on a different kilonova model~\citep{Bulla:2019muo, Dietrich:2020efo, Anand:2020eyg, Almualla:2021znj} with different priors for ejecta masses.
Additionally, \cite{Kunert:2023vqd} also uses an inclination prior that is dependent on the sampled jet opening angle.
Hence, the inferred parameters slightly differ, most notably for the ejecta masses and the interstellar medium density. 
However, the total ejecta masses $m_{\rm ej} = m_{\rm ej,dyn} + m_{\rm ej, wind}$ obtained from the two different kilonova models are consistent. 
Our analyses both result in $m_{\rm ej} = {0.028}_{-0.012}^{+0.045}$, while the analysis of \cite{Kunert:2023vqd} finds $m_{\rm ej} = {0.021}_{-0.011}^{+0.043}$.
This indicates that even if different assumptions in KN modeling lead to quantitatively different results for the specific model parameters, the total ejecta mass can be inferred consistently.

\section{Discussion and conclusion}
\label{sec:conclusion}
The \fiesta package provides ML surrogates of KN and GRB afterglow models to enable efficient likelihood evaluation when sampling a posterior from photometric transient data.
It offers a flexible API for training these surrogates, enabling the use of effectively three different architectures.
Since \fiesta utilizes the \jax~framework, training can be hardware-accelerated by execution on a GPU.
Additionally, \fiesta includes functionalities to sample the light curve posterior with \flowMC~\citep{Wong:2022xvh}, an adaptive MCMC sampler that uses gradient-based samplers, and normalizing flow proposals that can also be run on GPUs.

To demonstrate \fiesta's features, we present surrogates for the \afterglowpy~\citep{Ryan:2019fhz} and \pbag~\citep{Nedora:2024vrv} GRB afterglow models, as well as for the KN model from \possis~\citep{Bulla:2019muo, Bulla:2022mwo}.
We find that the surrogates perform well against the respective test data sets, as the prediction error is usually bound within 0.3--0.5~mag. 
When used during inference, this finite prediction error needs to be offset with a systematic uncertainty, $\sigma_{\rm sys}$, which can also be sampled in a flexible manner within \fiesta's implementation.
We applied these surrogates to two events with joint KN and GRB afterglow emission, namely AT2017gfo/GRB170817A and GRB211211A.
Using our surrogates, we find similar results to previous analyses of these events, even when using the new \pbag model for the GRB afterglow.
However, our posteriors can be evaluated within minutes; whereas previous analyses, for instance, those of \cite{Pang:2022rzc, Koehn:2024set}, could take several hours to days, mainly due to the more expensive likelihood evaluation when computing the light curve directly from \afterglowpy.

Despite the advantages regarding speed when relying on ML surrogates in the likelihood evaluation, this approach also comes with some drawbacks.
For one, the parameter ranges in which our surrogates interpolate the training data are fixed and, therefore, it is not possible to extend priors without retraining the surrogates.
However, these surrogates can easily be fine-tuned on new training datasets for future use in other places in the parameter space.
Further, the prediction error from the surrogate compared to the physical base model could introduce biases in the posteriors.
We assessed these potential biases by creating \afterglowpy injection recoveries, where the posterior can be obtained with the surrogate and with the base model.
We find that the values for the KL divergence match theoretical expectations within the surrogate uncertainty~\citep{Bevins:2025srk}.
Furthermore, while we did find biases in the recovery of certain parameters (based on our P-P plots in Figs.~\ref{fig:pp_plot_GRB} and \ref{fig:pp_plot_KN}), we do not find that this bias is necessarily caused by the use of our surrogates.
Still, failure of the surrogate in certain parameter regions cannot be excluded and sanity checks are recommended to validate the results.
For instance, the physical base model could be run with the best-fit parameters from the posterior to verify that they actually provide a good fit to the data.

The use of ML techniques for accelerated likelihood evaluation and efficient sampling in Bayesian inference is a common practice in various areas of astrophysics.
For instance, the analysis of X-ray spectra has recently been accelerated by the use of advanced stochastic samplers written in \textsc{jax}~\citep{jaxspec} or with the help of ML surrogate models~\citep{Barret:2024kvc, Tutone:2025hwh, Dupourque:2025vqe}. 
For supernovae, Gaussian process surrogate models were developed for inference in~\cite{Simongini:2024chh, Simongini:2025gse}.
\cite{Leeney:2025pwe} created a \textsc{jax}-compatible version of \textsc{sncosmo}~\citep{sncosmo}.
In \fiesta, we focused on surrogates for electromagnetic transients of BNS mergers; however, in principle, surrogates for other types of transients could also be incorporated.
Additionally, as \fiesta surrogates utilize the \jax framework, this opens up the possibility for them to be combined with other \jax software, such as \jim~\citep{Wong:2023lgb, Wouters:2024oxj} for the parameter estimation on GW signals from BNSs. 
This would enable fast sampling for the joint analysis of GW signals and electromagnetic counterparts and could be linked further to the inference of the EOS for NS matter through the \jester~\citep{Wouters:2025zju} package, which also leverages \jax to efficiently evaluate the likelihood of an EOS candidate with a given set of NS observations.
Additionally, the backward compatibility of our surrogates to \NMMA provides the opportunity to perform efficient Bayesian model comparison to either compare transient types~\citep{Breschi:2021tbm, Kunert:2023vqd, Hussenot-Desenonges:2023oll} or to investigate different systematics in the modeling. 
Such investigations could be related to KN ejecta morphology~\citep{Heinzel:2020qlt, King:2025tqo, Hussenot-Desenonges:2025gik}, KN heating rates and opacities~\citep{Tanaka:2019iqp, Bulla:2022mwo, Sarin:2024tja, Brethauer:2024zxg}, GRB jet structure~\citep{Kunert:2023vqd, Hayes:2019hso, Lin:2021auf} or environment~\citep{Aksulu:2021crt}, or mappings from BNS parameters to ejecta parameters~\citep{Ristic:2025bvt}.
Given the launch of new transient observatories such as the Vera C. Rubin Observatory~\citep{Andreoni:2024pkp} or ULTRASAT~\citep{Shvartzvald:2023ofi}, the surrogates and accelerated inference techniques presented here could also be used to study different observing strategies to increase the efficiency of parameter estimation~\citep{Ragosta:2024bfr, Andrade:2025gme} and maximize the science output of future detections.

\begin{acknowledgements}
We would like to thank the anonymous referee for their helpful suggestions.
We would like to thank Kaze Wong for useful discussions and assistance with the \flowMC package.
We would like to thank Nina Kunert for providing the light curve data of GRB211211A.
We would like to thank Anna Neuweiler for her assistance with setting up the \possis code.
Computations were in part performed on the national supercomputer HPE Apollo Hawk at the High Performance Computing Center Stuttgart (HLRS) under the grant number GWanalysis/44189 and on the DFG-funded research cluster jarvis at the University of Potsdam (INST 336/173-1; project number: 502227537). 
This work used the Dutch national e-infrastructure with the support of the SURF Cooperative using grant no. EINF-8596.
H.~K., H.~R., and T.~D. acknowledge funding from the Daimler and Benz Foundation for the project “NUMANJI” and from the European Union (ERC, SMArt, 101076369). 
T.W. is supported by the research program of the Netherlands Organization for Scientific Research (NWO) under grant number OCENW.XL21.XL21.038. P.T.H.P. is supported by the research program of the Netherlands Organization for Scientific Research (NWO) under grant number VI.Veni.232.021.
M.B. acknowledges the Department of Physics and Earth Science of the University of Ferrara for the financial support through the FIRD 2024 grant.
Views and opinions expressed are those of the authors only and do not necessarily reflect those of the European Union or the European Research Council. Neither the European Union nor the granting authority can be held responsible for them.
\end{acknowledgements}

\bibliographystyle{aa.bst}
\bibliography{bibliography} 

@article{Ahumada:2025ubp,
    author = "Ahumada, Tom{\'a}s and others",
    title = "{LIGO/Virgo/KAGRA neutron star merger candidate S250206dm: Zwicky Transient Facility observations}",
    eprint = "2507.00357",
    archivePrefix = "arXiv",
    primaryClass = "astro-ph.HE",
    journal = "ArXiv e-prints",
    month = "7",
    year = "2025"
}

@article{Almualla:2021znj,
    author = "Almualla, Mouza and Ning, Yuhong and Salehi, Pouyan and Bulla, Mattia and Dietrich, Tim and Coughlin, Michael W. and Guessoum, Nidhal",
    title = "{Using Neural Networks to Perform Rapid High-Dimensional Kilonova Parameter Inference}",
    eprint = "2112.15470",
    archivePrefix = "arXiv",
    primaryClass = "astro-ph.HE",
    journal = "ArXiv e-prints",
    month = "12",
    year = "2021"
}

@article{Anand:2020eyg,
    author = "Anand, Shreya and others",
    title = "{Optical follow-up of the neutron star{\textendash}black hole mergers S200105ae and S200115j}",
    eprint = "2009.07210",
    archivePrefix = "arXiv",
    primaryClass = "astro-ph.HE",
    doi = "10.1038/s41550-020-1183-3",
    journal = "Nature Astron.",
    volume = "5",
    number = "1",
    pages = "46--53",
    year = "2021"
}

@article{Anand:2023jbz,
    author = "Anand, Shreya and others",
    title = "{Chemical Distribution of the Dynamical Ejecta in the Neutron Star Merger GW170817}",
    eprint = "2307.11080",
    archivePrefix = "arXiv",
    primaryClass = "astro-ph.HE",
    journal = "ArXiv e-prints",
    month = "7",
    year = "2023"
}

@article{Andrade:2025gme,
    author = "Andrade, Cristina and Alserkal, Raiyah and Manzano, Luis Salazar and Martin, Emma and Andreoni, Igor and Coughlin, Michael W. and Guessoum, Nidhal and Sandoval, Liliana Rivera",
    title = "{The Effect of Vera C. Rubin Observatory Cadence Selections on Kilonova Detectability}",
    eprint = "2502.14124",
    archivePrefix = "arXiv",
    primaryClass = "astro-ph.HE",
    doi = "10.1088/1538-3873/adbfbc",
    journal = "PASP",
    volume = "137",
    number = "3",
    pages = "034102",
    year = "2025"
}

@article{Andreoni:2017ppd,
    author = "Andreoni, I. and others",
    title = "{Follow up of GW170817 and its electromagnetic counterpart by Australian-led observing programs}",
    eprint = "1710.05846",
    archivePrefix = "arXiv",
    primaryClass = "astro-ph.HE",
    doi = "10.1017/pasa.2017.65",
    journal = "Publ. Astron. Soc. Austral.",
    volume = "34",
    pages = "e069",
    year = "2017"
}

@article{Andreoni:2024pkp,
    author = "Andreoni, Igor and others",
    title = "{Rubin ToO 2024: Envisioning the Vera C. Rubin Observatory LSST Target of Opportunity program}",
    eprint = "2411.04793",
    archivePrefix = "arXiv",
    primaryClass = "astro-ph.IM",
    month = "11",
    year = "2024"
}

@article{Annala:2021gom,
    author = {Annala, Eemeli and Gorda, Tyler and Katerini, Evangelia and Kurkela, Aleksi and N\"attil\"a, Joonas and Paschalidis, Vasileios and Vuorinen, Aleksi},
    title = "{Multimessenger Constraints for Ultradense Matter}",
    eprint = "2105.05132",
    archivePrefix = "arXiv",
    primaryClass = "astro-ph.HE",
    reportNumber = "HIP-2021-11/TH",
    doi = "10.1103/PhysRevX.12.011058",
    journal = "PRX",
    volume = "12",
    number = "1",
    pages = "011058",
    year = "2022"
}

@article{Ascenzi:2018mbh,
    author = "Ascenzi, Stefano and others",
    title = "{A luminosity distribution for kilonovae based on short gamma-ray burst afterglows}",
    eprint = "1811.05506",
    archivePrefix = "arXiv",
    primaryClass = "astro-ph.HE",
    doi = "10.1093/mnras/stz891",
    journal = "MNRAS",
    volume = "486",
    number = "1",
    pages = "672--690",
    year = "2019"
}

@article{Boersma:2022qtg,
    author = "Boersma, Oliver M. and van Leeuwen, Joeri",
    title = "{DeepGlow: an efficient neural-network emulator of physical afterglow models for gamma-ray bursts and gravitational-wave events}",
    eprint = "2212.10943",
    archivePrefix = "arXiv",
    primaryClass = "astro-ph.HE",
    doi = "10.1017/pasa.2023.32",
    journal = "Publ. Astron. Soc. Austral.",
    volume = "40",
    pages = "e30",
    month = "06",
    year = "2023"
}

@article{Bulla:2019muo,
    author = "Bulla, Mattia",
    title = "{POSSIS: predicting spectra, light curves and polarization for multi-dimensional models of supernovae and kilonovae}",
    eprint = "1906.04205",
    archivePrefix = "arXiv",
    primaryClass = "astro-ph.HE",
    doi = "10.1093/mnras/stz2495",
    journal = "MNRAS",
    volume = "489",
    number = "4",
    pages = "5037--5045",
    year = "2019"
}

@article{Bulla:2022mwo,
    author = "Bulla, Mattia",
    title = "{The critical role of nuclear heating rates, thermalization efficiencies, and opacities for kilonova modelling and parameter inference}",
    eprint = "2211.14348",
    archivePrefix = "arXiv",
    primaryClass = "astro-ph.HE",
    doi = "10.1093/mnras/stad232",
    journal = "MNRAS",
    volume = "520",
    number = "2",
    pages = "2558--2570",
    year = "2023"
}

@article{Breschi:2021tbm,
    author = "Breschi, Matteo and Perego, Albino and Bernuzzi, Sebastiano and Del Pozzo, Walter and Nedora, Vsevolod and Radice, David and Vescovi, Diego",
    title = "{AT2017gfo: Bayesian inference and model selection of multicomponent kilonovae and constraints on the neutron star equation of state}",
    eprint = "2101.01201",
    archivePrefix = "arXiv",
    primaryClass = "astro-ph.HE",
    doi = "10.1093/mnras/stab1287",
    journal = "MNRAS",
    volume = "505",
    number = "2",
    pages = "1661--1677",
    year = "2021"
}

@article{Breschi:2024qlc,
    author = "Breschi, Matteo and Gamba, Rossella and Carullo, Gregorio and Godzieba, Daniel and Bernuzzi, Sebastiano and Perego, Albino and Radice, David",
    title = "{Bayesian inference of multimessenger astrophysical data: Joint and coherent inference of gravitational waves and kilonovae}",
    eprint = "2401.03750",
    archivePrefix = "arXiv",
    primaryClass = "astro-ph.HE",
    doi = "10.1051/0004-6361/202449173",
    journal = "A\&A",
    volume = "689",
    pages = "A51",
    year = "2024"
}

@article{Brethauer:2024zxg,
    author = "Brethauer, Daniel and Kasen, Daniel and Margutti, Raffaella and Chornock, Ryan",
    title = "{Impact of Systematic Modeling Uncertainties on Kilonova Property Estimation}",
    eprint = "2408.02731",
    archivePrefix = "arXiv",
    primaryClass = "astro-ph.HE",
    doi = "10.3847/1538-4357/ad7d83",
    journal = "ApJ",
    volume = "975",
    number = "2",
    pages = "213",
    year = "2024"
}

@article{Coulter:2017wya,
    author = "Coulter, D. A. and others",
    title = "{Swope Supernova Survey 2017a (SSS17a), the Optical Counterpart to a Gravitational Wave Source}",
    eprint = "1710.05452",
    archivePrefix = "arXiv",
    primaryClass = "astro-ph.HE",
    doi = "10.1126/science.aap9811",
    journal = "Science",
    volume = "358",
    pages = "1556",
    year = "2017"
}

@article{Chornock:2017sdf,
    author = "Chornock, R. and others",
    title = "{The Electromagnetic Counterpart of the Binary Neutron Star Merger LIGO/VIRGO GW170817. IV. Detection of Near-infrared Signatures of r-process Nucleosynthesis with Gemini-South}",
    eprint = "1710.05454",
    archivePrefix = "arXiv",
    primaryClass = "astro-ph.HE",
    reportNumber = "FERMILAB-PUB-17-497-A-AE-CD",
    doi = "10.3847/2041-8213/aa905c",
    journal = "ApJL",
    volume = "848",
    number = "2",
    pages = "L19",
    year = "2017"
}

@article{Collins:2023btn,
    author = "Collins, Christine E. and others",
    title = "{Towards inferring the geometry of kilonovae}",
    eprint = "2309.05579",
    archivePrefix = "arXiv",
    primaryClass = "astro-ph.HE",
    doi = "10.1093/mnras/stae571",
    journal = "MNRAS",
    volume = "529",
    number = "2",
    pages = "1333--1346",
    year = "2024"
}

@article{Curtis:2021guz,
    author = {Curtis, Sanjana and M\"osta, Philipp and Wu, Zhenyu and Radice, David and Roberts, Luke and Ricigliano, Giacomo and Perego, Albino},
    title = "{r-process nucleosynthesis and kilonovae from hypermassive neutron star post-merger remnants}",
    eprint = "2112.00772",
    archivePrefix = "arXiv",
    primaryClass = "astro-ph.HE",
    doi = "10.1093/mnras/stac3128",
    journal = "MNRAS",
    volume = "518",
    number = "4",
    pages = "5313--5322",
    year = "2022"
}

@article{Dax:2024mcn,
    author = {Dax, Maximilian and Green, Stephen R. and Gair, Jonathan and Gupte, Nihar and P{\"u}rrer, Michael and Raymond, Vivien and Wildberger, Jonas and Macke, Jakob H. and Buonanno, Alessandra and Sch{\"o}lkopf, Bernhard},
    title = "{Real-time inference for binary neutron star mergers using machine learning}",
    eprint = "2407.09602",
    archivePrefix = "arXiv",
    primaryClass = "gr-qc",
    reportNumber = "LIGO-P2400294",
    doi = "10.1038/s41586-025-08593-z",
    journal = "Nature",
    volume = "639",
    number = "8053",
    pages = "49--53",
    year = "2025"
}

@article{DES:2017kbs,
    author = "Soares-Santos, M. and others",
    collaboration = "DES, Dark Energy Camera GW-EM",
    title = "{The Electromagnetic Counterpart of the Binary Neutron Star Merger LIGO/Virgo GW170817. I. Discovery of the Optical Counterpart Using the Dark Energy Camera}",
    eprint = "1710.05459",
    archivePrefix = "arXiv",
    primaryClass = "astro-ph.HE",
    reportNumber = "FERMILAB-PUB-17-454-AE-CD-PPD",
    doi = "10.3847/2041-8213/aa9059",
    journal = "ApJL",
    volume = "848",
    number = "2",
    pages = "L16",
    year = "2017"
}

@article{Dietrich:2020efo,
    author = "Dietrich, Tim and Coughlin, Michael W. and Pang, Peter T. H. and Bulla, Mattia and Heinzel, Jack and Issa, Lina and Tews, Ingo and Antier, Sarah",
    title = "{Multimessenger constraints on the neutron-star equation of state and the Hubble constant}",
    eprint = "2002.11355",
    archivePrefix = "arXiv",
    primaryClass = "astro-ph.HE",
    reportNumber = "LA-UR-20-21470",
    doi = "10.1126/science.abb4317",
    journal = "Science",
    volume = "370",
    number = "6523",
    pages = "1450--1453",
    year = "2020"
}

@software{flax2020github,
  author = {Jonathan Heek and Anselm Levskaya and Avital Oliver and Marvin Ritter and Bertrand Rondepierre and Andreas Steiner and Marc van {Z}ee},
  title = {{F}lax: A neural network library and ecosystem for {JAX}},
  url = {http://github.com/google/flax},
  version = {0.10.2},
  year = {2024},
}

@article{Ford:2023bdi,
    author = "Ford, N. M. and Vieira, Nicholas and Ruan, John J. and Haggard, Daryl",
    title = "{KilonovAE: Exploring Kilonova Spectral Features with Autoencoders}",
    eprint = "2308.15657",
    archivePrefix = "arXiv",
    primaryClass = "astro-ph.HE",
    doi = "10.3847/1538-4357/ad0b7d",
    journal = "ApJ",
    volume = "961",
    number = "1",
    pages = "119",
    year = "2024"
}

@article{Gabrie:2021tlu,
    author = "Gabri\'e, Marylou and Rotskoff, Grant M. and Vanden-Eijnden, Eric",
    title = "{Adaptive Monte Carlo augmented with normalizing flows}",
    eprint = "2105.12603",
    archivePrefix = "arXiv",
    primaryClass = "physics.data-an",
    doi = "10.1073/pnas.2109420119",
    journal = "PNAS",
    volume = "119",
    number = "10",
    pages = "e2109420119",
    year = "2022"
}

@article{Guillochon:2017bmg,
    author = "Guillochon, James and Nicholl, Matt and Villar, V. Ashley and Mockler, Brenna and Narayan, Gautham and Mandel, Kaisey S. and Berger, Edo and Williams, Peter K. G.",
    title = "{MOSFiT: Modular Open-Source Fitter for Transients}",
    eprint = "1710.02145",
    archivePrefix = "arXiv",
    primaryClass = "astro-ph.IM",
    doi = "10.3847/1538-4365/aab761",
    journal = "ApJS",
    volume = "236",
    number = "1",
    pages = "6",
    year = "2018"
}

@article{Gianfagna:2023cgk,
    author = "Gianfagna, Giulia and Piro, Luigi and Pannarale, Francesco and Van Eerten, Hendrik and Ricci, Fulvio and Ryan, Geoffrey",
    title = "{Potential biases and prospects for the Hubble constant estimation via electromagnetic and gravitational-wave joint analyses}",
    eprint = "2309.17073",
    archivePrefix = "arXiv",
    primaryClass = "astro-ph.HE",
    doi = "10.1093/mnras/stae198",
    journal = "MNRAS",
    volume = "528",
    number = "2",
    pages = "2600--2613",
    year = "2024"
}

@article{Ghirlanda:2018uyx,
    author = "Ghirlanda, G. and others",
    title = "{Compact radio emission indicates a structured jet was produced by a binary neutron star merger}",
    eprint = "1808.00469",
    archivePrefix = "arXiv",
    primaryClass = "astro-ph.HE",
    doi = "10.1126/science.aau8815",
    journal = "Science",
    volume = "363",
    pages = "968",
    year = "2019"
}

@article{Gillanders:2023jpd,
    author = "Gillanders, J. H. and Sim, S. A. and Smartt, S. J. and Goriely, S. and Bauswein, A.",
    title = "{Modelling the spectra of the kilonova AT2017gfo \textendash{} II. Beyond the photospheric epochs}",
    eprint = "2306.15055",
    archivePrefix = "arXiv",
    primaryClass = "astro-ph.HE",
    doi = "10.1093/mnras/stad3688",
    journal = "MNRAS",
    volume = "529",
    number = "3",
    pages = "2918--2945",
    year = "2024"
}

@article{Goldstein:2017mmi,
    author = "Goldstein, A. and others",
    title = "{An Ordinary Short Gamma-Ray Burst with Extraordinary Implications: Fermi-GBM Detection of GRB 170817A}",
    eprint = "1710.05446",
    archivePrefix = "arXiv",
    primaryClass = "astro-ph.HE",
    doi = "10.3847/2041-8213/aa8f41",
    journal = "ApJL",
    volume = "848",
    number = "2",
    pages = "L14",
    year = "2017"
}

@article{Govreen-Segal:2023wvs,
    author = "Govreen-Segal, Taya and Nakar, Ehud",
    title = "{Analytic model for off-axis GRB afterglow images {\textendash} geometry measurement and implications for measuring H0}",
    eprint = "2302.10211",
    archivePrefix = "arXiv",
    primaryClass = "astro-ph.HE",
    doi = "10.1093/mnras/stad1628",
    journal = "MNRAS",
    volume = "524",
    number = "1",
    pages = "403--425",
    year = "2023"
}

@article{Guven:2020dok,
    author = {G\"uven, H. and Bozkurt, K. and Khan, E. and Margueron, J.},
    title = "{Multimessenger and multiphysics Bayesian inference for the GW170817 binary neutron star merger}",
    eprint = "2001.10259",
    archivePrefix = "arXiv",
    primaryClass = "nucl-th",
    doi = "10.1103/PhysRevC.102.015805",
    journal = "PRC",
    volume = "102",
    number = "1",
    pages = "015805",
    year = "2020"
}

@article{Hayes:2019hso,
    author = "Hayes, Fergus and Heng, Ik Siong and Veitch, John and Williams, Daniel",
    title = "{Comparing Short Gamma-Ray Burst Jet Structure Models}",
    eprint = "1911.04190",
    archivePrefix = "arXiv",
    primaryClass = "astro-ph.HE",
    doi = "10.3847/1538-4357/ab72fc",
    journal = "ApJ",
    volume = "891",
    pages = "124",
    year = "2020"
}

@article{Heinzel:2020qlt,
    author = "Heinzel, J. and Coughlin, M. W. and Dietrich, T. and Bulla, M. and Antier, S. and Christensen, N. and Coulter, D. A. and Foley, R. J. and Issa, L. and Khetan, N.",
    title = "{Comparing inclination dependent analyses of kilonova transients}",
    eprint = "2010.10746",
    archivePrefix = "arXiv",
    primaryClass = "astro-ph.HE",
    doi = "10.1093/mnras/stab221",
    journal = "MNRAS",
    volume = "502",
    number = "2",
    pages = "3057--3065",
    year = "2021"
}

@article{Hotokezaka:2018dfi,
    author = "Hotokezaka, Kenta and Nakar, Ehud and Gottlieb, Ore and Nissanke, Samaya and Masuda, Kento and Hallinan, Gregg and Mooley, Kunal P. and Deller, Adam. T.",
    title = "{A Hubble constant measurement from superluminal motion of the jet in GW170817}",
    eprint = "1806.10596",
    archivePrefix = "arXiv",
    primaryClass = "astro-ph.CO",
    doi = "10.1038/s41550-019-0820-1",
    journal = "Nature Astron.",
    volume = "3",
    number = "10",
    pages = "940--944",
    year = "2019"
}

@article{Hu:2024lrj,
    author = "Hu, Qian and Irwin, Jessica and Sun, Qi and Messenger, Christopher and Suleiman, Lami and Heng, Ik Siong and Veitch, John",
    title = "{Decoding Long-duration Gravitational Waves from Binary Neutron Stars with Machine Learning: Parameter Estimation and Equations of State}",
    eprint = "2412.03454",
    archivePrefix = "arXiv",
    primaryClass = "gr-qc",
    reportNumber = "LIGO-P2400567, ET-0666B-24",
    doi = "10.3847/2041-8213/ade42f",
    journal = "ApJL",
    volume = "987",
    pages = "L17",
    year = "2025"
}

@article{Humphrey:2008xz,
    author = "Humphrey, Philip J. and Liu, Wenhao and Buote, David A.",
    title = "{Chi-square and Poissonian Data: Biases Even in the High-Count Regime and How to Avoid them}",
    eprint = "0811.2796",
    archivePrefix = "arXiv",
    primaryClass = "astro-ph",
    doi = "10.1088/0004-637X/693/1/822",
    journal = "ApJ",
    volume = "693",
    pages = "822",
    year = "2009"
}

@article{Hussenot-Desenonges:2023oll,
    author = "Hussenot-Desenonges, T. and others",
    title = "{Multiband analyses of the bright GRB 230812B and the associated SN2023pel}",
    eprint = "2310.14310",
    archivePrefix = "arXiv",
    primaryClass = "astro-ph.HE",
    doi = "10.1093/mnras/stae503",
    journal = "MNRAS",
    volume = "530",
    number = "1",
    pages = "1--19",
    year = "2024"
}

@article{Hussenot-Desenonges:2025gik,
    author = "Hussenot-Desenonges, Thomas and Pillas, Marion and Antier, Sarah and Hello, Patrice and Pang, Peter T. H.",
    title = "{Kilonova modelling and parameter inference: Understanding uncertainties and evaluating compatibility between observations and models}",
    eprint = "2505.21392",
    archivePrefix = "arXiv",
    primaryClass = "astro-ph.HE",
    journal = "ArXiv e-prints",
    month = "5",
    year = "2025"
}

@article{Jhawar:2024ezm,
    author = "Jhawar, Sahil and Wouters, Thibeau and Pang, Peter T. H. and Bulla, Mattia and Coughlin, Michael W. and Dietrich, Tim",
    title = "{Data-driven approach for modeling the temporal and spectral evolution of kilonova systematic uncertainties}",
    eprint = "2410.21978",
    archivePrefix = "arXiv",
    primaryClass = "astro-ph.HE",
    doi = "10.1103/PhysRevD.111.043046",
    journal = "PRD",
    volume = "111",
    number = "4",
    pages = "043046",
    year = "2025"
}

@article{J-GEM:2017tyx,
    author = "Utsumi, Yousuke and others",
    collaboration = "J-GEM",
    title = "{J-GEM observations of an electromagnetic counterpart to the neutron star merger GW170817}",
    eprint = "1710.05848",
    archivePrefix = "arXiv",
    primaryClass = "astro-ph.HE",
    doi = "10.1093/pasj/psx118",
    journal = "Publ. Astron. Soc. Jap.",
    volume = "69",
    number = "6",
    pages = "101",
    year = "2017"
}

@article{JWST:2023jqa,
    author = "Levan, Andrew J. and others",
    collaboration = "JWST",
    title = "{Heavy-element production in a compact object merger observed by JWST}",
    eprint = "2307.02098",
    archivePrefix = "arXiv",
    primaryClass = "astro-ph.HE",
    doi = "10.1038/s41586-023-06759-1",
    journal = "Nature",
    volume = "626",
    number = "8000",
    pages = "737--741",
    year = "2024"
}

@article{Kasen:2017sxr,
    author = "Kasen, Daniel and Metzger, Brian and Barnes, Jennifer and Quataert, Eliot and Ramirez-Ruiz, Enrico",
    title = "{Origin of the heavy elements in binary neutron-star mergers from a gravitational wave event}",
    eprint = "1710.05463",
    archivePrefix = "arXiv",
    primaryClass = "astro-ph.HE",
    doi = "10.1038/nature24453",
    journal = "Nature",
    volume = "551",
    pages = "80",
    year = "2017"
}

@article{Kasen:2018drm,
    author = "Kasen, Daniel and Barnes, Jennifer",
    title = "{Radioactive Heating and Late Time Kilonova Light Curves}",
    eprint = "1807.03319",
    archivePrefix = "arXiv",
    primaryClass = "astro-ph.HE",
    doi = "10.3847/1538-4357/ab06c2",
    journal = "ApJ",
    volume = "876",
    number = "2",
    pages = "128",
    year = "2019"
}

@article{Kawaguchi:2018ptg,
    author = "Kawaguchi, Kyohei and Shibata, Masaru and Tanaka, Masaomi",
    title = "{Radiative transfer simulation for the optical and near-infrared electromagnetic counterparts to GW170817}",
    eprint = "1806.04088",
    archivePrefix = "arXiv",
    primaryClass = "astro-ph.HE",
    doi = "10.3847/2041-8213/aade02",
    journal = "ApJL",
    volume = "865",
    number = "2",
    pages = "L21",
    year = "2018"
}

@article{Kawaguchi:2019nju,
    author = "Kawaguchi, Kyohei and Shibata, Masaru and Tanaka, Masaomi",
    title = "{Diversity of kilonova light curves}",
    eprint = "1908.05815",
    archivePrefix = "arXiv",
    primaryClass = "astro-ph.HE",
    doi = "10.3847/1538-4357/ab61f6",
    journal = "ApJ",
    volume = "889",
    number = "2",
    pages = "171",
    year = "2020"
}

@article{Kedia:2022onl,
    author = "Kedia, Atul and Ristic, Marko and O'Shaughnessy, Richard and Yelikar, Anjali B. and Wollaeger, Ryan T. and Korobkin, Oleg and Chase, Eve A. and Fryer, Christopher L. and Fontes, Christopher J.",
    title = "{Surrogate light curve models for kilonovae with comprehensive wind ejecta outflows and parameter estimation for AT2017gfo}",
    eprint = "2211.04363",
    archivePrefix = "arXiv",
    primaryClass = "astro-ph.HE",
    reportNumber = "LA-UR-22-31562, P2200324",
    doi = "10.1103/PhysRevResearch.5.013168",
    journal = "Phys. Rev. Res.",
    volume = "5",
    number = "1",
    pages = "013168",
    year = "2023"
}

@article{King:2025tqo,
    author = "King, Brendan L. and De, Soumi and Korobkin, Oleg and Coughlin, Michael W. and Pang, Peter T. H.",
    title = "{Inferring neutron star merger ejecta morphologies with kilonovae}",
    eprint = "2505.16876",
    archivePrefix = "arXiv",
    primaryClass = "astro-ph.HE",
    reportNumber = "LA-UR-25-24747",
    journal = "ArXiv e-prints",
    month = "5",
    year = "2025",
    note = "submitted to PASP"
}

@article{Kingma:2013hel,
    author = "Kingma, Diederik P. and Welling, Max",
    title = "{Auto-Encoding Variational Bayes}",
    eprint = "1312.6114",
    archivePrefix = "arXiv",
    primaryClass = "stat.ML",
    journal = "ArXiv e-prints",
    month = "12",
    year = "2013"
}

@article{Kiuchi:2017pte,
    author = "Kiuchi, Kenta and Kawaguchi, Kyohei and Kyutoku, Koutarou and Sekiguchi, Yuichiro and Shibata, Masaru and Taniguchi, Keisuke",
    title = "{Sub-radian-accuracy gravitational waveforms of coalescing binary neutron stars in numerical relativity}",
    eprint = "1708.08926",
    archivePrefix = "arXiv",
    primaryClass = "astro-ph.HE",
    doi = "10.1103/PhysRevD.96.084060",
    journal = "PRD",
    volume = "96",
    number = "8",
    pages = "084060",
    year = "2017"
}

@article{Koehn:2024set,
    author = "Koehn, Hauke and others",
    title = "{From existing and new nuclear and astrophysical constraints to stringent limits on the equation of state of neutron-rich dense matter}",
    eprint = "2402.04172",
    archivePrefix = "arXiv",
    primaryClass = "astro-ph.HE",
    reportNumber = "LA-UR-24-20420",
    doi = "10.1103/PhysRevX.15.021014",
    journal = "PRX",
    volume = "15",
    number = "2",
    pages = "021014",
    year = "2025"
}

@article{Kunert:2023vqd,
    author = "Kunert, N. and others",
    title = "{Bayesian model selection for GRB 211211A through multiwavelength analyses}",
    eprint = "2301.02049",
    archivePrefix = "arXiv",
    primaryClass = "astro-ph.HE",
    reportNumber = "LA-UR-22-32364",
    doi = "10.1093/mnras/stad3463",
    journal = "MNRAS",
    volume = "527",
    number = "2",
    pages = "3900--3911",
    year = "2024"
}

@article{Lamb:2018ohw,
    author = "Lamb, Gavin P. and Mandel, Ilya and Resmi, Lekshmi",
    title = "{Late-time Evolution of Afterglows from Off-Axis Neutron-Star Mergers}",
    eprint = "1806.03843",
    archivePrefix = "arXiv",
    primaryClass = "astro-ph.HE",
    doi = "10.1093/mnras/sty2196",
    journal = "MNRAS",
    volume = "481",
    number = "2",
    pages = "2581--2589",
    year = "2018"
}

@article{Levan:2023qdh,
    author = "Levan, Andrew J. and others",
    title = "{A long-duration gamma-ray burst of dynamical origin from the nucleus of an ancient galaxy}",
    eprint = "2303.12912",
    archivePrefix = "arXiv",
    primaryClass = "astro-ph.HE",
    doi = "10.1038/s41550-023-01998-8",
    journal = "Nature Astron.",
    volume = "7",
    number = "8",
    pages = "976--985",
    year = "2023"
}

@article{LIGOScientific:2017vwq,
    author = "Abbott, B. P. and others",
    collaboration = "LIGO Scientific, Virgo",
    title = "{GW170817: Observation of Gravitational Waves from a Binary Neutron Star Inspiral}",
    eprint = "1710.05832",
    archivePrefix = "arXiv",
    primaryClass = "gr-qc",
    reportNumber = "LIGO-P170817",
    doi = "10.1103/PhysRevLett.119.161101",
    journal = "PRL",
    volume = "119",
    number = "16",
    pages = "161101",
    year = "2017"
}

@article{LIGOScientific:2017ync,
    author = "Abbott, B. P. and others",
    collaboration = "LIGO Scientific, Virgo, Fermi GBM, INTEGRAL, IceCube, AstroSat Cadmium Zinc Telluride Imager Team, IPN, Insight-Hxmt, ANTARES, Swift, AGILE Team, 1M2H Team, Dark Energy Camera GW-EM, DES, DLT40, GRAWITA, Fermi-LAT, ATCA, ASKAP, Las Cumbres Observatory Group, OzGrav, DWF (Deeper Wider Faster Program), AST3, CAASTRO, VINROUGE, MASTER, J-GEM, GROWTH, JAGWAR, CaltechNRAO, TTU-NRAO, NuSTAR, Pan-STARRS, MAXI Team, TZAC Consortium, KU, Nordic Optical Telescope, ePESSTO, GROND, Texas Tech University, SALT Group, TOROS, BOOTES, MWA, CALET, IKI-GW Follow-up, H.E.S.S., LOFAR, LWA, HAWC, Pierre Auger, ALMA, Euro VLBI Team, Pi of Sky, Chandra Team at McGill University, DFN, ATLAS Telescopes, High Time Resolution Universe Survey, RIMAS, RATIR, SKA South Africa/MeerKAT",
    title = "{Multi-messenger Observations of a Binary Neutron Star Merger}",
    eprint = "1710.05833",
    archivePrefix = "arXiv",
    primaryClass = "astro-ph.HE",
    reportNumber = "LIGO-P1700294, VIR-0802A-17, FERMILAB-PUB-17-478-A-AE-CD",
    doi = "10.3847/2041-8213/aa91c9",
    journal = "ApJL",
    volume = "848",
    number = "2",
    pages = "L12",
    year = "2017"
}

@article{LIGOScientific:2017zic,
    author = "Abbott, B. P. and others",
    collaboration = "LIGO Scientific, Virgo, Fermi-GBM, INTEGRAL",
    title = "{Gravitational Waves and Gamma-rays from a Binary Neutron Star Merger: GW170817 and GRB 170817A}",
    eprint = "1710.05834",
    archivePrefix = "arXiv",
    primaryClass = "astro-ph.HE",
    reportNumber = "LIGO-P1700308",
    doi = "10.3847/2041-8213/aa920c",
    journal = "ApJL",
    volume = "848",
    number = "2",
    pages = "L13",
    year = "2017"
}

@article{LIGOScientific:2017adf,
    author = "Abbott, B. P. and others",
    collaboration = "LIGO Scientific, Virgo, 1M2H, Dark Energy Camera GW-E, DES, DLT40, Las Cumbres Observatory, VINROUGE, MASTER",
    title = "{A gravitational-wave standard siren measurement of the Hubble constant}",
    eprint = "1710.05835",
    archivePrefix = "arXiv",
    primaryClass = "astro-ph.CO",
    reportNumber = "LIGO-P1700296, FERMILAB-PUB-17-472-A-AE",
    doi = "10.1038/nature24471",
    journal = "Nature",
    volume = "551",
    number = "7678",
    pages = "85--88",
    year = "2017"
}

@article{Lin:2021auf,
    author = "Lin, En-Tzu and Hayes, Fergus and Lamb, Gavin P. and Heng, Ik Siong and Kong, Albert K. H. and Williams, Michael J. and Saha, Surojit and Veitch, John",
    title = "{A Bayesian Inference Framework for Gamma-ray Burst Afterglow Properties}",
    eprint = "2109.14993",
    archivePrefix = "arXiv",
    primaryClass = "astro-ph.HE",
    doi = "10.3390/universe7090349",
    journal = "Universe",
    volume = "7",
    number = "9",
    pages = "349",
    year = "2021"
}

@article{Lipunov:2017dwd,
    author = "Lipunov, V. M. and others",
    title = "{MASTER Optical Detection of the First LIGO/Virgo Neutron Star Binary Merger GW170817}",
    eprint = "1710.05461",
    archivePrefix = "arXiv",
    primaryClass = "astro-ph.HE",
    doi = "10.3847/2041-8213/aa92c0",
    journal = "ApJL",
    volume = "850",
    number = "1",
    pages = "L1",
    year = "2017"
}

@article{Lukosiute:2022tmd,
    author = "Luko\v{s}i\={u}t\.{e}, Kamil\.{e} and Raaijmakers, Geert and Doctor, Zoheyr and Soares-Santos, Marcelle and Nord, Brian",
    title = "{KilonovaNet: Surrogate models of kilonova spectra with conditional variational autoencoders}",
    eprint = "2204.00285",
    archivePrefix = "arXiv",
    primaryClass = "astro-ph.IM",
    reportNumber = "FERMILAB-PUB-21-393-E-SCD",
    doi = "10.1093/mnras/stac2342",
    journal = "MNRAS",
    volume = "516",
    number = "1",
    pages = "1137--1148",
    year = "2022"
}

@article{Marinari:1992qd,
    author = "Marinari, Enzo and Parisi, Giorgio",
    title = "{Simulated tempering: A New Monte Carlo scheme}",
    eprint = "hep-lat/9205018",
    archivePrefix = "arXiv",
    reportNumber = "ROM2F-92-06, SCCS-241",
    doi = "10.1209/0295-5075/19/6/002",
    journal = "EPL",
    volume = "19",
    pages = "451--458",
    year = "1992"
}

@article{Mei:2022ncd,
    author = "Mei, Alessio and others",
    title = "{Gigaelectronvolt emission from a compact binary merger}",
    eprint = "2205.08566",
    archivePrefix = "arXiv",
    primaryClass = "astro-ph.HE",
    doi = "10.1038/s41586-022-05404-7",
    journal = "Nature",
    volume = "612",
    number = "7939",
    pages = "236--239",
    year = "2022"
}

@article{Metzger:2014ila,
    author = "Metzger, Brian D. and Fern\'andez, Rodrigo",
    title = "{Red or blue? A potential kilonova imprint of the delay until black hole formation following a neutron star merger}",
    eprint = "1402.4803",
    archivePrefix = "arXiv",
    primaryClass = "astro-ph.HE",
    doi = "10.1093/mnras/stu802",
    journal = "MNRAS",
    volume = "441",
    pages = "3444--3453",
    year = "2014"
}

@article{Metzger:2019zeh,
    author = "Metzger, Brian D.",
    title = "{Kilonovae}",
    eprint = "1910.01617",
    archivePrefix = "arXiv",
    primaryClass = "astro-ph.HE",
    doi = "10.1007/s41114-019-0024-0",
    journal = "Living Rev. Rel.",
    volume = "23",
    number = "1",
    pages = "1",
    year = "2020"
}

@article{Miceli:2022efx,
    author = "Miceli, Davide and Nava, Lara",
    title = "{Gamma-Ray Bursts Afterglow Physics and the VHE Domain}",
    eprint = "2205.12146",
    archivePrefix = "arXiv",
    primaryClass = "astro-ph.HE",
    doi = "10.3390/galaxies10030066",
    journal = "Galaxies",
    volume = "10",
    number = "3",
    pages = "66",
    year = "2022"
}

@article{Mooley:2018qfh,
    author = "Mooley, K. P. and Deller, A. T. and Gottlieb, O. and Nakar, E. and Hallinan, G. and Bourke, S. and Frail, D. A. and Horesh, A. and Corsi, A. and Hotokezaka, K.",
    title = "{Superluminal motion of a relativistic jet in the neutron-star merger GW170817}",
    eprint = "1806.09693",
    archivePrefix = "arXiv",
    primaryClass = "astro-ph.HE",
    doi = "10.1038/s41586-018-0486-3",
    journal = "Nature",
    volume = "561",
    number = "7723",
    pages = "355--359",
    year = "2018"
}

@article{Mooley:2022uqa,
    author = "Mooley, Kunal P. and Anderson, Jay and Lu, Wenbin",
    title = "{Optical superluminal motion measurement in the neutron-star merger GW170817}",
    eprint = "2210.06568",
    archivePrefix = "arXiv",
    primaryClass = "astro-ph.HE",
    doi = "10.1038/s41586-022-05145-7",
    journal = "Nature",
    volume = "610",
    number = "7931",
    pages = "273--276",
    year = "2022"
}

@article{Mukherjee:2019qmm,
    author = "Mukherjee, Suvodip and Lavaux, Guilhem and Bouchet, Fran\c{c}ois R. and Jasche, Jens and Wandelt, Benjamin D. and Nissanke, Samaya M. and Leclercq, Florent and Hotokezaka, Kenta",
    title = "{Velocity correction for Hubble constant measurements from standard sirens}",
    eprint = "1909.08627",
    archivePrefix = "arXiv",
    primaryClass = "astro-ph.CO",
    doi = "10.1051/0004-6361/201936724",
    journal = "A\&A",
    volume = "646",
    pages = "A65",
    year = "2021"
}

@article{Nava:2012hq,
    author = "Nava, L. and Sironi, L. and Ghisellini, G. and Celotti, A. and Ghirlanda, G.",
    title = "{Afterglow emission in Gamma-Ray Bursts: I. Pair-enriched ambient medium and radiative blast waves}",
    eprint = "1211.2806",
    archivePrefix = "arXiv",
    primaryClass = "astro-ph.HE",
    doi = "10.1093/mnras/stt872",
    journal = "MNRAS",
    volume = "433",
    pages = "2107",
    year = "2013"
}

@article{Nedora:2020hxc,
    author = "Nedora, Vsevolod and Bernuzzi, Sebastiano and Radice, David and Daszuta, Boris and Endrizzi, Andrea and Perego, Albino and Prakash, Aviral and Safarzadeh, Mohammadtaher and Schianchi, Federico and Logoteta, Domenico",
    title = "{Numerical Relativity Simulations of the Neutron Star Merger GW170817: Long-Term Remnant Evolutions, Winds, Remnant Disks, and Nucleosynthesis}",
    eprint = "2008.04333",
    archivePrefix = "arXiv",
    primaryClass = "astro-ph.HE",
    doi = "10.3847/1538-4357/abc9be",
    journal = "ApJ",
    volume = "906",
    number = "2",
    pages = "98",
    year = "2021"
}

@article{Nedora:2024vrv,
    author = "Nedora, Vsevolod and Menegazzi, Ludovica Crosato and Peretti, Enrico and Dietrich, Tim and Shibata, Masaru",
    title = "{Multi-physics framework for fast modeling of gamma-ray burst afterglows}",
    eprint = "2409.16852",
    archivePrefix = "arXiv",
    primaryClass = "astro-ph.HE",
    doi = "10.1093/mnras/staf302",
    journal = "MNRAS", 
    volume = "538",
    number = "3",
    pages = "2089–2115",
    year = "2025"
}

@article{Newton:1994,
    author = "Newton, M. and Raftery, A.",
    title = "{Approximate Bayesian Inference with the Weighted Likelihood Bootstrap}",
    journal = "Journal of the Royal Statistical Society. Series B (Methodological)",
    volume = "56",
    number = "1",
    pages = "3-48",
    year = "1994",
}

@article{Nicholl:2021rcr,
    author = "Nicholl, Matt and Margalit, Ben and Schmidt, Patricia and Smith, Graham P. and Ridley, Evan J. and Nuttall, James",
    title = "{Tight multimessenger constraints on the neutron star equation of state from GW170817 and a forward model for kilonova light-curve synthesis}",
    eprint = "2102.02229",
    archivePrefix = "arXiv",
    primaryClass = "astro-ph.HE",
    doi = "10.1093/mnras/stab1523",
    journal = "MNRAS",
    volume = "505",
    number = "2",
    pages = "3016--3032",
    year = "2021"
}

@article{Pang:2022rzc,
    author = "Pang, Peter T. H. and others",
    title = "{An updated nuclear-physics and multi-messenger astrophysics framework for binary neutron star mergers}",
    eprint = "2205.08513",
    archivePrefix = "arXiv",
    primaryClass = "astro-ph.HE",
    reportNumber = "LA-UR-22-23872, LIGO-P2200150",
    doi = "10.1038/s41467-023-43932-6",
    journal = "Nature Commun.",
    volume = "14",
    number = "1",
    pages = "8352",
    year = "2023"
}

@article{Papamakarios:2019fms,
    author = "Papamakarios, George and Nalisnick, Eric and Rezende, Danilo Jimenez and Mohamed, Shakir and Lakshminarayanan, Balaji",
    title = "{Normalizing Flows for Probabilistic Modeling and Inference}",
    eprint = "1912.02762",
    archivePrefix = "arXiv",
    primaryClass = "stat.ML",
    doi = "10.5555/3546258.3546315",
    journal = "J. Machine Learning Res.",
    volume = "22",
    number = "1",
    pages = "2617--2680",
    year = "2021"
}

@article{Planck:2018vyg,
    author = "Aghanim, N. and others",
    collaboration = "Planck",
    title = "{Planck 2018 results. VI. Cosmological parameters}",
    eprint = "1807.06209",
    archivePrefix = "arXiv",
    primaryClass = "astro-ph.CO",
    doi = "10.1051/0004-6361/201833910",
    journal = "A\&A",
    volume = "641",
    pages = "A6",
    year = "2020",
    note = "[Erratum: A\&A 652, C4 (2021)]"
}

@article{Pellouin:2024gqj,
    author = "Pellouin, Cl\'ement and Daigne, Fr\'ed\'eric",
    title = "{Very high energy afterglow of structured jets: GW 170817 and prospects for future detections}",
    eprint = "2406.08254",
    archivePrefix = "arXiv",
    primaryClass = "astro-ph.HE",
    doi = "10.1051/0004-6361/202347516",
    journal = "A\&A",
    volume = "690",
    pages = "A281",
    year = "2024"
}

@article{Peng:2024jqe,
    author = "Peng, Yinglei and Risti\'c, Marko and Kedia, Atul and O'Shaughnessy, Richard and Fontes, Christopher J. and Fryer, Chris L. and Korobkin, Oleg and Mumpower, Matthew R. and Villar, V. Ashley and Wollaeger, Ryan T.",
    title = "{Kilonova light-curve interpolation with neural networks}",
    eprint = "2402.05871",
    archivePrefix = "arXiv",
    primaryClass = "astro-ph.HE",
    reportNumber = "LA-UR-24-20638",
    doi = "10.1103/PhysRevResearch.6.033078",
    journal = "Phys. Rev. Res.",
    volume = "6",
    number = "3",
    pages = "033078",
    year = "2024"
}

@article{Pognan:2022pix,
    author = "Pognan, Quentin and Jerkstrand, Anders and Grumer, Jon",
    title = "{NLTE effects on kilonova expansion opacities}",
    eprint = "2202.09245",
    archivePrefix = "arXiv",
    primaryClass = "astro-ph.HE",
    doi = "10.1093/mnras/stac1253",
    journal = "MNRAS",
    volume = "513",
    number = "4",
    pages = "5174--5197",
    year = "2022"
}

@article{Polanska:2024arc,
    author = "Polanska, Alicja and Price, Matthew A. and Piras, Davide and Spurio Mancini, Alessio and McEwen, Jason D.",
    title = "{Learned harmonic mean estimation of the Bayesian evidence with normalizing flows}",
    eprint = "2405.05969",
    archivePrefix = "arXiv",
    primaryClass = "astro-ph.IM",
    doi = "10.33232/001c.146026",
    journal = "OJAp",
    volume = "8",
    pages = "-",
    year = "2025"
}

@article{Raaijmakers:2021uju,
    author = "Raaijmakers, G. and Greif, S. K. and Hebeler, K. and Hinderer, T. and Nissanke, S. and Schwenk, A. and Riley, T. E. and Watts, A. L. and Lattimer, J. M. and Ho, W. C. G.",
    title = "{Constraints on the Dense Matter Equation of State and Neutron Star Properties from NICER\textquoteright{}s Mass\textendash{}Radius Estimate of PSR J0740+6620 and Multimessenger Observations}",
    eprint = "2105.06981",
    archivePrefix = "arXiv",
    primaryClass = "astro-ph.HE",
    doi = "10.3847/2041-8213/ac089a",
    journal = "ApJL",
    volume = "918",
    number = "2",
    pages = "L29",
    year = "2021"
}

@article{Radice:2017lry,
    author = "Radice, David and Perego, Albino and Zappa, Francesco and Bernuzzi, Sebastiano",
    title = "{GW170817: Joint Constraint on the Neutron Star Equation of State from Multimessenger Observations}",
    eprint = "1711.03647",
    archivePrefix = "arXiv",
    primaryClass = "astro-ph.HE",
    reportNumber = "LIGO-P1700421, VIR-0894A-17",
    doi = "10.3847/2041-8213/aaa402",
    journal = "ApJL",
    volume = "852",
    number = "2",
    pages = "L29",
    year = "2018"
}

@article{Radice:2018pdn,
    author = "Radice, David and Perego, Albino and Hotokezaka, Kenta and Fromm, Steven A. and Bernuzzi, Sebastiano and Roberts, Luke F.",
    title = "{Binary Neutron Star Mergers: Mass Ejection, Electromagnetic Counterparts and Nucleosynthesis}",
    eprint = "1809.11161",
    archivePrefix = "arXiv",
    primaryClass = "astro-ph.HE",
    doi = "10.3847/1538-4357/aaf054",
    journal = "ApJ",
    volume = "869",
    number = "2",
    pages = "130",
    year = "2018"
}

@article{Ragosta:2024bfr,
    author = "Ragosta, Fabio and Ahumada, Tomas and Piranomonte, Silvia and Andreoni, Igor and Melandri, Andrea and Colombo, Alberto and Coughlin, Michael W.",
    title = "{Kilonova Parameter Estimation with LSST at Vera C. Rubin Observatory}",
    eprint = "2403.14016",
    archivePrefix = "arXiv",
    primaryClass = "astro-ph.HE",
    doi = "10.3847/1538-4357/ad35c1",
    journal = "ApJ",
    volume = "966",
    number = "2",
    pages = "214",
    year = "2024"
}

@article{Rastinejad:2022zbg,
    author = "Rastinejad, Jillian C. and others",
    title = "{A kilonova following a long-duration gamma-ray burst at 350 Mpc}",
    eprint = "2204.10864",
    archivePrefix = "arXiv",
    primaryClass = "astro-ph.HE",
    doi = "10.1038/s41586-022-05390-w",
    journal = "Nature",
    volume = "612",
    number = "7939",
    pages = "223--227",
    year = "2022"
}

@inproceedings{Rezende:2014azh,
    author = "Rezende, Danilo Jimenez and Mohamed, Shakir and Wierstra, Daan",
    title = "{Stochastic Backpropagation and Approximate Inference in Deep Generative Models}",
    booktitle = "{Proceedings of the 31st International Conference on Machine Learning (ICML)}",
    eprint = "1401.4082",
    archivePrefix = "arXiv",
    primaryClass = "stat.ML",
    month = "1",
    year = "2014", 
    adsurl="",
}

@misc{Rezende:2015ocs,
    author = "Rezende, Danilo Jimenez and Mohamed, Shakir",
    title = "{Variational Inference with Normalizing Flows}",
    eprint = "1505.05770",
    archivePrefix = "arXiv",
    primaryClass = "stat.ML",
    month = "5",
    year = "2015"
}

@article{Rinaldi:2024xjz,
    author = "Rinaldi, Enrico and Fraija, Nissim and Dainotti, Maria Giovanna",
    title = "{Parameter Inference of a State-of-the-Art Physical Afterglow Model for GRB 190114C}",
    doi = "10.3390/galaxies12010005",
    journal = "Galaxies",
    volume = "12",
    number = "1",
    pages = "5",
    year = "2024"
}

@article{Ristic:2021ksz,
    author = "Ristic, M. and Champion, E. and O'Shaughnessy, R. and Wollaeger, R. and Korobkin, O. and Chase, E. A. and Fryer, C. L. and Hungerford, A. L. and Fontes, C. J.",
    title = "{Interpolating detailed simulations of kilonovae: Adaptive learning and parameter inference applications}",
    eprint = "2105.07013",
    archivePrefix = "arXiv",
    primaryClass = "astro-ph.HE",
    reportNumber = "LA-UR-21-24289",
    doi = "10.1103/PhysRevResearch.4.013046",
    journal = "Phys. Rev. Res.",
    volume = "4",
    number = "1",
    pages = "013046",
    year = "2022"
}

@article{Ryan:2019fhz,
    author = "Ryan, Geoffrey and van Eerten, Hendrik and Piro, Luigi and Troja, Eleonora",
    title = "{Gamma-Ray Burst Afterglows in the Multimessenger Era: Numerical Models and Closure Relations}",
    eprint = "1909.11691",
    archivePrefix = "arXiv",
    primaryClass = "astro-ph.HE",
    doi = "10.3847/1538-4357/ab93cf",
    journal = "ApJ",
    volume = "896",
    number = "2",
    pages = "166",
    year = "2020"
}

@article{Ryan:2023pzk,
    author = "Ryan, Geoffrey and van Eerten, Hendrik and Troja, Eleonora and Piro, Luigi and O'Connor, Brendan and Ricci, Roberto",
    title = "{Modeling of Long-term Afterglow Counterparts to Gravitational Wave Events: The Full View of GRB 170817A}",
    eprint = "2310.02328",
    archivePrefix = "arXiv",
    primaryClass = "astro-ph.HE",
    doi = "10.3847/1538-4357/ad6a14",
    journal = "ApJ",
    volume = "975",
    number = "1",
    pages = "131",
    year = "2024"
}

@article{Saha:2023zse,
    author = "Saha, Surojit and others",
    title = "{Rapid Generation of Kilonova Light Curves Using Conditional Variational Autoencoder}",
    eprint = "2310.17450",
    archivePrefix = "arXiv",
    primaryClass = "astro-ph.HE",
    doi = "10.3847/1538-4357/ad02f4",
    journal = "ApJ",
    volume = "961",
    number = "2",
    pages = "165",
    year = "2024"
}

@article{Salafia:2022dkz,
    author = "Salafia, Om Sharan and Ghirlanda, Giancarlo",
    title = "{The Structure of Gamma Ray Burst Jets}",
    eprint = "2206.11088",
    archivePrefix = "arXiv",
    primaryClass = "astro-ph.HE",
    doi = "10.3390/galaxies10050093",
    journal = "Galaxies",
    volume = "10",
    number = "5",
    pages = "93",
    year = "2022"
}

@article{Sarin:2021yya,
    author = "Sarin, Nikhil and Ashton, Gregory and Lasky, Paul D. and Ackley, Kendall and Mong, Yik-Lun and Galloway, Duncan K.",
    title = "{CDF-S XT1: The off-axis afterglow of a neutron star merger at $z=2.23$}",
    eprint = "2105.10108",
    archivePrefix = "arXiv",
    primaryClass = "astro-ph.HE",
    journal = "ArXiv e-prints",
    month = "5",
    year = "2021"
}

@article{Sarin:2023khf,
    author = "Sarin, Nikhil and others",
    title = "{redback: a Bayesian inference software package for electromagnetic transients}",
    eprint = "2308.12806",
    archivePrefix = "arXiv",
    primaryClass = "astro-ph.HE",
    doi = "10.1093/mnras/stae1238",
    journal = "MNRAS",
    volume = "531",
    number = "1",
    pages = "1203--1227",
    year = "2024"
}

@article{Sarin:2024tja,
    author = "Sarin, Nikhil and Rosswog, Stephan",
    title = "{Cautionary Tales on Heating-rate Prescriptions in Kilonovae}",
    eprint = "2404.07271",
    archivePrefix = "arXiv",
    primaryClass = "astro-ph.HE",
    doi = "10.3847/2041-8213/ad739d",
    journal = "ApJL",
    volume = "973",
    number = "1",
    pages = "L24",
    year = "2024"
}

@article{Savchenko:2017ffs,
    author = "Savchenko, V. and others",
    title = "{INTEGRAL Detection of the First Prompt Gamma-Ray Signal Coincident with the Gravitational-wave Event GW170817}",
    eprint = "1710.05449",
    archivePrefix = "arXiv",
    primaryClass = "astro-ph.HE",
    doi = "10.3847/2041-8213/aa8f94",
    journal = "ApJL",
    volume = "848",
    number = "2",
    pages = "L15",
    year = "2017"
}

@article{Shappee:2017zly,
    author = "Shappee, B. J. and others",
    title = "{Early Spectra of the Gravitational Wave Source GW170817: Evolution of a Neutron Star Merger}",
    eprint = "1710.05432",
    archivePrefix = "arXiv",
    primaryClass = "astro-ph.HE",
    doi = "10.1126/science.aaq0186",
    journal = "Science",
    volume = "358",
    pages = "1574",
    year = "2017"
}

@article{Shibata:2019wef,
    author = "Shibata, Masaru and Hotokezaka, Kenta",
    title = "{Merger and Mass Ejection of Neutron-Star Binaries}",
    eprint = "1908.02350",
    archivePrefix = "arXiv",
    primaryClass = "astro-ph.HE",
    doi = "10.1146/annurev-nucl-101918-023625",
    journal = "Ann. Rev. Nucl. Part. Sci.",
    volume = "69",
    pages = "41--64",
    year = "2019"
}

@article{Shvartzvald:2023ofi,
    author = "Shvartzvald, Y. and others",
    title = "{ULTRASAT: A Wide-field Time-domain UV Space Telescope}",
    eprint = "2304.14482",
    archivePrefix = "arXiv",
    primaryClass = "astro-ph.IM",
    doi = "10.3847/1538-4357/ad2704",
    journal = "ApJ",
    volume = "964",
    number = "1",
    pages = "74",
    year = "2024"
}

@article{Stratta:2024kbs,
    author = "Stratta, G. and others",
    title = "{The Puzzling Long GRB 191019A: Evidence for Kilonova Light}",
    eprint = "2412.04059",
    archivePrefix = "arXiv",
    primaryClass = "astro-ph.HE",
    doi = "10.3847/1538-4357/ad9b7b",
    journal = "ApJ",
    volume = "979",
    number = "2",
    pages = "159",
    year = "2025"
}

@article{Tanaka:2019iqp,
    author = "Tanaka, Masaomi and Kato, Daiji and Gaigalas, Gediminas and Kawaguchi, Kyohei",
    title = "{Systematic Opacity Calculations for Kilonovae}",
    eprint = "1906.08914",
    archivePrefix = "arXiv",
    primaryClass = "astro-ph.HE",
    doi = "10.1093/mnras/staa1576",
    journal = "MNRAS",
    volume = "496",
    number = "2",
    pages = "1369--1392",
    year = "2020"
}

@article{Tanvir:2013pia,
    author = "Tanvir, N. R. and Levan, A. J. and Fruchter, A. S. and Hjorth, J. and Wiersema, K. and Tunnicliffe, R. and de Ugarte Postigo, A.",
    title = "{A ''kilonova'' associated with short-duration gamma-ray burst 130603B}",
    eprint = "1306.4971",
    archivePrefix = "arXiv",
    primaryClass = "astro-ph.HE",
    doi = "10.1038/nature12505",
    journal = "Nature",
    volume = "500",
    pages = "547",
    year = "2013"
}

@article{Tanvir:2017pws,
    author = "Tanvir, N. R. and others",
    title = "{The Emergence of a Lanthanide-Rich Kilonova Following the Merger of Two Neutron Stars}",
    eprint = "1710.05455",
    archivePrefix = "arXiv",
    primaryClass = "astro-ph.HE",
    doi = "10.3847/2041-8213/aa90b6",
    journal = "ApJL",
    volume = "848",
    number = "2",
    pages = "L27",
    year = "2017"
}

@ARTICLE{Thielemann:2011,
       author = {{Thielemann}, F. -K. and {Arcones}, A. and {K{\"a}ppeli}, R. and {Liebend{\"o}rfer}, M. and {Rauscher}, T. and {Winteler}, C. and {Fr{\"o}hlich}, C. and {Dillmann}, I. and {Fischer}, T. and {Martinez-Pinedo}, G. and {Langanke}, K. and {Farouqi}, K. and {Kratz}, K. -L. and {Panov}, I. and {Korneev}, I.~K.},
        title = "{What are the astrophysical sites for the r-process and the production of heavy elements?}",
      journal = {Progress in Particle and Nuclear Physics},
         year = 2011,
        month = apr,
       volume = {66},
       number = {2},
        pages = {346-353},
          doi = {10.1016/j.ppnp.2011.01.032},
       adsurl = {https://ui.adsabs.harvard.edu/abs/2011PrPNP..66..346T}
}

@article{TOROS:2017pqe,
    author = "D\'\i{}az, M. C. and others",
    collaboration = "TOROS",
    title = "{Observations of the first electromagnetic counterpart to a gravitational wave source by the TOROS collaboration}",
    eprint = "1710.05844",
    archivePrefix = "arXiv",
    primaryClass = "astro-ph.HE",
    doi = "10.3847/2041-8213/aa9060",
    journal = "ApJL",
    volume = "848",
    number = "2",
    pages = "L29",
    year = "2017"
}

@article{Troja:2018ybt,
    author = "Troja, E. and others",
    title = "{A luminous blue kilonova and an off-axis jet from a compact binary merger at z=0.1341}",
    eprint = "1806.10624",
    archivePrefix = "arXiv",
    primaryClass = "astro-ph.HE",
    doi = "10.1038/s41467-018-06558-7",
    journal = "Nature Commun.",
    volume = "9",
    pages = "4089",
    year = "2018"
}

@article{Troja:2018uns,
    author = "Troja, E. and van Eerten, H. and Ryan, G. and Ricci, R. and Burgess, J. M. and Wieringa, M. H. and Piro, L. and Cenko, S. B. and Sakamoto, T.",
    title = "{A year in the life of GW 170817: the rise and fall of a structured jet from a binary neutron star merger}",
    eprint = "1808.06617",
    archivePrefix = "arXiv",
    primaryClass = "astro-ph.HE",
    doi = "10.1093/mnras/stz2248",
    journal = "MNRAS",
    volume = "489",
    number = "2",
    pages = "1919--1926",
    year = "2019"
}

@article{Troja:2019ccb,
    author = "Troja, E. and others",
    title = "{The afterglow and kilonova of the short GRB 160821B}",
    eprint = "1905.01290",
    archivePrefix = "arXiv",
    primaryClass = "astro-ph.HE",
    doi = "10.1093/mnras/stz2255",
    journal = "MNRAS",
    volume = "489",
    number = "2",
    pages = "2104--2116",
    year = "2019",
    note = "[Erratum: MNRAS 490, 4367 (2019)]"
}

@article{Troja:2022yya,
    author = "Troja, E. and others",
    title = "{A nearby long gamma-ray burst from a merger of compact objects}",
    eprint = "2209.03363",
    archivePrefix = "arXiv",
    primaryClass = "astro-ph.HE",
    doi = "10.1038/s41586-022-05327-3",
    journal = "Nature",
    volume = "612",
    number = "7939",
    pages = "228--231",
    year = "2022"
}

@article{Valenti:2017ngx,
    author = "Valenti, Stefano and Sand, David J. and Yang, Sheng and Cappellaro, Enrico and Tartaglia, Leonardo and Corsi, Alessandra and Jha, Saurabh W. and Reichart, Daniel E. and Haislip, Joshua and Kouprianov, Vladimir",
    title = "{The discovery of the electromagnetic counterpart of GW170817: kilonova AT 2017gfo/DLT17ck}",
    eprint = "1710.05854",
    archivePrefix = "arXiv",
    primaryClass = "astro-ph.HE",
    doi = "10.3847/2041-8213/aa8edf",
    journal = "ApJL",
    volume = "848",
    number = "2",
    pages = "L24",
    year = "2017"
}

@article{vanEerten:2011yn,
    author = "van Eerten, H. J. and van der Horst, A. J. and MacFadyen, A. I.",
    title = "{Gamma-ray burst afterglow broadband fitting based directly on hydrodynamics simulations}",
    eprint = "1110.5089",
    archivePrefix = "arXiv",
    primaryClass = "astro-ph.HE",
    doi = "10.1088/0004-637X/749/1/44",
    journal = "ApJ",
    volume = "749",
    pages = "44",
    year = "2012"
}

@article{Villar:2017wcc,
    author = "Villar, V. Ashley and others",
    title = "{The Combined Ultraviolet, Optical, and Near-Infrared Light Curves of the Kilonova Associated with the Binary Neutron Star Merger GW170817: Unified Data Set, Analytic Models, and Physical Implications}",
    eprint = "1710.11576",
    archivePrefix = "arXiv",
    primaryClass = "astro-ph.HE",
    doi = "10.3847/2041-8213/aa9c84",
    journal = "ApJL",
    volume = "851",
    number = "1",
    pages = "L21",
    year = "2017"
}

@article{Wallace:2024pmx,
    author = "Wallace, Wendy F. and Sarin, Nikhil",
    title = "{A detailed dive into fitting strategies for GRB afterglows with contamination: A case study with kilonovae}",
    eprint = "2409.07539",
    archivePrefix = "arXiv",
    primaryClass = "astro-ph.HE",
    doi = "10.1093/mnras/staf623",
    journal = "MNRAS",
    volume = "539",
    number = "4",
    pages = "3319–3335",
    year = "2025"
}

@article{Wang:2020vgr,
    author = "Wang, Hao and Giannios, Dimitrios",
    title = "{Multimessenger parameter estimation of GW170817: from jet structure to the Hubble constant}",
    eprint = "2009.04427",
    archivePrefix = "arXiv",
    primaryClass = "astro-ph.HE",
    doi = "10.3847/1538-4357/abd39c",
    journal = "ApJ",
    volume = "908",
    number = "2",
    pages = "200",
    year = "2021"
}

@article{Wang:2024wbt,
    author = "Wang, Hao and Dastidar, Ranadeep G. and Giannios, Dimitrios and Duffell, Paul C.",
    title = "{jetsimpy: A Highly Efficient Hydrodynamic Code for Gamma-Ray Burst Afterglow}",
    eprint = "2402.19359",
    archivePrefix = "arXiv",
    primaryClass = "astro-ph.HE",
    doi = "10.3847/1538-4365/ad4d9d",
    journal = "ApJS",
    volume = "273",
    number = "1",
    pages = "17",
    year = "2024"
}

@article{Wang:2025ccz,
    author = "Wang, Yihan and Chen, Connery and Zhang, Bing",
    title = "{VegasAfterglow: A high-performance framework for gamma-ray burst afterglows}",
    eprint = "2507.10829",
    archivePrefix = "arXiv",
    primaryClass = "astro-ph.HE",
    doi = "10.1016/j.jheap.2025.100490",
    journal = "JHEAp",
    volume = "50",
    pages = "100490",
    year = "2026"
}

@article{Warren:2021whb,
    author = "Warren, Donald C. and Dainotti, Maria and Barkov, Maxim V. and Ahlgren, Bjorn and Ito, Hirotaka and Nagataki, Shigehiro",
    title = "{A Semianalytic Afterglow with Thermal Electrons and Synchrotron Self-Compton Emission}",
    eprint = "2109.07687",
    archivePrefix = "arXiv",
    primaryClass = "astro-ph.HE",
    reportNumber = "RIKEN-iTHEMS-Report-21",
    doi = "10.3847/1538-4357/ac2f43",
    journal = "ApJ",
    volume = "924",
    number = "1",
    pages = "40",
    year = "2022"
}

@article{Waxman:2019png,
    author = "Waxman, Eli and Ofek, Eran O. and Kushnir, Doron",
    title = "{Late-time Kilonova Light Curves and Implications to GW170817}",
    eprint = "1902.01197",
    archivePrefix = "arXiv",
    primaryClass = "astro-ph.HE",
    doi = "10.3847/1538-4357/ab1f71",
    journal = "ApJ",
    volume = "878",
    number = "2",
    pages = "93",
    year = "2019"
}

@article{Wollaeger:2021qgf,
    author = "Wollaeger, R. T. and Fryer, C. L. and Chase, E. A. and Fontes, C. J. and Ristic, M. and Hungerford, A. L. and Korobkin, O. and O'Shaughnessy, R. and Herring, A. M.",
    title = "{A Broad Grid of 2D Kilonova Emission Models}",
    eprint = "2105.11543",
    archivePrefix = "arXiv",
    primaryClass = "astro-ph.HE",
    reportNumber = "LA-UR-20-30338",
    doi = "10.3847/1538-4357/ac0d03",
    journal = "ApJ",
    volume = "918",
    number = "1",
    pages = "10",
    year = "2021"
}

@article{Wong:2022xvh,
    author = "Wong, Kaze W. k. and Gabri\'e, Marylou and Foreman-Mackey, Daniel",
    title = "{flowMC: Normalizing flow enhanced sampling package for probabilistic inference in JAX}",
    eprint = "2211.06397",
    archivePrefix = "arXiv",
    primaryClass = "astro-ph.IM",
    doi = "10.21105/joss.05021",
    journal = "JOSS",
    volume = "8",
    number = "83",
    pages = "5021",
    year = "2023"
}

@article{Wysocki:2019grj,
    author = "Wysocki, D. and O'Shaughnessy, R. and Lange, Jacob and Fang, Yao-Lung L.",
    title = "{Accelerating parameter inference with graphics processing units}",
    eprint = "1902.04934",
    archivePrefix = "arXiv",
    primaryClass = "astro-ph.IM",
    reportNumber = "LIGO DCC LIGO-P1900036",
    doi = "10.1103/PhysRevD.99.084026",
    journal = "PRD",
    volume = "99",
    number = "8",
    pages = "084026",
    year = "2019"
}

@article{Yang:2022qmy,
    author = {Yang, Jun and Ai, Shunke and Zhang, Bin-Bin and Zhang, Bing and Liu, Zi-Ke and Wang, Xiangyu Ivy and Yang, Yu-Han and Yin, Yi-Han and Li, Ye and L\"u, Hou-Jun},
    title = "{A long-duration gamma-ray burst with a peculiar origin}",
    eprint = "2204.12771",
    archivePrefix = "arXiv",
    primaryClass = "astro-ph.HE",
    doi = "10.1038/s41586-022-05403-8",
    journal = "Nature",
    volume = "612",
    number = "7939",
    pages = "232--235",
    year = "2022"
}

@article{Yang:2023mqt,
    author = "Yang, Yu-Han and others",
    title = "{A lanthanide-rich kilonova in the aftermath of a long gamma-ray burst}",
    eprint = "2308.00638",
    archivePrefix = "arXiv",
    primaryClass = "astro-ph.HE",
    doi = "10.1038/s41586-023-06979-5",
    journal = "Nature",
    volume = "626",
    pages = "742–745",
    year = "2024"
}

@article{Zhang:2020tem,
    author = "Zhang, B. Theodore and Murase, Kohta and Yuan, Chengchao and Kimura, Shigeo S. and M\'esz\'aros, Peter",
    title = "{External Inverse-Compton Emission Associated with Extended and Plateau Emission of Short Gamma-Ray Bursts: Application to GRB 160821B}",
    eprint = "2012.09143",
    archivePrefix = "arXiv",
    primaryClass = "astro-ph.HE",
    doi = "10.3847/2041-8213/abe0b0",
    journal = "ApJL",
    volume = "908",
    number = "2",
    pages = "L36",
    year = "2021"
}

@article{jaxspec,
       author = {{Dupourqu{\'e}}, S. and {Barret}, D. and {Diez}, C.~M. and {Guillot}, S. and {Quintin}, E.},
        title = "{{\textsc{jaxspec}: A fast and robust Python library for X-ray spectral fitting}}",
      journal = {\aap},
         year = 2024,
        month = oct,
       volume = {690},
          eid = {A317},
        pages = {A317},
          doi = {10.1051/0004-6361/202451736},
       adsurl = {https://ui.adsabs.harvard.edu/abs/2024A\&A...690A.317D}
}

@article{kobyzev2020normalizing,
  title={Normalizing flows: An introduction and review of current methods},
  author={Kobyzev, Ivan and Prince, Simon JD and Brubaker, Marcus A},
  journal={IEEE transactions on pattern analysis and machine intelligence},
  volume={43},
  number={11},
  pages={3964--3979},
  year={2020},
  publisher={IEEE}
}

@article{grenander1994representations,
  title={Representations of knowledge in complex systems},
  author={Grenander, Ulf and Miller, Michael I},
  journal={Journal of the Royal Statistical Society: Series B (Methodological)},
  volume={56},
  number={4},
  pages={549--581},
  year={1994},
  publisher={Wiley Online Library}
}

@article{Ristic:2025bvt,
    author = "Risti\'c, Marko and O'Shaughnessy, Richard and Wagner, Kate and Fontes, Christopher J. and Fryer, Chris L. and Korobkin, Oleg and Mumpower, Matthew R. and Wollaeger, Ryan T.",
    title = "{Joint Electromagnetic and Gravitational Wave Inference of Binary Neutron Star Merger GW170817 Using Forward-Modeling Ejecta Predictions}",
    eprint = "2503.12320",
    archivePrefix = "arXiv",
    primaryClass = "astro-ph.HE",
    journal = "ArXiv e-prints",
    reportNumber = "LA-UR-25-22248",
    month = "3",
    year = "2025"
}

@article{Tutone:2025hwh,
    author = "Tutone, A. and others",
    title = "{X-ray spectral fitting with Monte Carlo dropout neural networks}",
    eprint = "2503.09224",
    archivePrefix = "arXiv",
    primaryClass = "astro-ph.IM",
    doi = "10.1051/0004-6361/202553690",
    journal = "A\&A",
    volume = "696",
    pages = "A77",
    year = "2025"
}

@article{Bevins:2025srk,
    author = "Bevins, H. T. J. and Gessey-Jones, T. and Handley, W. J.",
    title = "{On the accuracy of posterior recovery with neural network emulators}",
    eprint = "2503.13263",
    archivePrefix = "arXiv",
    primaryClass = "astro-ph.CO",
    doi = "10.1093/mnras/staf1590",
    journal = "MNRAS",
    note = "in press",
    month = "11",
    year = "2025"
}

@article{Simongini:2024chh,
    author = "Simongini, Andrea and Ragosta, Fabio and Piranomonte, Silvia and Di Palma, Irene",
    title = "{Building spectral templates and reconstructing parameters for core-collapse supernovae with CASTOR}",
    eprint = "2408.03916",
    archivePrefix = "arXiv",
    primaryClass = "astro-ph.HE",
    doi = "10.1093/mnras/stae1911",
    journal = "MNRAS",
    volume = "533",
    number = "3",
    pages = "3053--3067",
    year = "2024"
}

@article{Simongini:2025gse,
    author = "Simongini, Andrea and Ragosta, Fabio and Piranomonte, Silvia and Di Palma, Irene",
    title = "{Core collaApse Supernovae parameTers estimatOR a novel software for data analysis}",
    doi = "10.1051/epjconf/202531913002",
    journal = "EPJ Web Conf.",
    volume = "319",
    pages = "13002",
    year = "2025"
}

@article{Skilling:2006gxv,
    author = "Skilling, John",
    title = "{Nested sampling for general Bayesian computation}",
    doi = "10.1214/06-BA127",
    journal = "Bayesian Analysis",
    volume = "1",
    number = "4",
    pages = "833--859",
    year = "2006"
}

@article{Skilling:2004pqw,
    author = "Skilling, John",
    title = "{Nested Sampling}",
    doi = "10.1063/1.1835238",
    journal = "AIP Conf. Proc.",
    volume = "735",
    number = "1",
    pages = "395",
    year = "2004"
}

@book{Neal:2011mrf,
    author = "Neal, Radford M.",
    title = "{Handbook of Markov Chain Monte Carlo}",
    eprint = "1206.1901",
    archivePrefix = "arXiv",
    primaryClass = "stat.CO",
    doi = "10.1201/b10905",
    month = "5",
    year = "2011"
}

@article{Feroz:2008xx,
    author = "Feroz, F. and Hobson, M. P. and Bridges, M.",
    title = "{MultiNest: an efficient and robust Bayesian inference tool for cosmology and particle physics}",
    eprint = "0809.3437",
    archivePrefix = "arXiv",
    primaryClass = "astro-ph",
    doi = "10.1111/j.1365-2966.2009.14548.x",
    journal = "MNRAS",
    volume = "398",
    pages = "1601--1614",
    year = "2009"
}

@article{Ristic:2023ywr,
    author = "Ristic, Marko and O'Shaughnessy, Richard and Villar, V. Ashley and Wollaeger, Ryan T. and Korobkin, Oleg and Fryer, Chris L. and Fontes, Christopher J. and Kedia, Atul",
    title = "{Interpolated kilonova spectra models: Examining the effects of a phenomenological, blue component in the fitting of AT2017gfo spectra}",
    eprint = "2304.06699",
    archivePrefix = "arXiv",
    primaryClass = "astro-ph.HE",
    doi = "10.1103/PhysRevResearch.5.043106",
    journal = "Phys. Rev. Res.",
    volume = "5",
    number = "4",
    pages = "043106",
    year = "2023"
}

@article{Wong:2023lgb,
    author = "Wong, Kaze W. K. and Isi, Maximiliano and Edwards, Thomas D. P.",
    title = "{Fast Gravitational-wave Parameter Estimation without Compromises}",
    eprint = "2302.05333",
    archivePrefix = "arXiv",
    primaryClass = "astro-ph.IM",
    doi = "10.3847/1538-4357/acf5cd",
    journal = "ApJ",
    volume = "958",
    number = "2",
    pages = "129",
    year = "2023"
}

@article{Wouters:2024oxj,
    author = "Wouters, Thibeau and Pang, Peter T. H. and Dietrich, Tim and Van Den Broeck, Chris",
    title = "{Robust parameter estimation within minutes on gravitational wave signals from binary neutron star inspirals}",
    eprint = "2404.11397",
    archivePrefix = "arXiv",
    primaryClass = "astro-ph.IM",
    doi = "10.1103/PhysRevD.110.083033",
    journal = "PRD",
    volume = "110",
    number = "8",
    pages = "083033",
    year = "2024"
}

@article{Wouters:2025zju,
    author = "Wouters, Thibeau and Pang, Peter T. H. and Koehn, Hauke and Rose, Henrik and Somasundaram, Rahul and Tews, Ingo and Dietrich, Tim and Van Den Broeck, Chris",
    title = "{Leveraging differentiable programming in the inverse problem of neutron stars}",
    eprint = "2504.15893",
    archivePrefix = "arXiv",
    primaryClass = "astro-ph.HE",
    reportNumber = "LA-UR-25-23486",
    doi = "10.1103/v2y8-kxvx",
    journal = "PRD",
    volume = "112",
    number = "4",
    pages = "043037",
    year = "2025"
}

@article{Barret:2024kvc,
    author = "Barret, Didier and Dupourqu\'e, Simon",
    title = "{Simulation-based inference with neural posterior estimation applied to X-ray spectral fitting - Demonstration of working principles down to the Poisson regime}",
    eprint = "2401.06061",
    archivePrefix = "arXiv",
    primaryClass = "astro-ph.IM",
    doi = "10.1051/0004-6361/202449214",
    journal = "A\&A",
    volume = "686",
    pages = "A133",
    year = "2024"
}

@article{Dupourque:2025vqe,
    author = "Dupourqu{\'e}, Simon and Barret, Didier",
    title = "{Simulation-based inference with neural posterior estimation applied to X-ray spectral fitting - II. High-resolution spectroscopy with Athena X-IFU}",
    eprint = "2506.05911",
    archivePrefix = "arXiv",
    primaryClass = "astro-ph.IM",
    doi = "10.1051/0004-6361/202555215",
    journal = "A\&A",
    volume = "699",
    pages = "A179",
    year = "2025"
}

@article{Aksulu:2021crt,
    author = "Aksulu, M. D. and Wijers, R. A. M. J. and van Eerten, H. J. and van der Horst, A. J.",
    title = "{Exploring the GRB population: robust afterglow modelling}",
    eprint = "2106.14921",
    archivePrefix = "arXiv",
    primaryClass = "astro-ph.HE",
    doi = "10.1093/mnras/stac246",
    journal = "MNRAS",
    volume = "511",
    number = "2",
    pages = "2848--2867",
    year = "2022"
}

@article{Aksulu:2020hnl,
    author = "Aksulu, M. D. and Wijers, R. A. M. J. and van Eerten, H. J. and van der Horst, A. J.",
    title = "{A new approach to modelling gamma-ray burst afterglows: Using Gaussian processes to account for the systematics}",
    eprint = "2004.04166",
    archivePrefix = "arXiv",
    primaryClass = "astro-ph.HE",
    doi = "10.1093/mnras/staa2297",
    journal = "MNRAS",
    volume = "497",
    number = "4",
    pages = "4672--4683",
    year = "2020"
}

@article{Ryan:2014nea,
    author = "Ryan, Geoffrey and van Eerten, Hendrik and MacFadyen, Andrew and Zhang, Bin-Bin",
    title = "{Gamma Ray Bursts Are Observed Off-Axis}",
    eprint = "1405.5516",
    archivePrefix = "arXiv",
    primaryClass = "astro-ph.HE",
    doi = "10.1088/0004-637X/799/1/3",
    journal = "ApJ",
    volume = "799",
    number = "1",
    pages = "3",
    year = "2015"
}

@article{Leeney:2025pwe,
    author = "Leeney, Samuel Alan Kossoff",
    title = "{JAX-bandflux: differentiable supernovae SALT modelling for cosmological analysis on GPUs}",
    eprint = "2504.08081",
    archivePrefix = "arXiv",
    journal = "ArXiv e-prints",
    primaryClass = "astro-ph.IM",
    month = "4",
    year = "2025",
    notes = "submitted to JOSS"
}

@software{sncosmo,
  author       = {Barbary, Kyle and
                  Bailey, Stephen and
                  Barentsen, Geert and
                  Barclay, Tom and
                  Biswas, Rahul and
                  Boone, Kyle and
                  Craig, Matt and
                  Feindt, Ulrich and
                  Friesen, Brian and
                  Goldstein, Danny and
                  Jha, Saurabh W. and
                  Jones, David O. and
                  Mondon, Florian and
                  Papadogiannakis, Seméli and
                  Perrefort, Daniel and
                  Pierel, Justin and
                  Rodney, Steve and
                  Rose, Benjamin and
                  Saunders, Clare and
                  Sipőcz, Brigitta and
                  Sofiatti, Caroline and
                  Thomas, Rollin C. and
                  van Santen, Jakob and
                  Vincenzi, Maria and
                  Wang, David and
                  Wood-Vasey, Michael},
  title        = {SNCosmo},
  month        = mar,
  year         = 2025,
  publisher    = {Zenodo},
  version      = {v2.12.1},
  doi          = {10.5281/zenodo.15019859},
  url          = {https://doi.org/10.5281/zenodo.15019859},
  swhid        = {swh:1:dir:335092d3d6716aebad6d052fc6219ae600062484
                   ;origin=https://doi.org/10.5281/zenodo.592747;visi
                   t=swh:1:snp:22689df264588628a8d487675f3a1a7715c77b
                   26;anchor=swh:1:rel:dccc5b60d0073ddc18c513bc1969c0
                   9cba697fdb;path=sncosmo-sncosmo-6c34401
                  },
}

@ARTICLE{Setzer2023,
       author = {{Setzer}, Christian N. and {Peiris}, Hiranya V. and {Korobkin}, Oleg and {Rosswog}, Stephan},
        title = "{Modelling populations of kilonovae}",
      journal = {\mnras},
         year = 2023,
        month = apr,
       volume = {520},
       number = {2},
        pages = {2829-2842},
          doi = {10.1093/mnras/stad257},
archivePrefix = {arXiv},
       eprint = {2205.12286},
 primaryClass = {astro-ph.HE},
       adsurl = {https://ui.adsabs.harvard.edu/abs/2023MNRAS.520.2829S}
}

@article{Speagle:2019ivv,
    author = "Speagle, Joshua S.",
    title = "{dynesty: a dynamic nested sampling package for estimating Bayesian posteriors and evidences}",
    eprint = "1904.02180",
    archivePrefix = "arXiv",
    primaryClass = "astro-ph.IM",
    doi = "10.1093/mnras/staa278",
    journal = "MNRAS",
    volume = "493",
    number = "3",
    pages = "3132--3158",
    year = "2020"
}

@misc{mcewen2023machinelearningassistedbayesian,
      title={Machine learning assisted Bayesian model comparison: learnt harmonic mean estimator}, 
      author={Jason D. McEwen and Christopher G. R. Wallis and Matthew A. Price and Alessio Spurio Mancini},
      year={2023},
      eprint={2111.12720},
      archivePrefix={arXiv},
      primaryClass={stat.ME},
      url={https://arxiv.org/abs/2111.12720}, 
}

@misc{nvidia_h100_2024,
    title = {{NVIDIA H100 Tensor Core GPU}},
    author = {{NVIDIA Corporation}},
    year = {2024},
    url = {https://www.nvidia.com/en-us/data-center/h100/},
    note = {Accessed: 2025-07-17}
  }

@misc{nvidia_rtx6000,
    title = {{NVIDIA RTX 6000 Ada Generation Datasheet}},
    author = {{NVIDIA Corporation}},
    year = {2023},
    url = {https://resources.nvidia.com/en-us-briefcase-for-datasheets/proviz-print-rtx6000-1?ncid=no-ncid},
    note = {Accessed: 2025-07-17}
  }

@misc{intel_xeon_silver_4310_2022,
    title = {{Intel® Xeon® Silver 4310 Processor: 18M Cache, 2.10 GHz}},
    author = {{Intel Corporation}},
    year = {2022},
    url = {https://www.intel.com/content/www/us/en/products/sku/215277/intel-xeon-silver-4310-processor-18m-cache-2-10-ghz/specifications.html},
    note = {Accessed: 2025-07-17}
  }

\begin{appendix}
\section{GRB afterglow tophat jet surrogates}
\label{app:tophat_surrogates}
\begin{figure}
    \centering
    \includegraphics[width=\linewidth]{"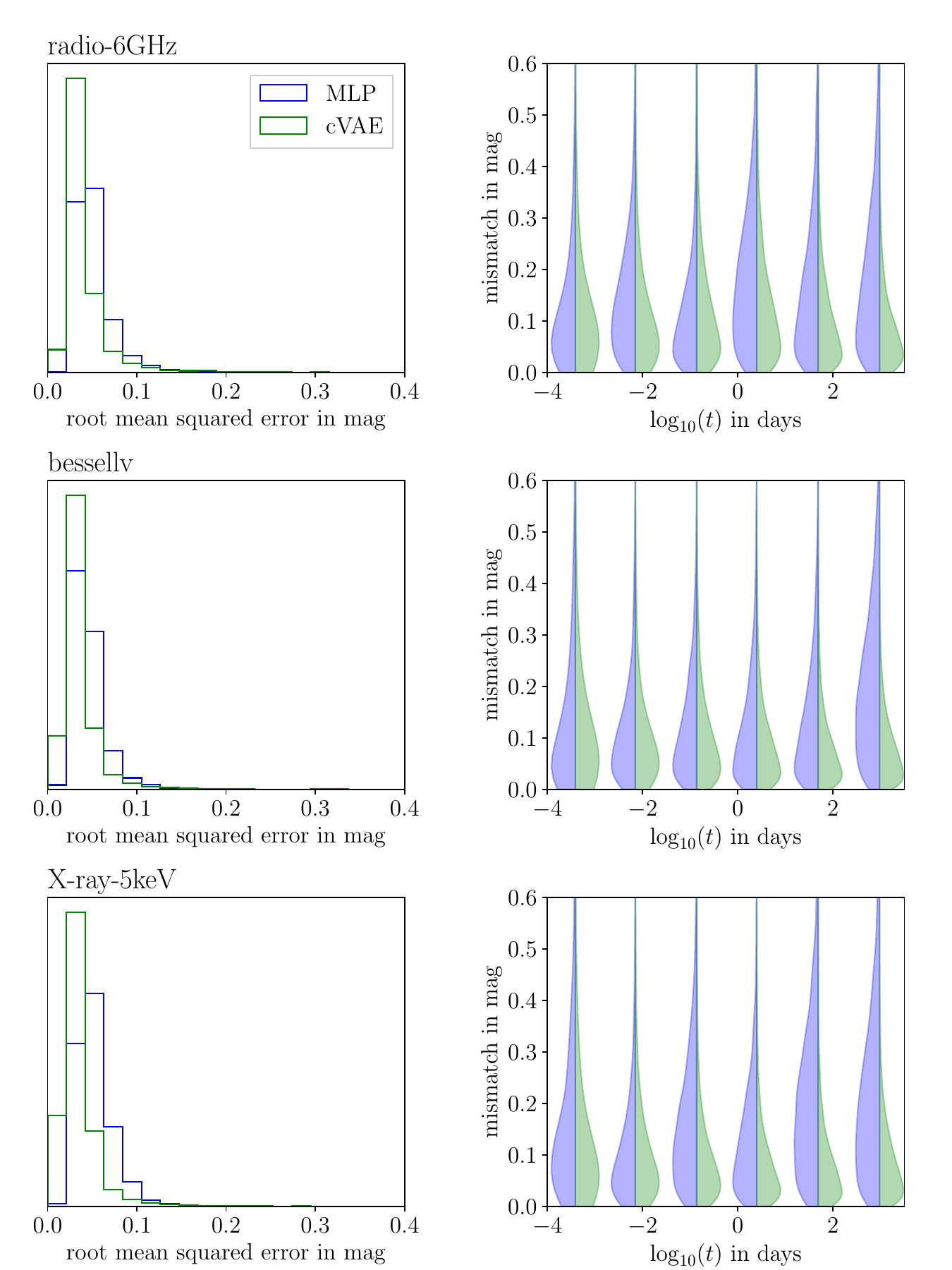"}
    \caption{Benchmarks of the two surrogates for the \afterglowpy tophat jet model. 
    We show the deviations of surrogate predictions against a test data set of size $n_{\text{test}}=7500$.
    Figure layout is the same as in Fig.~\ref{fig:benchmark_afgpy_gaussian}.}
    \label{fig:benchmark_afgpy_tophat}
\end{figure}

\begin{figure}
    \centering
    \includegraphics[width=\linewidth]{"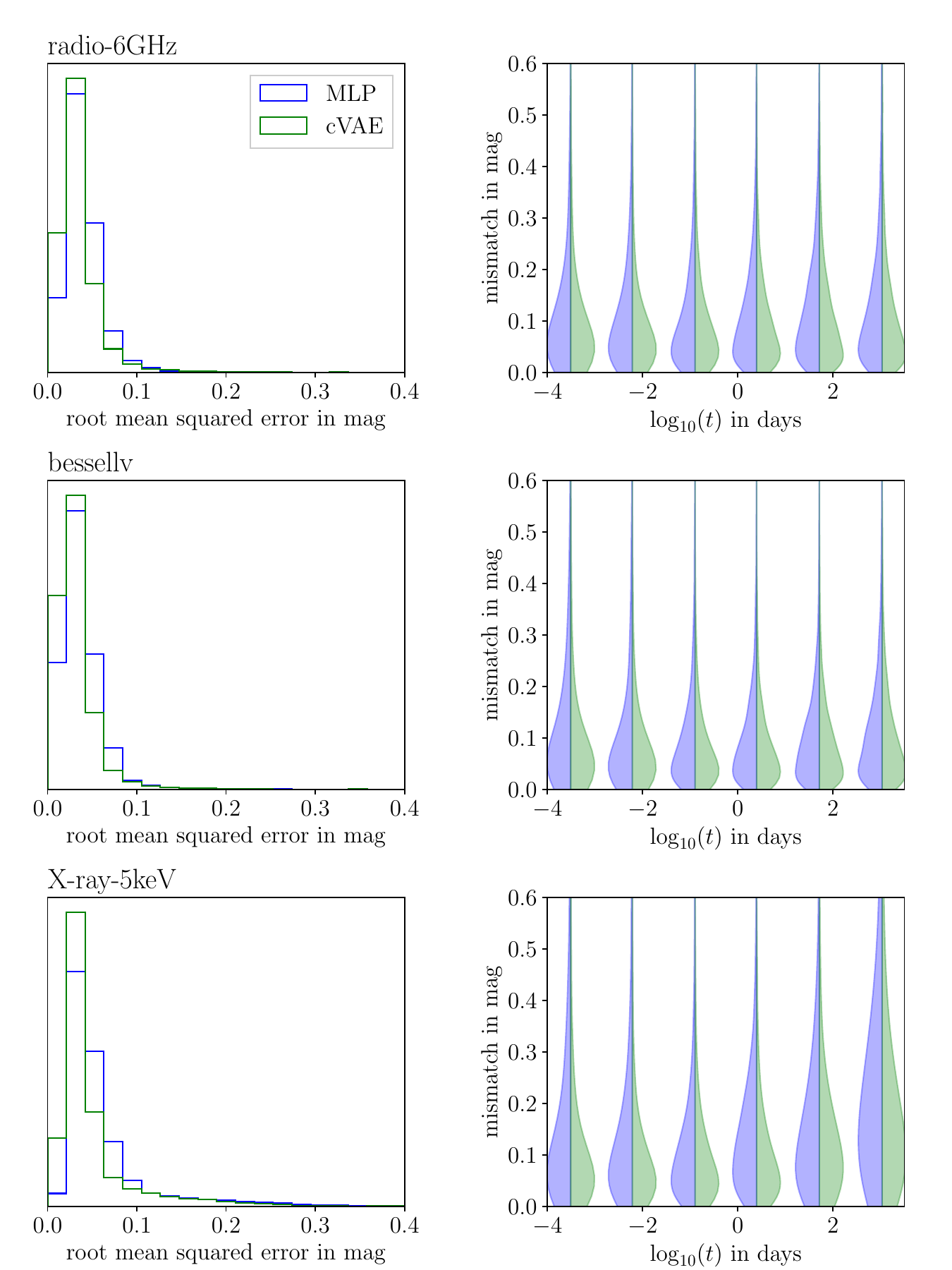"}
    \caption{Benchmarks of the two surrogates for the \pbag tophat jet model. 
    We show the deviations of surrogate predictions against a test data set of size $n_{\text{test}}=7680$.
    Figure layout is the same as in Fig.~\ref{fig:benchmark_afgpy_gaussian}.}
    \label{fig:benchmark_pbag_tophat}
\end{figure}

In this appendix, we present the benchmarks of the surrogates for the tophat jet models of \afterglowpy and \pbag.
Figures~\ref{fig:benchmark_afgpy_tophat} and \ref{fig:benchmark_pbag_tophat} show the distribution of the prediction error of the surrogate against a test data set.
The different architectures are shown again as different colors. 
In general, all trained surrogates perform well, as the typical mean squared error across the entire light curve is $\lesssim 0.1$~mag across all filters.
The only exception is the X-ray filter for \pbag, but this can be attributed again to the hard cut below $2\cdot 10^{-22}$~mJy we imposed on the training data (see Sect.~\ref{subsec:GRB_afterglow_surrogates}).

For \afterglowpy, the MLP surrogate performs noticeably worse than the surrogate with a cVAE architecture, especially at later times.
Considering some example light curves, we find that the reason for this is likely due to the fact that the MLP architecture performs worse when a counterjet causes a late bump in the light curve, whereas the cVAE does not struggle with this feature as much. 
However, the counterjet contribution is also incorporated in the Gaussian models above and in the \pbag tophat model, where the MLP and cVAE perform similarly.
For both \afterglowpy and \pbag, the absolute mismatch across the light curve of the cVAE surrogate is typically confined within 0.2~mag, except again in the X-ray filter for \pbag.
For the \afterglowpy cVAE, 95\% of test predictions are always within 1 mag of the proper light curve; for the \pbag cVAE it is even 97\%.

\section{Injection recoveries for the \pbag and \possis surrogates}
\label{app:pbag_possis_injections}

\begin{figure*}
    \centering
    \includegraphics[width=1\linewidth]{"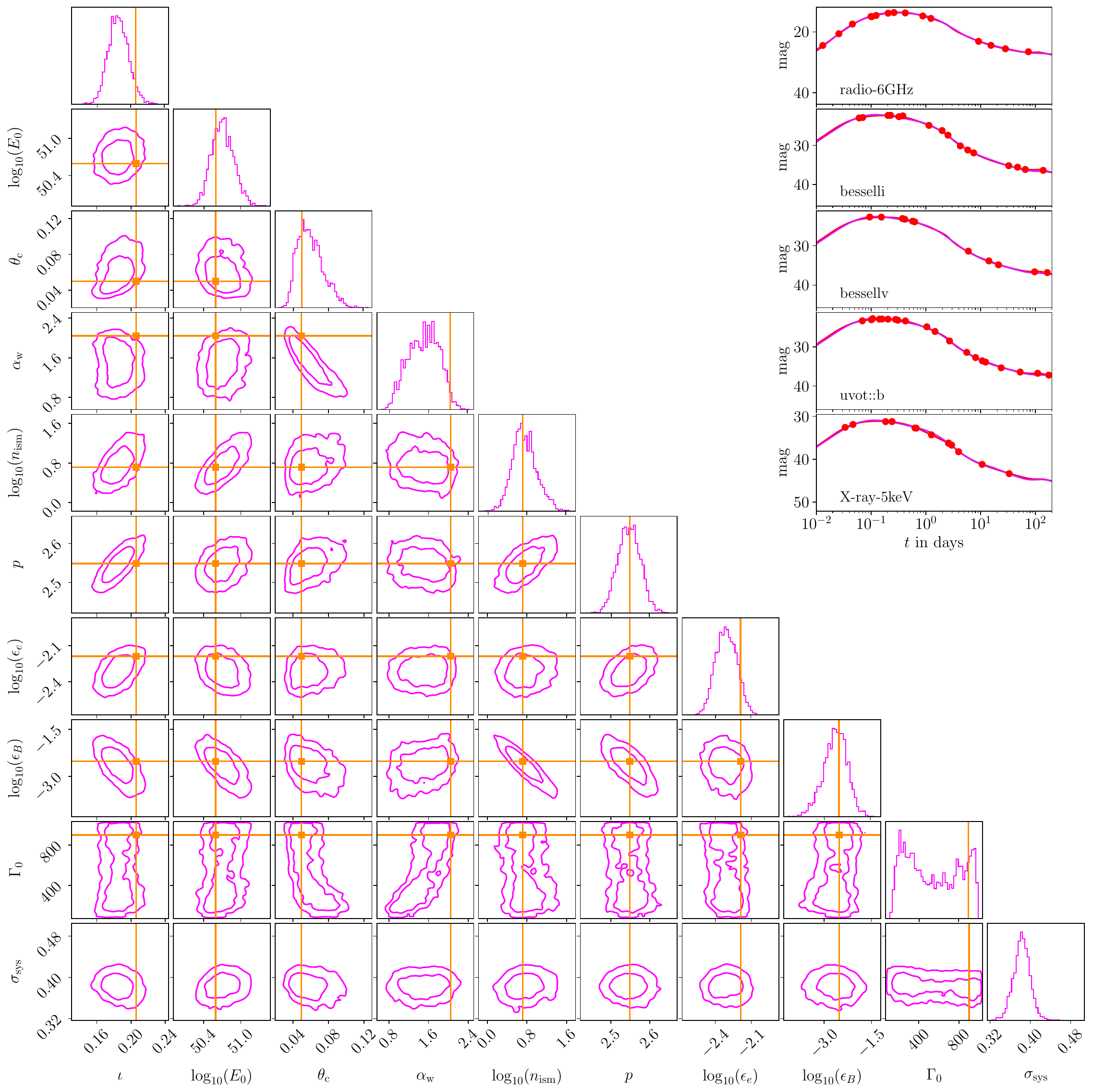"}
    \caption{Parameter recovery for an injected mock light curve from the Gaussian \pbag jet model. Figure layout as in Fig.~\ref{fig:injection_afgpy_gaussian}. 
    The injection is from the \pbag base model, the recovery is with the \fiesta surrogate and the \flowMC sampler.}
    \label{fig:injection_pbag_gaussian}
\end{figure*}

\begin{figure*}
\includegraphics[width=1\linewidth]{"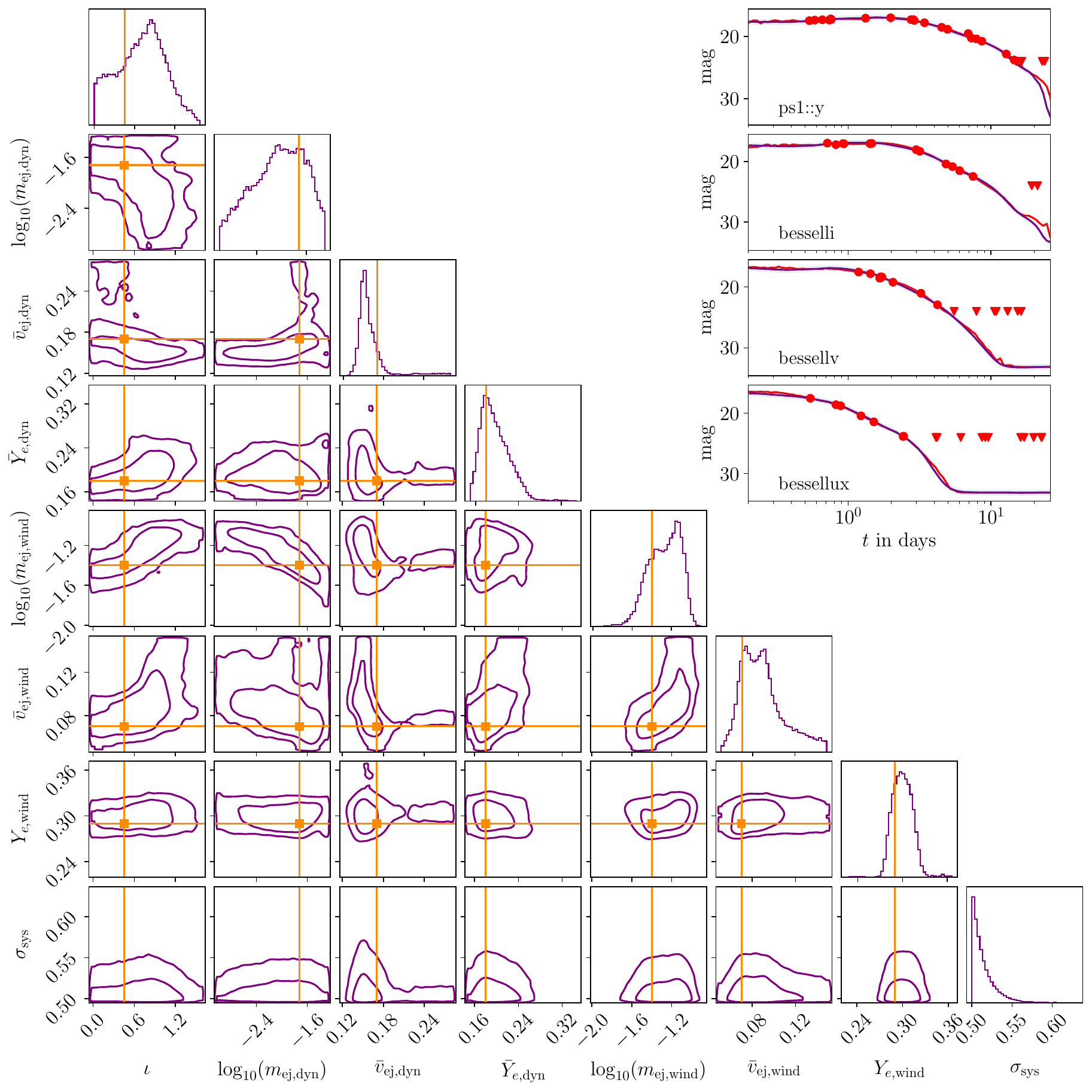"}
    \caption{Parameter recovery for an injected mock light curve from the Gaussian pyblastafterglow jet model. Figure layout as in Fig.~\ref{fig:injection_afgpy_gaussian}. 
    The injection is from the \possis directly, the recovery is with the fiesta surrogate and the flowMC sampler. 
    The insets on the upper right side show the injection data with the best-fit light curve (purple) and the true \possis light curve (red). Triangles indicate upper mock detection limits at 24~mag placed to prevent the surrogate being used on parts of the light curve where the training data is subject to Monte Carlo noise.}
    \label{fig:injection_Bu2025}
\end{figure*}

In Fig.~\ref{fig:injection_pbag_gaussian}, we show the posterior corner plots from an injection with the \pbag afterglow model. 
Likewise, in Fig.~\ref{fig:injection_Bu2025} we show the posterior from an injection with \possis.
The meaning of the respective parameters is given in Table~\ref{tab:surrogate_models}, the priors are uniform within the ranges specified there.
We set the prior for $\sigma_{\rm sys}$ to be uniform from 0.3 to 1~mag in case of the \pbag inference and uniform from 0.5 to 1~mag in case of the \possis inference, since for the latter the surrogate is expected to have slightly larger prediction errors.
We find that the true injected values are well recovered in both cases.
For the recovery of the \pbag injection, we find that $\sigma_{\rm sys}$, sampled from a uniform prior $\mathcal{U}(0.3, 1)$ converges to a value slightly larger than the lower prior bound.
We attribute this to the late bump in the \pbag light curve at 100--150 days that is not accurately accounted for by the surrogate in the optical.
On this interval, the surrogate deviates from the true light curve by about 0.4~mag, which the systematic uncertainty takes into account.
However, this only works because the surrogate matches the other parts of the light curve within 0.3~mag and the larger deviation is confined to a small segment. 
In general, sampling $\sigma_{\rm sys}$ without a sufficiently high prior bound may fail to adequately capture the surrogate prediction error.

\end{appendix}

\begin{acronym}
    \acro{AD}[AD]{automatic differentiation}
    \acro{JIT}[JIT]{just-in-time}
    \acro{PE}[PE]{parameter estimation}
    \acro{MCMC}[MCMC]{Markov chain Monte Carlo}
    \acro{GW}[GW]{gravitational wave}
    \acrodefplural{GWs}{gravitational waves}
    \acro{EM}[EM]{electromagnetic}
    \acro{KN}[KN]{kilonova}
    \acrodefplural{KNe}{kilonovae}
    \acro{GRB}[GRB]{gamma-ray burst}
    \acrodefplural{GRBs}{gamma-ray bursts}
    \acro{CBC}[CBC]{compact binary coalescences}
    \acro{NS}[NS]{neutron star}
    \acrodefplural{NSs}{neutron stars}
    \acro{KDE}[KDE]{kernel density estimate}
    \acro{NF}[NF]{normalizing flow}
    \acro{BBH}[BBH]{binary black hole}
    \acro{BNS}[BNS]{binary neutron star}
    \acro{NSBH}[NSBH]{neutron star-black hole}
    \acro{EOS}[EOS]{equation of state}
    \acro{EFT}[EFT]{effective field theory}
    \acro{chiEFT}[$\chi$EFT]{chiral effective field theory}
    \acro{NEP}[NEP]{nuclear empirical parameter}
    \acro{HIC}[HIC]{heavy-ion collision}
    \acrodefplural{NEPs}{nuclear empirical parameters}
    \acro{MM}[MM]{metamodel}
    \acro{CSE}[CSE]{speed-of-sound extension scheme}
    \acro{TOV}[TOV]{Tolman-Oppenheimer-Volkoff}
    \acro{JS}[JS]{Jensen-Shannon}
    \acro{CPU}[CPU]{central processing unit}
    \acro{GPU}[GPU]{graphical processing unit}
    \acro{TPU}[TPU]{tensor processing unit}
    \acro{ML}[ML]{machine learning}
    \acro{SNR}[SNR]{signal-to-noise ratio}
    \acro{PSD}[PSD]{power spectral density}
    \acro{NICER}[NICER]{Neutron star Interior Composition ExploreR}
    
\end{acronym}
\label{LastPage}
\end{document}